\documentclass[twocolumn]{aastex63}
\usepackage[utf8]{inputenc}
\usepackage{natbib}
\setcitestyle{round}
\usepackage{amsmath,amssymb}
\usepackage{afterpage}
\usepackage{xcolor}
\usepackage[caption=false]{subfig}
\usepackage{soul}
\usepackage{enumitem}
\usepackage{xcolor}

\newcommand{\hi}{{\rm H\,{\small I}}}

\newcommand{\kms}{\ensuremath{\,{\rm km\,s^{-1}}}}

\newcommand{\percc}{\ensuremath{\,{\rm cm^{-3}}}}
\newcommand{\ts}{\ensuremath{T_S}}
\newcommand{\nhi}{\ensuremath{N(\hi{})}}
\newcommand{\persc}{\ensuremath{\,{\rm cm^{-2}}}}
\newcommand{\hcop}{\text{HCO\textsuperscript{+}}}
\newcommand{\hcn}{\text{HCN}}
\newcommand{\hnc}{\text{HNC}}
\newcommand{\cch}{\text{C\textsubscript{2}H}}
\newcommand{\htwo}{\text{H\textsubscript{2}}}
\newcommand{\papertwo}{\text{Paper II}}

\shorttitle{Atomic conditions suitable for molecule formation}
\shortauthors{Rybarczyk et al.}

\begin{document}

\title{The role of neutral hydrogen in setting the abundances of molecular species in the Milky Way's diffuse interstellar medium. I. Observational constraints from ALMA and NOEMA}

\correspondingauthor{Daniel R. Rybarczyk}
\email{rybarczyk@astro.wisc.edu}

\author[0000-0003-3351-6831]{Daniel R. Rybarczyk}
\affiliation{University of Wisconsin--Madison, Department of Astronomy, 475 N Charter St, Madison, WI 53703, USA}

\author[0000-0002-3418-7817]{Sne\v zana Stanimirovi\'c}
\affiliation{University of Wisconsin--Madison, Department of Astronomy, 475 N Charter St, Madison, WI 53703, USA}

\author[0000-0003-1613-6263]{Munan Gong}
\affiliation{Max-Planck-Institut f\"ur Extraterrestrische Physik, Garching by Munich, D-85748, Germany}

\author[0000-0002-6984-5752]{Brian Babler}
\affiliation{University of Wisconsin--Madison, Department of Astronomy, 475 N Charter St, Madison, WI 53703, USA}

\author[0000-0002-7743-8129]{Claire E. Murray}
\affiliation{Department of Physics \& Astronomy, Johns Hopkins University, 3400 N. Charles Street, Baltimore, MD 21218, USA}

\author[0000-0002-2418-7952]{Maryvonne Gerin}
\affiliation{LERMA, Observatoire de Paris, PSL Research University, CNRS, Ecole Normale Sup\'erieure, Sorbonne Universit\'e, 75005 Paris,
France}

\author[0000-0001-6114-9173]{Jan Martin Winters}
\affiliation{Institut de Radioastronomie Millim\'etrique (IRAM), 300 rue de la Piscine,  F-38406 St. Martin d’H\'eres, France}

\author[0000-0002-1583-8514]{Gan Luo}
\affiliation{School of Astronomy and Space Science, Nanjing University, Nanjing 210093, People’s Republic of China}
\affiliation{Key Laboratory of Modern Astronomy and Astrophysics (Nanjing University), Ministry of Education, Nanjing 210093, People’s Republic of China}

\author[0000-0003-0109-2392]{T. M. Dame}
\affiliation{Center for Astrophysics | Harvard \& Smithsonian, 60 Garden St
Cambridge, MA 02138, USA}

\author[0000-0002-9512-5492]{Lucille Steffes}
\affiliation{University of Wisconsin--Madison, Department of Astronomy, 475 N Charter St, Madison, WI 53703, USA}

\begin{abstract}

We have complemented existing observations of \hi{} absorption with new observations of \hcop{}, \cch{}, \hcn{}, and \hnc{} absorption from the Atacama Large Millimeter/submillimeter Array (ALMA) and the Northern Extended Millimeter Array (NOEMA) in the direction of 20 background radio continuum sources with $4^\circ\leq|b|\leq81^\circ$ to constrain the atomic gas conditions that are suitable for the formation of diffuse molecular gas. We find that these molecular species form along sightlines where $A_V\gtrsim0.25$, consistent with the threshold for the \hi{}-to-\htwo{} transition at solar metallicity. 
Moreover, we find that molecular gas is associated only with structures that have an \hi{} optical depth $>0.1$, a spin temperature $<80$ K, and a turbulent Mach number $\gtrsim 2$. 
We also identify a broad, faint component to the \hcop{} absorption in a majority of sightlines. Compared to the velocities where strong, narrow \hcop{} absorption is observed, the \hi{} at these velocities has a lower cold neutral medium (CNM) fraction and negligible CO emission.
The relative column densities and linewidths of the different molecular species observed here are similar to those observed in previous experiments over a range of Galactic latitudes, suggesting that gas in the solar neighborhood and gas in the Galactic plane are chemically similar.
For a select sample of previously-observed sightlines, we show that the absorption line profiles of \hcop{}, \hcn{}, \hnc{}, and \cch{} are stable over periods of $\sim3$ years and $\sim25$ years, likely indicating that molecular gas structures in these directions are at least $\gtrsim100$ AU in size.

\end{abstract}

\section{Introduction} \label{sec:intro}

The formation of molecular gas in the diffuse interstellar medium (ISM) marks the first stages of molecular cloud formation and interstellar chemistry. A variety of processes are believed to mediate molecule formation, including, for example, shielding \citep{2009ApJ...693..216K,Sternberg2014}, shock-driven turbulence \citep{2012ApJ...759...35I}, and turbulent dissipation \citep{2009A&A...495..847G,Lesaffre2020}, but the relative importance of these different effects remains uncertain, from both observational and theoretical perspectives. 

Molecular gas forms within the turbulent, multi-phase atomic ISM. Since at least the early work of \citet{1968SvA....11..737P} and \citet{1969ApJ...155L.149F}, it has been understood that atomic hydrogen (\hi{}) can exist in  a warm, diffuse phase (the warm neutral medium, WNM) and a cold, clumpy phase (the cold neutral medium, CNM). The WNM and CNM have been well characterized theoretically \citep{2003ApJ...587..278W} and observationally in the case of the Milky Way \citep{2003ApJ...586.1067H,2018ApJS..238...14M}. However, observations have also established that $\sim20\%$ of the \hi{} in the Milky Way is in a thermally unstable phase (the unstable neutral medium, UNM), with intermediate temperature and density \citep{2018ApJS..238...14M}. Hydrodynamical and magnetohydrodynamical simulations have shown that CNM structures can form out of the diffuse WNM as the result of dynamical processes like shocks \citep{2002ApJ...564L..97K,2007A&A...465..431H,2012ApJ...759...35I}. Thermal instabilities in shocked gas lead to the condensation of  CNM structures with densities several orders of magnitude higher than typical WNM densities. The smallest ($\lesssim0.1$ pc) and densest of these structures may represent tiny scale atomic structure \citep{2007A&A...465..431H,2020ApJ...893..152R}, an overdense, overpressured component of the atomic ISM observed in high resolution \hi{} absorption measurements \citep[``TSAS'';][and references therein]{1997ApJ...481..193H,2018ARA&A..56..489S}. Theoretical predictions suggest that molecular hydrogen (\htwo{}) forms if the \hi{} piled up behind a shock reaches a sufficiently high column density \citep{2015A&A...580A..49I} and the gas stays cold \citep{2006ApJ...648.1052H}, although this depends on the metallicity \citep[e.g.,][]{2011ApJ...741...12B} and the orientation of the local magnetic field with respect to the shock motion \citep{2012ApJ...759...35I}. The \hi{}-to-\htwo{} transition---where a large fraction of atomic hydrogen is converted to molecular hydrogen---has been observed to occur at a total column density $N\gtrsim5\times10^{20}$ \persc{} at solar metallicity \citep{1977ApJ...216..291S}.

While it is clear that the turbulent, multi-phase ISM is an important ingredient for the formation and survival of molecules, it is still not understood how local properties of \hi{} affect the molecular fraction. Several recent studies have provided hints that underlying physical properties of the \hi{}, such as the level of turbulence and the presence of thermally unstable \hi{}, play an important role.
Observationally, \citet{2014ApJ...793..132S} and \citet{Nguyen2019}  showed that molecular clouds are embedded within atomic gas that has a high CNM fraction relative to random diffuse regions. By modeling the \hi{}-to-\htwo{} transition in Perseus, \citet{2015Bialy} found that a mixture of CNM and UNM (and perhaps WNM) gas was important in controlling the \hi{}-to-\htwo{} transition. 
The width of the transition regions between warm \hi{}, cold \hi{}, and cold \htwo{}---and therefore the local properties of the \hi{}-to-\htwo{} transition---also depends on the level of turbulence \citep{Bialy2017,Lesaffre2007}, and turbulent mixing between the WNM and CNM can enhance the formation of certain molecular species \citep{Lesaffre2007,GloverClark2012}.

In this work, we investigate the early stages of molecule formation in the diffuse ISM and connect this with the underlying properties of atomic gas using new absorption line observations of \hcn{}, \cch{}, \hcop{}, and \hnc{} in the direction of 20 background radio continuum sources where the 21-SPONGE project \citep[21 cm Spectral Line Observations of Neutral Gas with the Karl G. Jansky Very Large Array;][]{2015ApJ...804...89M,2018ApJS..238...14M} previously observed \hi{} in emission and absorption.
In Section \ref{sec:observations}, we present these observations obtained using the Atacama Large Millimeter/submillimeter Array (ALMA) and the Northern Extended Millimeter Array (NOEMA), along with existing observations in these directions, including observations of \hi{} emission from the Arecibo Observatory and \hi{} absorption from the Very Large Array (VLA) obtained by the 21-SPONGE project \citep{2015ApJ...804...89M,2018ApJS..238...14M}, maps of interstellar redenning and dust temperature from the \textit{Planck} satellite \citep{2014A&A...571A..11P,2014A&A...571A..13P}, and observations of CO emission from the \citet{2001ApJ...547..792D} survey. 
21-SPONGE targeted mainly sources at Galactic latitude $>10^\circ$, where \hi{} spectra are simpler and radiative transfer calculations are easier than in the Galactic plane. The sensitivity of our molecular line spectra rival or exceed most pre-ALMA surveys \citep[e.g.,][]{1996A&A...307..237L,2000A&A...358.1069L,2001A&A...370..576L}, and our sample represents one of the largest homogeneous samples of Galactic absorption measurements at mm wavelengths to date.
We outline methods for extracting molecular column densities and decomposing absorption spectra into Gaussian components in Section \ref{sec:methods}. In Section \ref{sec:thresholds}, we test how the observed molecular column densities depend on the line of sight gas properties, including the extinction and the CNM and UNM column densities. We also investigate a broad, faint component of the \hcop{} absorption seen in the direction of most background sources. In Section \ref{sec:hi_molecules}, we establish thresholds for the \hi{} optical depth, spin temperature, and turbulent Mach number required for the onset of molecule formation. We then determine the relative abundances and linewidths of the four different molecular species observed in this work and compare our results with previous works in Section \ref{sec:species_comparison}. In Section \ref{sec:temporal_stability}, we investigate the temporal stability of absorption line profiles by comparing our results to previous observations of \hcn{}, \cch{}, \hcop{}, and \hnc{} for select lines of sight. This serves as a probe for AU scale structure. Finally, we discuss our results in Section \ref{sec:discussion} and present conclusions in Section \ref{sec:conclusions}. 

This is the first of two complementary papers. In Rybarczyk et al. (submitted; Paper II hereafter) we compare the observational results presented here to predictions from the \citet{Gong2017} photodissociation region (PDR) chemical model and the \citet{Gong2020} ISM simulations. 

\section{Observations} \label{sec:observations}

\begin{deluxetable*}{cccc}
\tablecaption{The molecular line transitions covered by the ALMA-SPONGE spectral setup. \hcn{} and \hcop{} were also covered by the NOEMA-SPONGE spectral setup. Column 1 lists the molecular species. Column 2 lists each transition. Column 3 lists the rest frequency for each transition. Column 4 gives the conversion between the optical depth integral and the column density for each species (see Section \ref{subsec:tau_to_N}); the values listed in Column 4 account for all of the observed transitions for each species and are calculated assuming an excitation temperature equal to the CMB temperature, 2.725 K. \label{tab:lines}}
\tablehead{\colhead{Species} & \colhead{Transition} & \colhead{Frequency} & \colhead{$N/\int\tau dv$} \\
& & GHz & \persc{}/\kms{}}
\startdata
\cch{} & $N=1-0$ $J=3/2-1/2$ $F=2-1$ & $87.3169$ & $4.34\times10^{13}$\\
\cch{} & $N=1-0$ $J=3/2-1/2$ $F=1-0$ & $87.3286$ & \\
\hcn{} & $J=1-0$ $F=1-1$ & $88.6304$ & $1.91\times10^{12}$\\
\hcn{} & $J=1-0$ $F=2-1$ & $88.6318$ &  \\
\hcn{} & $J=1-0$ $F=0-1$ & $88.6339$ &  \\
\hcop{} & $J=1-0$ & $89.1890$ & $1.11\times10^{12}$\\
\hnc{} & $J=1-0$ & $90.6336$ & $1.78\times10^{12}$\\
\enddata
\end{deluxetable*}

\subsection{Observations with ALMA (ALMA-SPONGE)} \label{subsec:alma-sponge}
We have observed \cch{}, \hcn{}, \hcop{}, and \hnc{} in absorption with ALMA during observing Cycles 6 and 7 (ALMA-SPONGE projects 2018.1.00585.S and 2019.1.01809.S, PI: Stanimirovic) in the direction of 20 bright background radio continuum sources previously observed at 21 cm wavelength using the VLA by the 21-SPONGE project. Observations were obtained between October 2018 and March 2020. The transitions covered by our observations are listed in Table \ref{tab:lines}. 89 GHz fluxes ranged from 0.03 Jy to 11.8 Jy. The \hcop{} and \hnc{} spectra were obtained at 30.5 kHz frequency channel spacing (0.1 \kms{} velocity channel spacing) while the \hcn{} and \cch{} spectra were obtained at 61 kHz frequency channel spacing (0.2 \kms{} velocity channel spacing).
In the analysis that follows, all spectra have been smoothed to $0.4$ \kms{} velocity resolution, comparable to that of the 21-SPONGE \hi{} absorption spectra.

With the exception of 3C111 and 3C123, all of the ALMA-SPONGE background radio continuum sources are unresolved and treated as single point sources. 3C111 is a three-component radio galaxy; we have independent observations of all three components, separated by $1\arcmin$--$3\arcmin$, each of which is unresolved. Although the 1420 MHz fluxes vary by less than a factor of 3 across the three components, the 89 GHz fluxes change by over two orders of magnitude. We do not consider the spectra in the direction of 3C111C hereafter, as they lack adequate sensitivity to accurately measure molecular abundances ($\sigma_\tau\gtrsim 1$). The optical depth noise of the \cch{} and \hnc{} spectra are $\sim0.3$ for 3C111B, the highest in our sample. As discussed in the next section, we have obtained more sensitive \hcop{} and \hcn{} absorption spectra in the direction of 3C111A and 3C111B with NOEMA. 
3C123 is also a multiple-component continuum source. \citet{2018ApJS..238...14M} resolved 3C123 into two components, 3C123A and 3C123B. At ALMA's higher resolution, we resolve 3C123 into four components. Because we make a direct comparison between the 21-SPONGE \hi{} absorption spectra and our molecular absorption spectra, though, we regrid our ALMA observations in the direction of 3C123 using the VLA  beam (20.3\arcsec{} $\times$ 5.3\arcsec{}) from the \citet{2018ApJS..238...14M} observations before extracting the absorption spectra. As in \citet{2018ApJS..238...14M}, we resolve only two components at this lower spatial resolution, which are here referred to as 3C123A and 3C123B to be consistent with their labeling. While the spectra toward 3C123A have a typical optical depth noise of 0.01, the optical depth noise in the 3C123B spectra is $\sim0.1$.

For sources without significant absorption, we extracted the spectra from the continuum-subtracted data cubes output by the default ALMA Science Pipeline at the pixel of peak continuum flux. To calculate upper limits to the optical depth, we added the continuum flux back to the continuum-subtracted spectra (correcting for the spectral index; for a majority of sightlines, the continuum is relatively flat). For sightlines that showed significant absorption,
we reran the ALMA Science Pipeline without continuum subtraction and extracted spectra from the pixel of peak continuum flux to ensure the most reliable estimate of the continuum flux and the optical depths. Except in the cases noted above, at $0.4$ \kms{} velocity resolution, we reach a typical optical depth noise of $\sim 0.01$ in the ALMA-SPONGE spectra.

Table \ref{tab:N_meas} lists the Galactic coordinates and integrated optical depths in the direction of the 19 ALMA-SPONGE bright background radio continuum sources (excluding 3C111C). 10 sightlines show no molecular absorption at a level of $3\sigma$. 6 sightlines show strong absorption across all four molecular species. In the direction of 3C111B, we detect \hnc{} absorption but no \cch{} absorption (\hcop{} and \hcn{} are observed with NOEMA; see next section). This most likely reflects the poor sensitivity in the direction of 3C111B---the peak \cch{} optical depth in the direction of 3C111A is 0.766, which is only 2.6 times the noise level in the 3C111B spectrum. In the direction of 3C78, we see weak \hcop{} absorption, but no \hcn{}, \cch{}, or \hnc{} absorption. In the direction of J2136, we see weak \cch{} absorption, but no \hcn{}, \hcop{}, or \hnc{} absorption. Although the detections in the direction of 3C123B, 3C78, and J2136 are marginal ($\sim3\sigma$), they appear significant at lower velocity resolution. Moreover, the absorption in these directions is aligned in velocity with \hi{} absorption features. We also see two hyperfine components for J2136 and we know from the 3C123A spectra that molecular gas is present toward 3C123 at the velocities where we detect marginal absorption in the direction of 3C123B. The spatial distribution of all ALMA-SPONGE sources is shown in Figure \ref{fig:FullSkyCO}, overlaid on a map of CO integrated emission measured by \citet{2014A&A...571A..13P}.

\subsection{Observations with NOEMA (NOEMA-SPONGE)} \label{subsec:noema-sponge}
We observed \hcop{} and \hcn{} in absorption at 62.5 kHz channel spacing (0.2 \kms{} velocity spacing) with NOEMA (NOEMA-SPONGE, projects W19AQ and S20AB) in the direction of 3C111---both A and B components---and BL Lac. 3C111 was observed in January 2020 and BL Lac was observed October 2020. The measured fluxes were 1.13 Jy for 3C111A, 0.06 Jy for 3C111B, and 2.10 Jy for BL Lac. Both \hcop{} and \hcn{} were placed in high resolution chunks in the lower side band, which was tuned to 90 GHz. Standard calibration was carried out using the CLIC and MAPPING software, part of the GILDAS software collection\footnote{https://www.iram.fr/IRAMFR/GILDAS} \citep{2005sf2a.conf..721P,2013ascl.soft05010G}. Because BL Lac and 3C111A were bright, we further performed self-calibration using the 2020 Self Calibration tool in MAPPING. We smooth the final spectra to 0.4 \kms{} velocity resolution, reaching an optical depth sensitivity of $\sim0.002$ for 3C111A and BL Lac and an optical depth sensitivity of $\sim0.05$ for 3C111B. The NOEMA-SPONGE sensitivity in the direction of 3C111B is significantly better than the ALMA-SPONGE sensitivity; only the NOEMA-SPONGE \hcop{} and \hcn{} data in this direction are considered hereafter. The \hcop{} and \hcn{} optical depth integrals in these directions are listed in Table \ref{tab:N_meas} (where the NOEMA-SPONGE fluxes are superscripted by a ``1'') and their positions are shown in Figure \ref{fig:FullSkyCO}. 

\subsection{Additional data sets}

\subsubsection{21-SPONGE} \label{subsec:21-sponge}
The 21-SPONGE project \citep[][]{2015ApJ...804...89M,2018ApJS..238...14M} obtained \hi{} absorption spectra using the VLA and \hi{} emission spectra using the Arecibo Observatory in the direction of 57 bright background radio continuum sources, ranging in latitude from $|b|=3.7^\circ$ to $|b|=81^\circ$. The absorption spectra had an optical depth noise of $\sim10^{-3}$ at the 0.4 \kms{} velocity resolution. \citet{2015ApJ...804...89M,2018ApJS..238...14M} decomposed the absorption and emission spectra into Gaussian components. They estimated the spin temperatures, \ts{}, and \hi{} column densities, \nhi{}, of the \hi{} structures seen in absorption. They also investigated the fraction of gas in the CNM, the WNM, and the UNM phases along each line of sight. We use their absorption spectra and adopt their Gaussian fits in this work. The 21-SPONGE \hi{} absorption spectra are shown in Figure \ref{fig:allspectra}.

\subsubsection{$E(B-V)$ and $T_d$ from Planck} \label{subsec:planck}

The total hydrogen column density, $N_\mathrm{H}=N(\hi{})+2N(\htwo{})$, can be inferred from the interstellar reddening, $E(B-V)$, along the line of sight. We use the dust radiance, $\mathcal{R}$, measured by \textit{Planck} to estimate $E(B-V)$. $\mathcal{R}$ is a more reliable predictor of $E(B-V)$ than the 353 \micron{} optical depth, $\tau_{353}$, at high Galactic latitude \citep{2014A&A...571A..11P}\footnote{We find only negligible differences between the $E(B-V)$ estimates made from $\mathcal{R}$ and $\tau_{353}$ for the sightlines in this work.}. We extract $E(B-V)$ at the pixel nearest to each background source using the dustmaps Python package \citep{2018JOSS....3..695M}. The \textit{Planck} dust maps have a resolution of 5\arcmin{} and a typical fractional uncertainty of $\sim10\%$ in $E(B-V)$. The results are listed in Table \ref{tab:N_meas}. We also extract the dust temperature, $T_d$, derived by \citet{2016A&A...596A.109P} using the generalized needlet internal linear combination (GNILC) method and a modified blackbody spectral model on \textit{Planck} temperature maps at 353, 545, 857, and 3000 GHz.
We use dust temperature as a proxy for the strength of the interstellar radiation field (ISRF)---the strength of the FUV radiation field scales as $T_d^{\beta+4}$, where $\beta$ is the power law index characterizing the dust emissivity cross-section as a function of frequency \citep[see Equation 7.15 of][]{2005ism..book.....L}. To estimate the ISRF we also extract the corresponding $\beta$ values at each of our positions.

In several places in this paper we use values of $E(B-V)$, $T_d$, and the ISRF derived from \textit{Planck} as estimates of the gas properties along the lines of sight. While this provides a uniform approach, because \textit{Planck}'s resolution is much lower than that of our \hi{} and molecular pencil-beam absorption spectra, we consider these only as rough estimates and not precise measures of the local physical conditions.

\begin{deluxetable*}{ccccccccc}
\tablecaption{
Observed properties of the sightlines in this study. \textbf{Col. 1}: Name of the background radio continuum source. \textbf{Col. 2}: The 90 GHz continuum flux.  \textbf{Col. 3}: Galactic longitude and latitude of the background source in degrees. \textbf{Col. 4}: Total \hi{} column density measured by 21-SPONGE along the line of sight \citep{2018ApJS..238...14M}. \textbf{Col. 5}: $E(B-V)$ inferred from the dust radiance measured by \textit{Planck} (Planck Collaboration et al. 2014). \textbf{Cols. 6--9}: The integrated optical depth for all four species for channels with $\tau>3\sigma_\tau$. Lower limits are indicated. As described in Section \ref{sec:observations}, BL Lac was not observed in \cch{} or \hnc{} absorption.  \label{tab:N_meas}
}
\tablehead{\colhead{Source} & $F_{90}$  & $l/b$ & \colhead{$N({\hi{}})$} & \colhead{$E(B-V)$} &
\colhead{$\int\tau_\hcn{}dv$} & \colhead{$\int\tau_\cch{}dv$} & 
\colhead{$\int\tau_\hcop{}dv$} & \colhead{$\int\tau_\hnc{}dv$} \\
\colhead{} & $\rm{Jy}$ & $^\circ$ & \colhead{$10^{20}$ \persc{}} & \colhead{mag} & \colhead{\kms{}} & \colhead{\kms{}} & \colhead{\kms{}} & \colhead{\kms{}}}
\startdata
3C154  & 0.390    & $185.6$/$4.0$ & $37.12$ & $0.425$ & $3.325 \pm 0.039$ & $1.079 \pm 0.025$ & $3.059 \pm 0.056$ & $0.819 \pm 0.023$ \\
3C111A  & 0.769/1.145\textsuperscript{1}       & $161.7$/$-8.8$ & $25.96$ & $0.827$ & $>12.710$ & $2.057 \pm 0.033$ & $>11.054$ & $3.606 \pm 0.129$ \\
3C111B & 0.034/0.060\textsuperscript{1}    &  $161.7$/$-8.8$ & $24.86$ & $0.827$ & $>11.345$ & $\cdot\cdot\cdot$  & $>6.780$ & $>2.028$ \\
BLLac & 2.189    & $92.6$/$-10.4$ & $18.68$ & $0.350$ & $4.316\pm0.007$ &  & $2.546\pm0.002$ & \\
3C138  & 0.470        & $187.4$/$-11.3$ & $20.51$ & $0.263$ & $\cdot\cdot\cdot$ & $\cdot\cdot\cdot$ & $\cdot\cdot\cdot$ & $\cdot\cdot\cdot$ \\
3C123A  & 0.451      & $170.6$/$-11.7$ & $17.33$ & $0.499$ & $6.029 \pm 0.171$ & $1.650 \pm 0.041$ & $6.280 \pm 0.234$ & $1.100 \pm 0.041$ \\
3C123B  & 0.270       & $170.6$/$-11.7$ & $18.66$ & $0.499$ & $>4.287$ & $0.331 \pm 0.189$ & $>4.071$ & $1.222 \pm 0.965$ \\
PKS0742 & 0.304    & $209.8$/$16.6$ & $3.37$ & $0.031$ & $\cdot\cdot\cdot$ & $\cdot\cdot\cdot$ & $\cdot\cdot\cdot$ & $\cdot\cdot\cdot$ \\
3C120  & 2.431        & $190.4$/$-27.4$ & $13.33$ & $0.275$ & $0.031 \pm 0.003$ & $0.178 \pm 0.006$ & $0.283 \pm 0.006$ & $0.006 \pm 0.002$ \\
J2136 & 2.063         & $55.5$/$-35.6$ & $4.34$ & $0.080$ & $\cdot\cdot\cdot$ & $0.012 \pm 0.002$ & $\cdot\cdot\cdot$ & $\cdot\cdot\cdot$ \\
3C346 & 0.107         & $35.3$/$35.8$ & $5.42$ & $0.082$ & $\cdot\cdot\cdot$ & $\cdot\cdot\cdot$ & $\cdot\cdot\cdot$ & $\cdot\cdot\cdot$ \\
3C454.3 & 11.838       & $86.1$/$-38.2$ & $7.03$ & $0.117$ & $0.112 \pm 0.004$ & $0.106 \pm 0.004$ & $0.276 \pm 0.004$ & $0.024 \pm 0.002$ \\
3C345 & 2.489        & $63.5$/$40.9$ & $0.75$ & $0.021$ & $\cdot\cdot\cdot$ & $\cdot\cdot\cdot$ & $\cdot\cdot\cdot$ & $\cdot\cdot\cdot$ \\
4C15.05 & 0.596      & $147.9$/$-44.0$ & $4.06$ & $0.057$ & $\cdot\cdot\cdot$ & $\cdot\cdot\cdot$ & $\cdot\cdot\cdot$ & $\cdot\cdot\cdot$ \\
3C78  & 0.347         & $174.9$/$-44.5$ & $10.65$ & $0.162$ & $\cdot\cdot\cdot$ & $\cdot\cdot\cdot$ & $0.043 \pm 0.012$ & $\cdot\cdot\cdot$ \\
1055+018 & 5.283      & $251.5$/$52.8$ & $3.05$ & $0.037$ & $\cdot\cdot\cdot$ & $\cdot\cdot\cdot$ & $\cdot\cdot\cdot$ & $\cdot\cdot\cdot$ \\
3C273  & 8.668        & $289.9$/$64.4$ & $2.16$ & $0.028$ & $\cdot\cdot\cdot$ & $\cdot\cdot\cdot$ & $\cdot\cdot\cdot$ & $\cdot\cdot\cdot$ \\
4C12.50 & 0.507     & $347.2$/$70.2$ & $2.18$ & $0.054$ & $\cdot\cdot\cdot$ & $\cdot\cdot\cdot$ & $\cdot\cdot\cdot$ & $\cdot\cdot\cdot$ \\
3C286  & 0.671       & $56.5$/$80.7$ & $1.21$ & $0.016$ & $\cdot\cdot\cdot$ & $\cdot\cdot\cdot$ & $\cdot\cdot\cdot$ & $\cdot\cdot\cdot$ \\
4C32.44  & 0.229     & $67.2$/$81.0$ & $1.36$ & $0.021$ & $\cdot\cdot\cdot$ & $\cdot\cdot\cdot$ & $\cdot\cdot\cdot$ & $\cdot\cdot\cdot$ \\

\enddata
$^1$ As described in Section \ref{sec:observations}, 3C111A and 3C111B were observed by both ALMA (\cch{}, \hnc{}) and NOEMA (\hcn{}, \hcop{}); both fluxes are listed, with the flux for the NOEMA data superscripted by a ``1.''
\end{deluxetable*}

\begin{figure*}
    \centering
    \includegraphics[width=\linewidth]{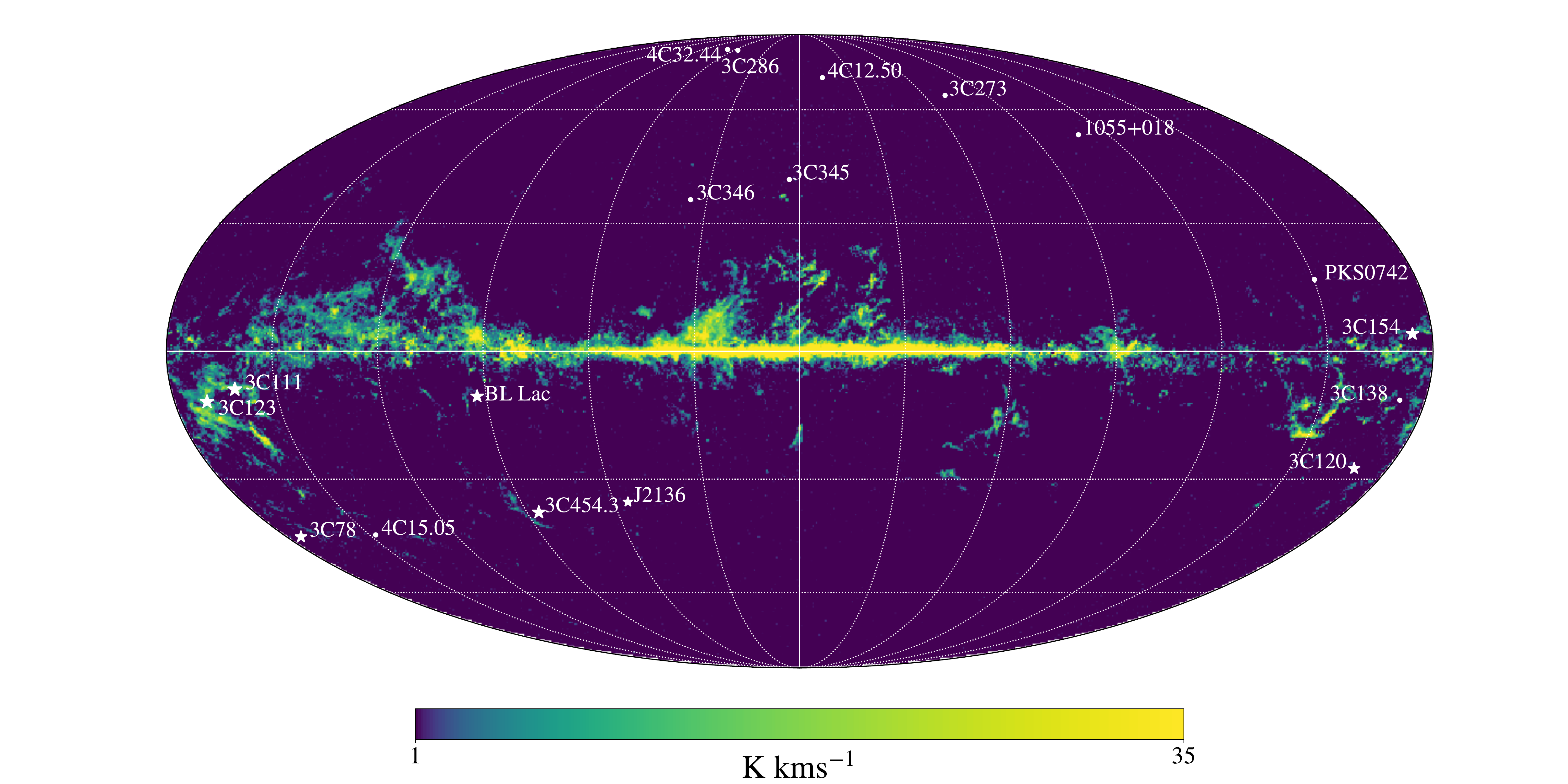}
    \caption{The distribution of background radio continuum sources observed in this work plotted over a full sky map of the CO integrated intensity from \citet{2014A&A...571A..13P}. Sources with molecular absorption detections at a level $\geq3\sigma$ are shown as filled stars; sources with no molecular absorption detections are shown as filled circles. The integrated intensity of the CO emission is shown with a logarithmic color scale, ranging from 1 K\,\kms{} to 35 K\,\kms{}. The different components of the multiple-component continuum sources 3C111 and 3C123 are indistinguishable at this resolution so are not labeled separately (see Table \ref{tab:N_meas}). Dashed lines show Galactic latitude and longitude with $30^\circ$ spacing.}
    \label{fig:FullSkyCO}
\end{figure*}

\subsection{CO emission spectra from the CfA 1.2m telescope}
\citet{2001ApJ...547..792D} obtained CO emission spectra across the entire Galactic plane and select clouds at higher latitudes using the CfA 1.2m telescope with a $8.4\arcmin$ beamwidth. For sources covered by this CO survey, we extract CO emission spectra from the nearest pixel in the \citet{2001ApJ...547..792D} maps, sampled between every other beamwidth ($0.25^\circ$) and every half beamwidth ($0.0625^\circ$) in the directions of our sources. The brightness temperature noise in these spectra ranges from 0.15 K to 0.30 K. For sources outside of the bounds covered by \citet{2001ApJ...547..792D}, we use unpublished spectra obtained with the CfA 1.2m telescope at $8.4\arcmin$ beamwidth and $0.25^\circ$ sampling.
The brightness temperature noise in these unpublished spectra is uniform, 0.18 K. All of the CO emission spectra taken from \citet{2001ApJ...547..792D} have a velocity resolution of $0.65$ \kms{}.

\section{Methods}
\label{sec:methods}

\subsection{Deriving Column Densities}
\label{subsec:tau_to_N}
For a ground state transition, the column density, $N$, and optical depth integral, $\int\tau dv$, for a particular species are related by 
\begin{equation} \label{eq:N}
    N = Q(T_{\mathrm{ex}})\frac{8\pi\nu^3}{c^3}\frac{1}{g_u A_{ul}}\Big[1-\exp(-h\nu/kT_{\mathrm{ex}})\Big]^{-1} \int\tau dv,
\end{equation}
where $\nu$ is the frequency of radiation resulting from the transition from the upper state $u$ to the lower state $l$, $g_u$ is the degeneracy of the upper state, $A_{ul}$ is the Einstein $A$ coefficient for the transition, $T_{\mathrm{ex}}$ is the excitation temperature, and $Q(T_{\mathrm{ex}})$ is the partition function.
We calculate $N/\int\tau dv$ using constants given in the Cologne Database for Molecular Spectroscopy \citep[CDMS;][]{2001A&A...370L..49M,2016JMoSp.327...95E} and the Leiden Atomic and Molecular Database \citep[LAMBDA;][]{2010ascl.soft10077S}. For \cch{}, we use $N/\int\tau dv = 1.6\times 2.71\times10^{13}$ $\persc{}/\kms{}$, provided by \cite{2000A&A...358.1069L}, where the factor of 1.6 accounts for the fact that we are measuring two of the six satellite \cch{} lines. All values of $N/\int\tau dv$ are calculated assuming an excitation temperature is equal to the temperature of the CMB, 2.725 K \citep[e.g.,][]{2010A&A...520A..20G,2020ApJ...889L...4L}\footnote{In \papertwo{} we investigate this assumption and show that the \hcop{} $(1-0)$ excitation temperature begins to rise above the CMB temperature when $n\gtrsim300$ \percc{}.}. All conversions are listed in Table \ref{tab:lines}; the \cch{} and \hcn{} conversions account for all hyperfine transitions listed in Table \ref{tab:lines}.

For saturated channels where the measured value of $I(v)\leq0$---and therefore $\tau(v)$ is infinite---we define the optical depth lower limit to be
\begin{equation} \label{eq:tau_ll}
    \tau_{ll}(v) = -\ln\Big(\frac{2\sigma_I}{I_0}\Big),
\end{equation}
where $\sigma_I$ is the noise in the spectrum of specific intensity. We then calculate a lower limit to the column density (Equation \ref{eq:N}) using $\tau(v)=\tau_{ll}(v)$ for the saturated channels.
We note that this systematically underestimates the optical depth for spectra with saturated absorption. In the analysis that follows, we therefore consider the optical depths and column densities only for spectra where the absorption is not saturated, unless otherwise noted. The optical depth integrals for saturated absorption spectra in Table \ref{tab:N_meas} should be considered as conservative lower limits.

In Section \ref{sec:thresholds}, we estimate the total hydrogen column density, $N_\mathrm{H}$, from the reddening, $E(B-V)$, according to \citet{2017MNRAS.471.3494Z}, 
\begin{equation} \label{eq:NH}
 N_\mathrm{H}=2.08\times10^{21}\times3.1 E(B-V) \text{ }\mathrm{cm^{-2}\,mag^{-1}},
\end{equation}
where $3.1E(B-V)$ is here used for the visual extinction, $A_V$. This estimate was derived from a compilation of Galactic sources with optical and X-ray observations, and is consistent with other previous estimates \citep{1974ApJ...187..243J,1978ApJ...224..132B,2009ApJ...693..216K,2009ApJS..180..125R} for sources spanning a wide range of Galactic latitudes and $N_\mathrm{H}$. Nevertheless, the ratio of $N_\mathrm{H}$ to $E(B-V)$ is not universal, and contributes additional uncertainty to the determination of $N_\mathrm{H}$---see Section \ref{sec:species_comparison} for discussion.

\subsection{Gaussian Decomposition} \label{subsec:gauss_decomp}

We decompose the molecular optical depth spectra into Gaussian functions,
\begin{equation}
    \label{eq:tau_g}
    \tau(v) = \sum_{i=1}^{M} \tau_i(v) = \sum_{i=1}^{M} \tau_{0,i}e^{-4\ln{2}(v-v_{0,i})^2/\Delta v_{0,i}^2},
\end{equation}
where $\tau_{0,i}$, $v_{0,i}$, and $\Delta v_{0,i}$ are the peak optical depth, the central velocity, and the full width at half maximum (FWHM) of the $i$\textsuperscript{th} Gaussian component, respectively, and there are $M$ Gaussian functions fit to the optical depth spectrum. 
Lucas, Liszt, Gerin, and collaborators have identified dozens of molecular absorption components in common across various molecular species, including \hcn{}, \cch{}, \hcop{}, and \hnc{}. Further, they have found that the central velocities of absorption components identified at these transitions are nearly identical, and that
FWHMs vary across different species by only $\sim10$--$30\%$.
Therefore, for each line of sight, we find the best solution to Equation \ref{eq:tau_g} for the \cch{}, \hcn{}, \hcop{}, and \hnc{} optical depth spectra using Python's \texttt{scipy.optimize.curve\_fit} from sets of initial guesses with the same number of components (modulo an integer factor accounting for the number of hyperfine transitions). The initial guesses for the FWHM and central velocity of each component are the same for all four optical depth spectra. The initial guesses for the central velocities of the different hyperfine \cch{} and \hcn{} components are offset according to the known frequency separation of the transitions, and the initial guesses for the optical depths are scaled to the LTE ratios ($2:1$ and $5:3:1$, respectively). For 3C111B and 3C123B, we use the same initial guesses at 3C111A and 3C123A, where the spectra are more sensitive. When fitting, we allow the peak optical depths to freely vary, the FWHMs to vary by a factor of two from the initial guess, and the central velocity to vary by $\pm1$ \kms{} from the initial guess. Only solutions with peak optical depths $>3\sigma_\tau$ are classified as detections\footnote{\hnc{} lines in the direction of 3C111B are detected at $<3\sigma$, but are included for comparison to 3C111A}. For \cch{} and \hcn{}, we require only that the strongest transition be detected at a level of $3\sigma$.

The Gaussian-fitted components to the absorption spectra are shown in Figure \ref{fig:allspectra}. Tables \ref{tab:cch_fits}, \ref{tab:hcn_fits}, \ref{tab:hcop_fits}, and \ref{tab:hnc_fits} list the fits for \cch{}, \hcn{}, \hcop{}, and \hnc{}, respectively. Peak optical depths range between  $0.008$ and $2.5$, excluding saturated lines with optical depths $\gtrsim3.5$. FWHMs range between $0.5$ \kms{} and $3.4$ \kms{}. For 10 sightlines, no features are detected (see Table \ref{tab:N_meas}). There is one feature identified in \cch{} absorption in the direction of J2136 for which there is no corresponding \hcn{}, \hcop{}, or \hnc{} absorption; and there is one feature identified in \hcop{} absorption in the direction of 3C78 for which there is no corresponding \hcn{}, \cch{}, or \hnc{} absorption. Two features are seen in \cch{} and \hcop{} absorption but not seen in \hnc{} or \hcn{} absorption. Three absorption features identified in \cch{}, \hcop{}, and \hcn{} are not detected in \hnc{}.

\begin{figure*}[h!]
\gridline{\fig{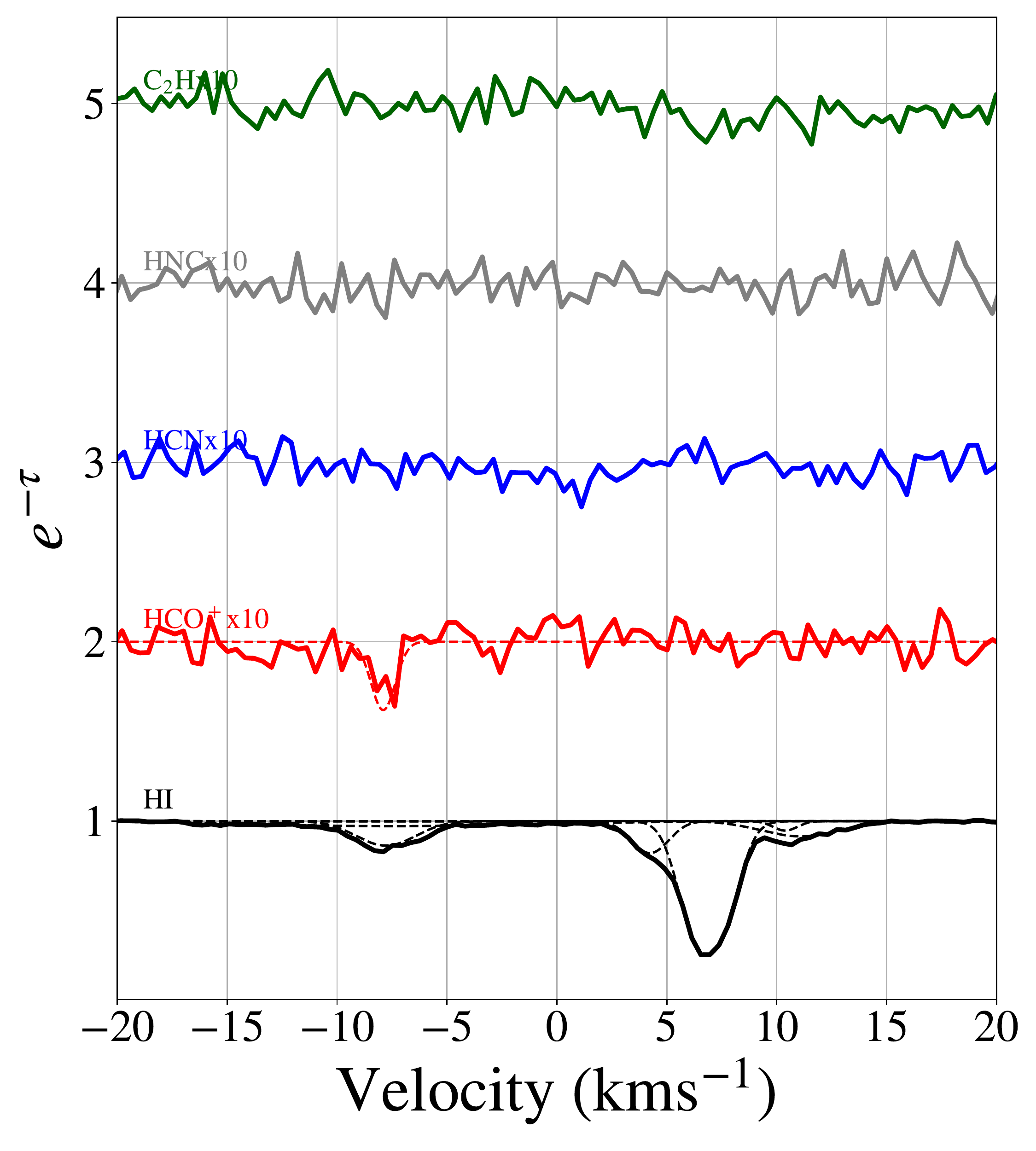}{0.3\textwidth}{3C78}
\fig{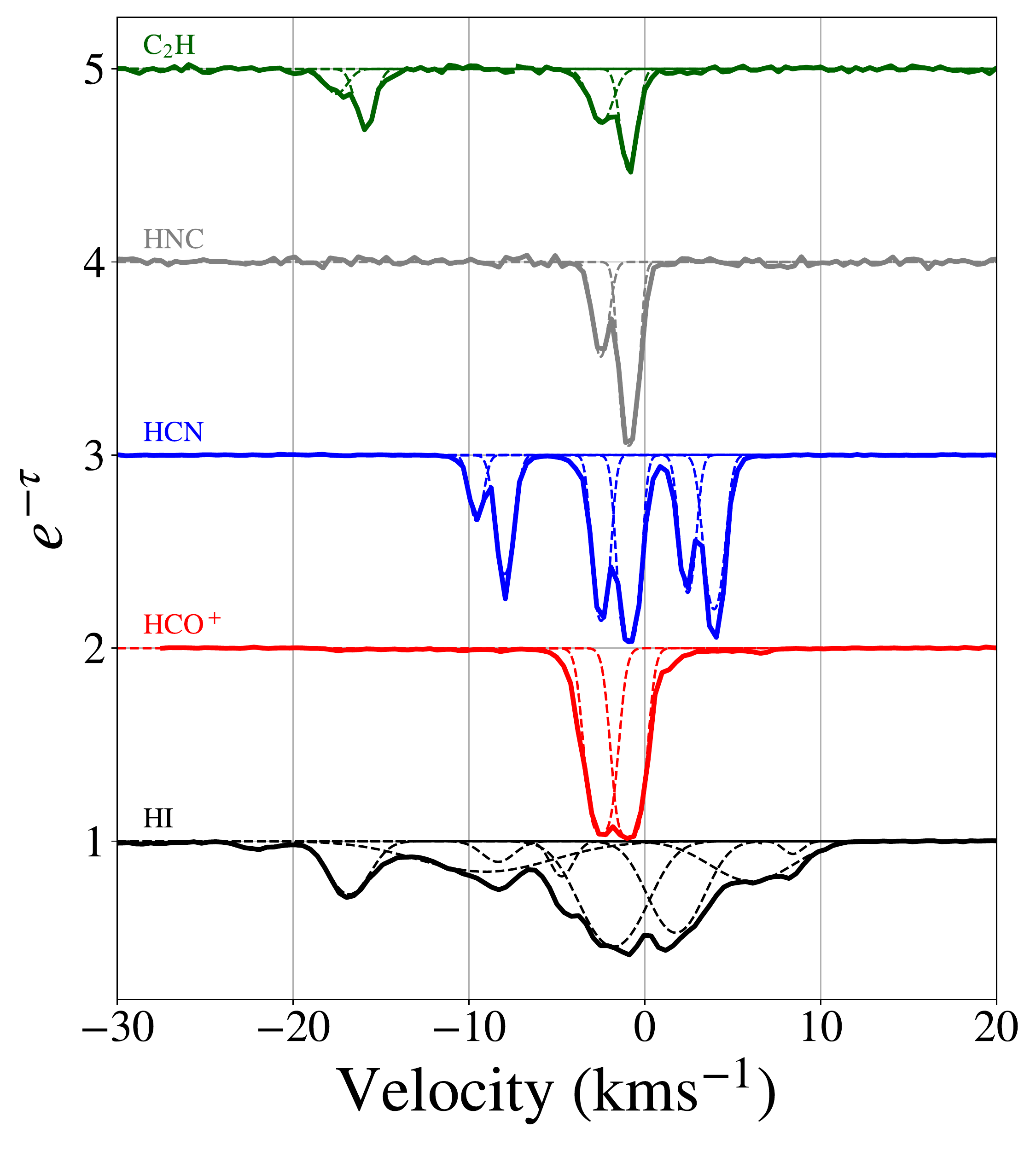}{0.3\textwidth}{3C111A}
\fig{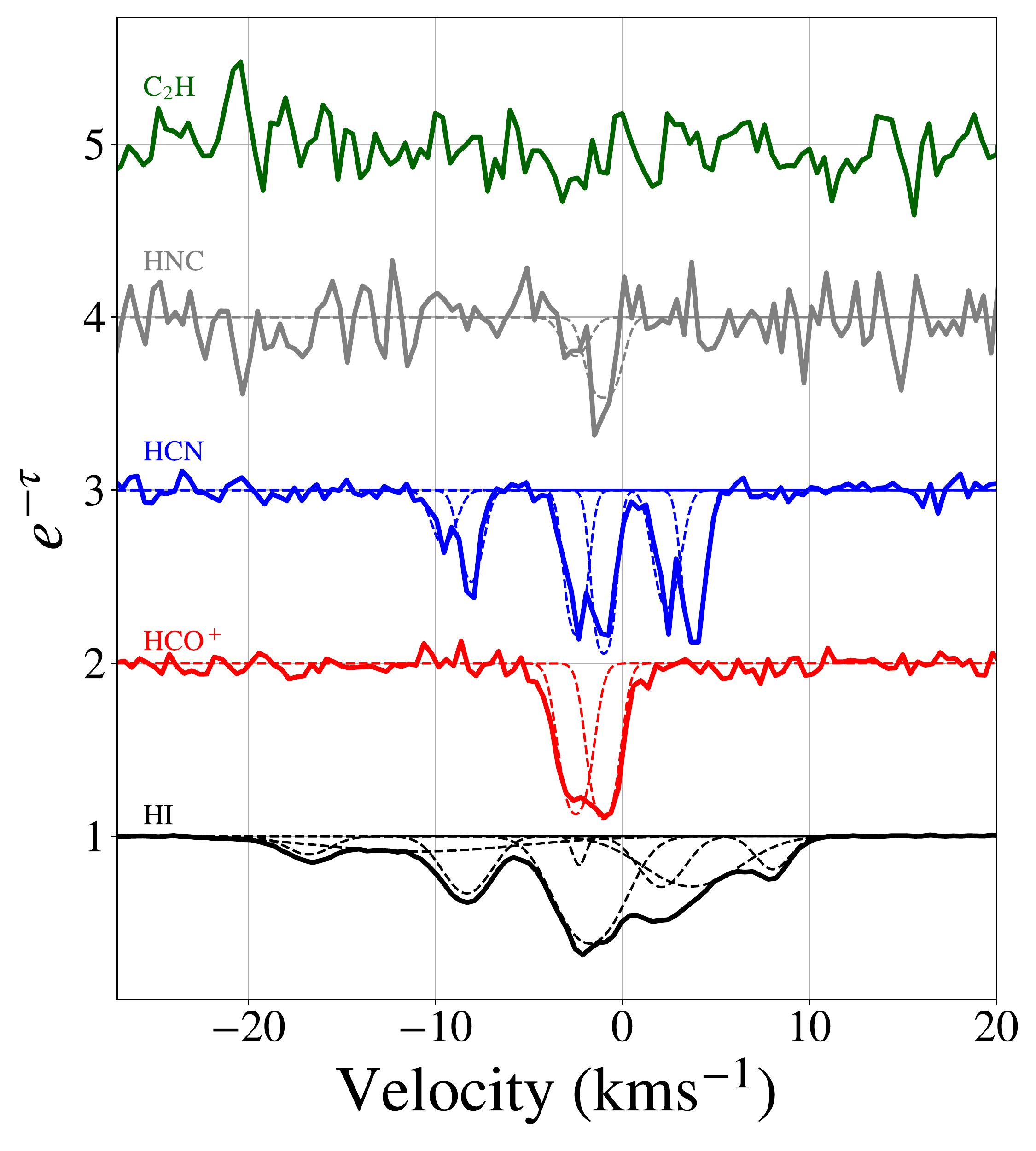}{0.3\textwidth}{3C111B}}
\gridline{\fig{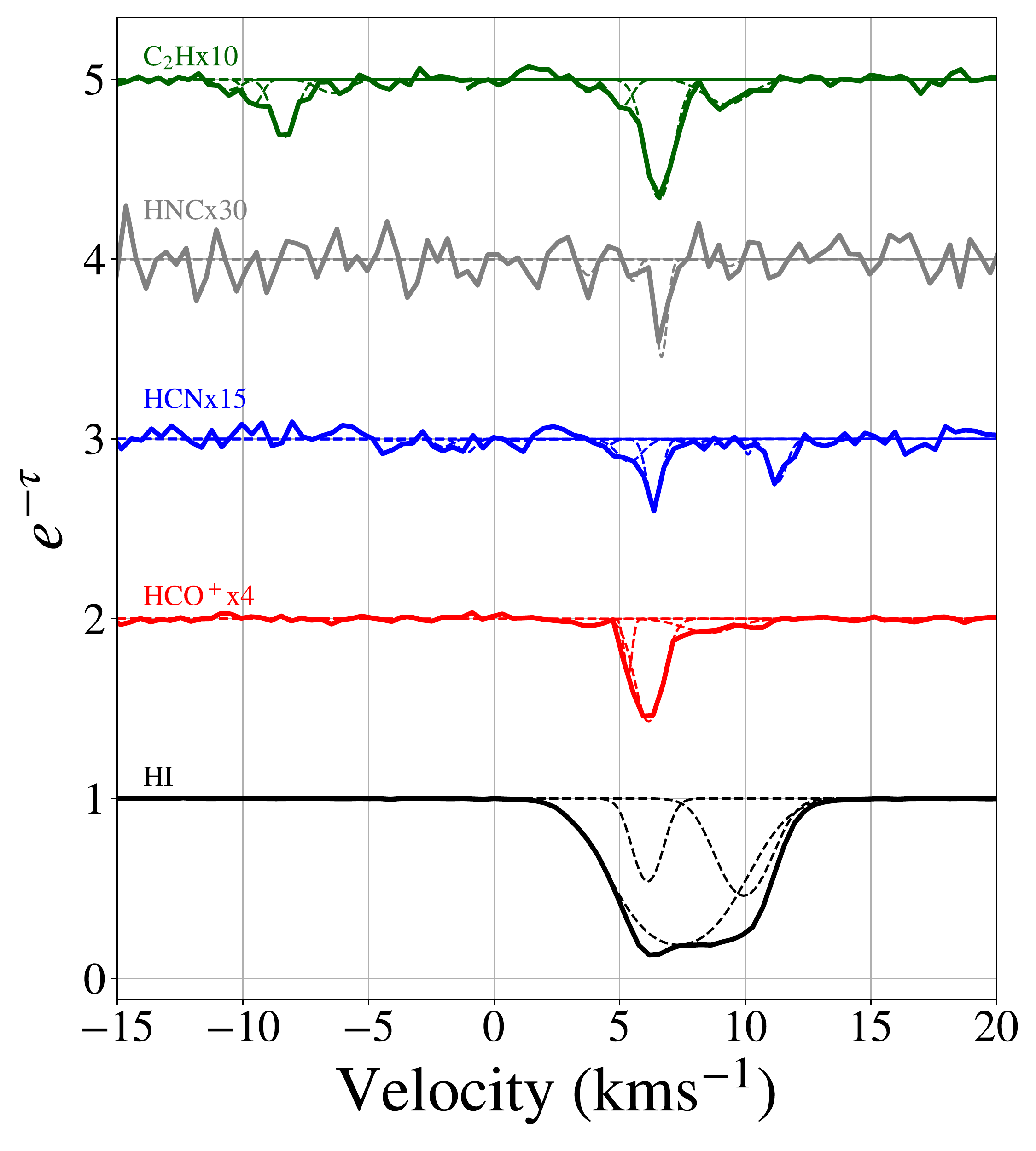}{0.3\textwidth}{3C120}
\fig{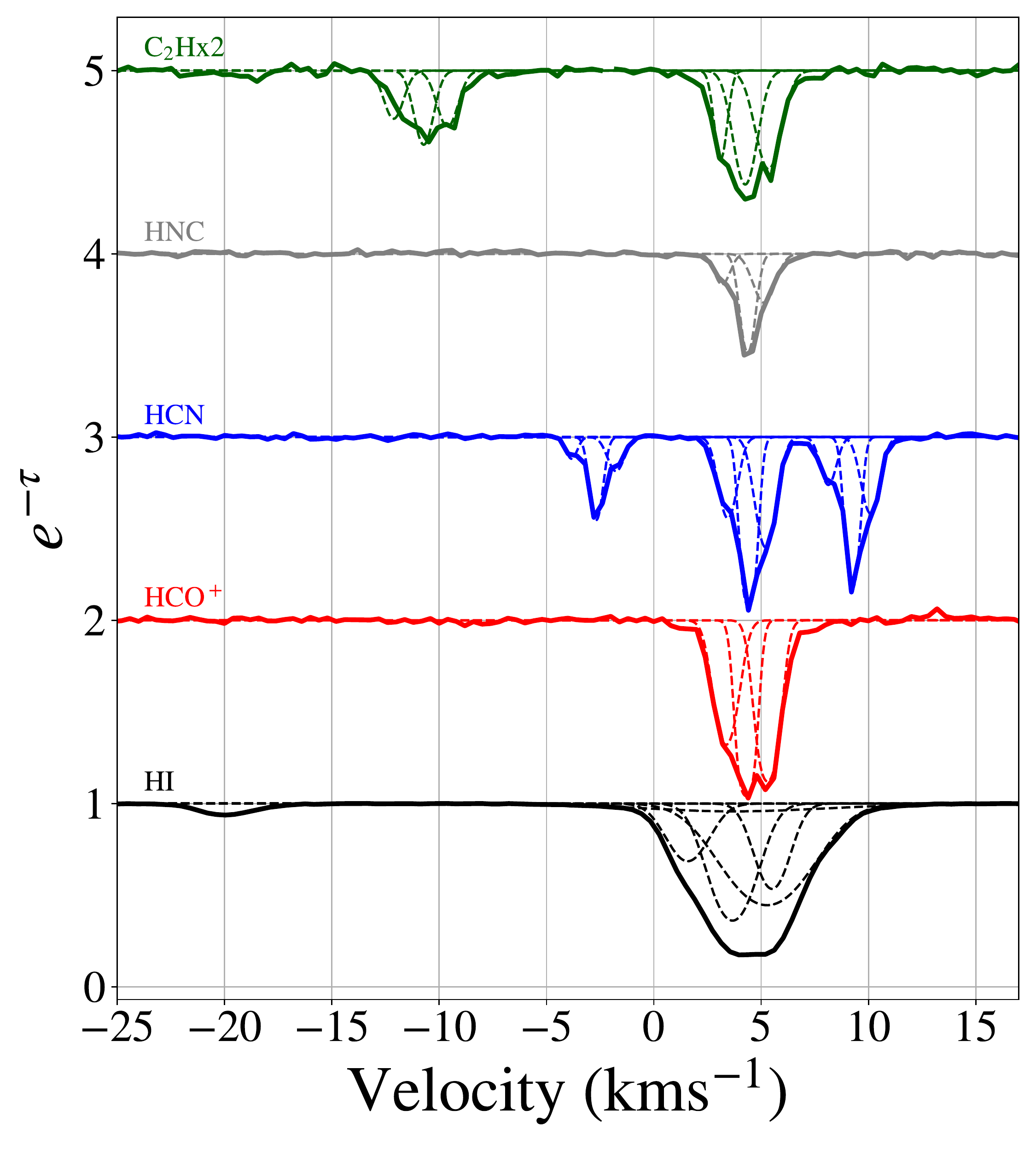}{0.3\textwidth}{3C123A}
\fig{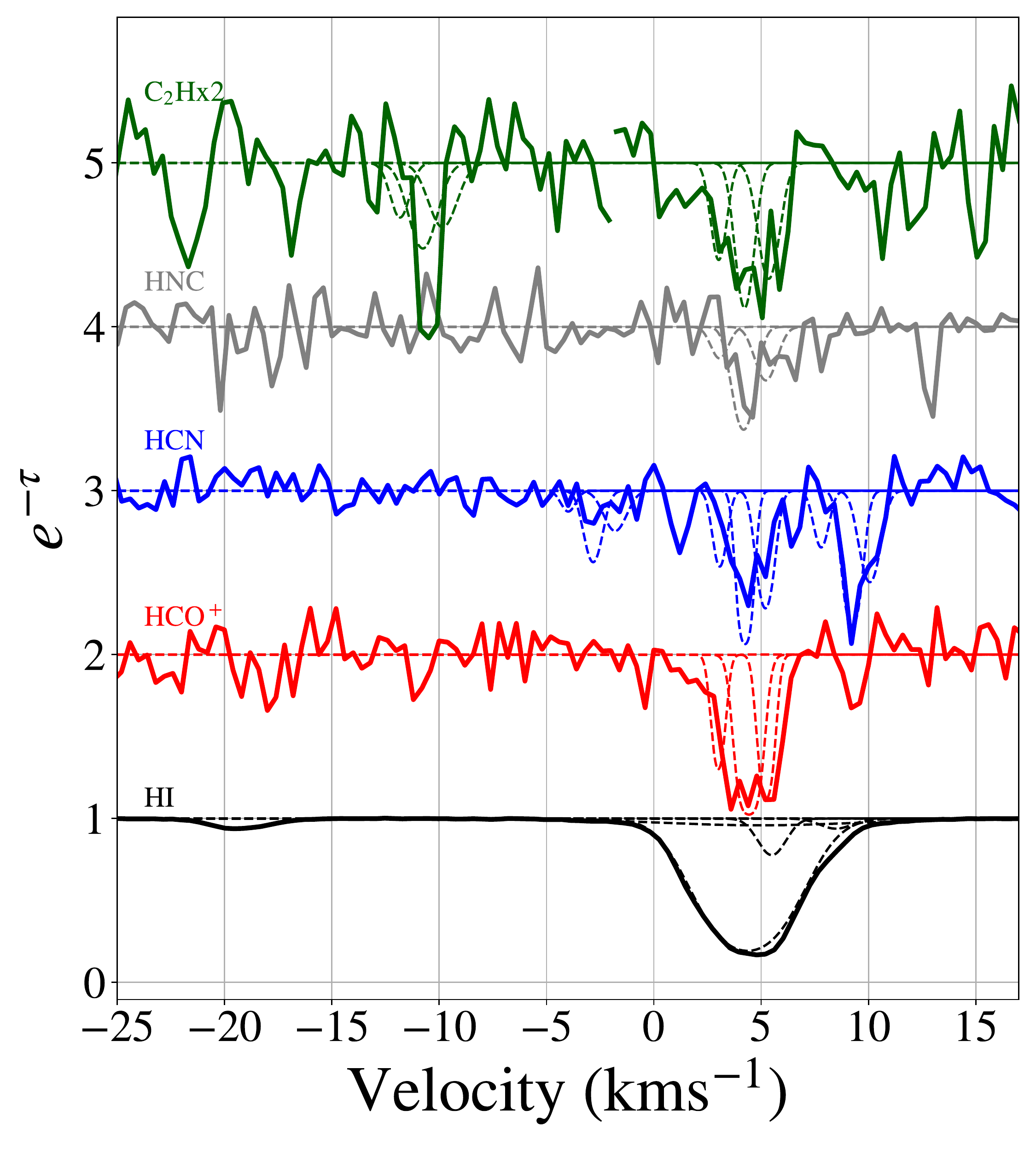}{0.3\textwidth}{3C123B}}
\gridline{\fig{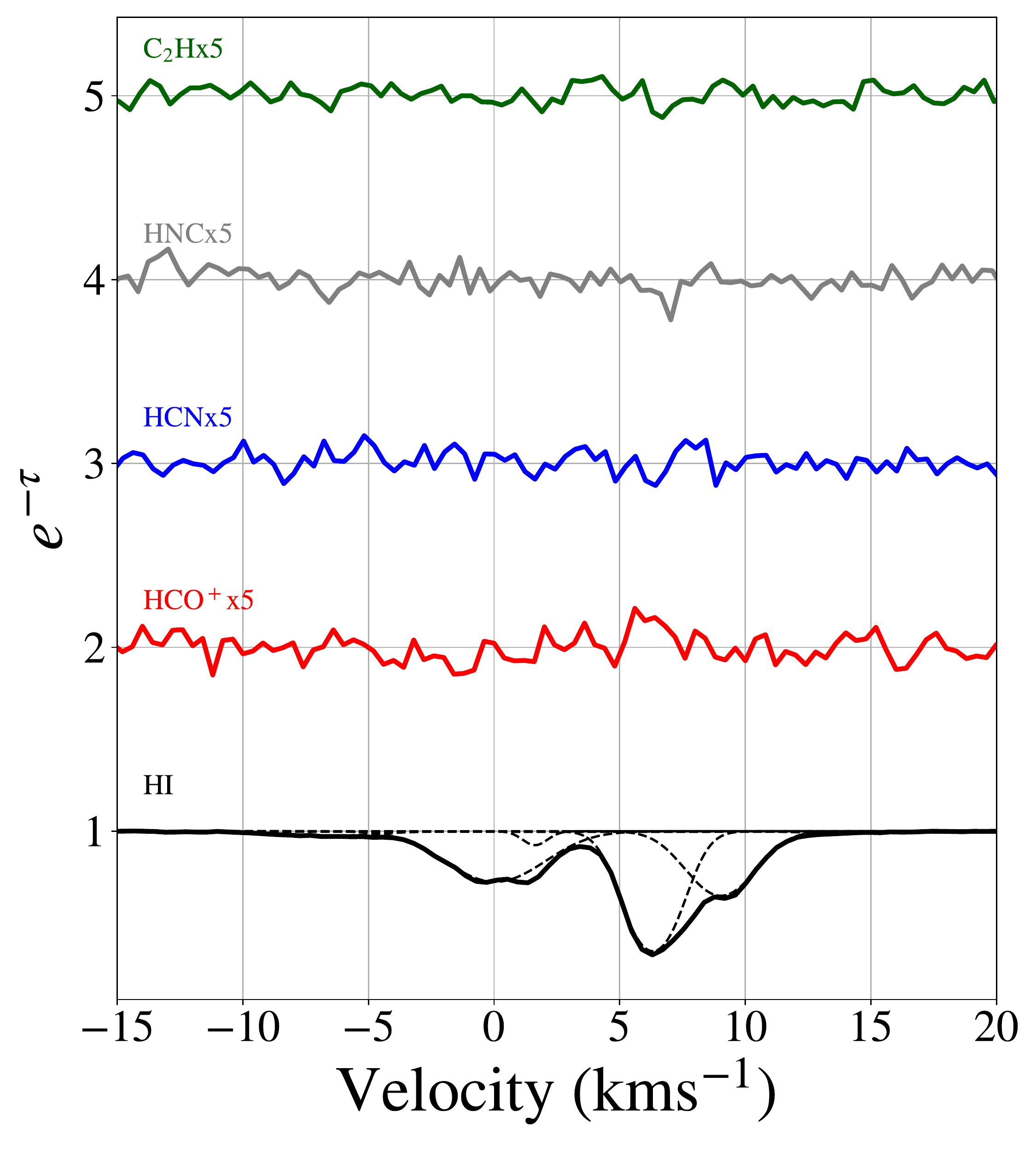}{0.3\textwidth}{3C138}
\fig{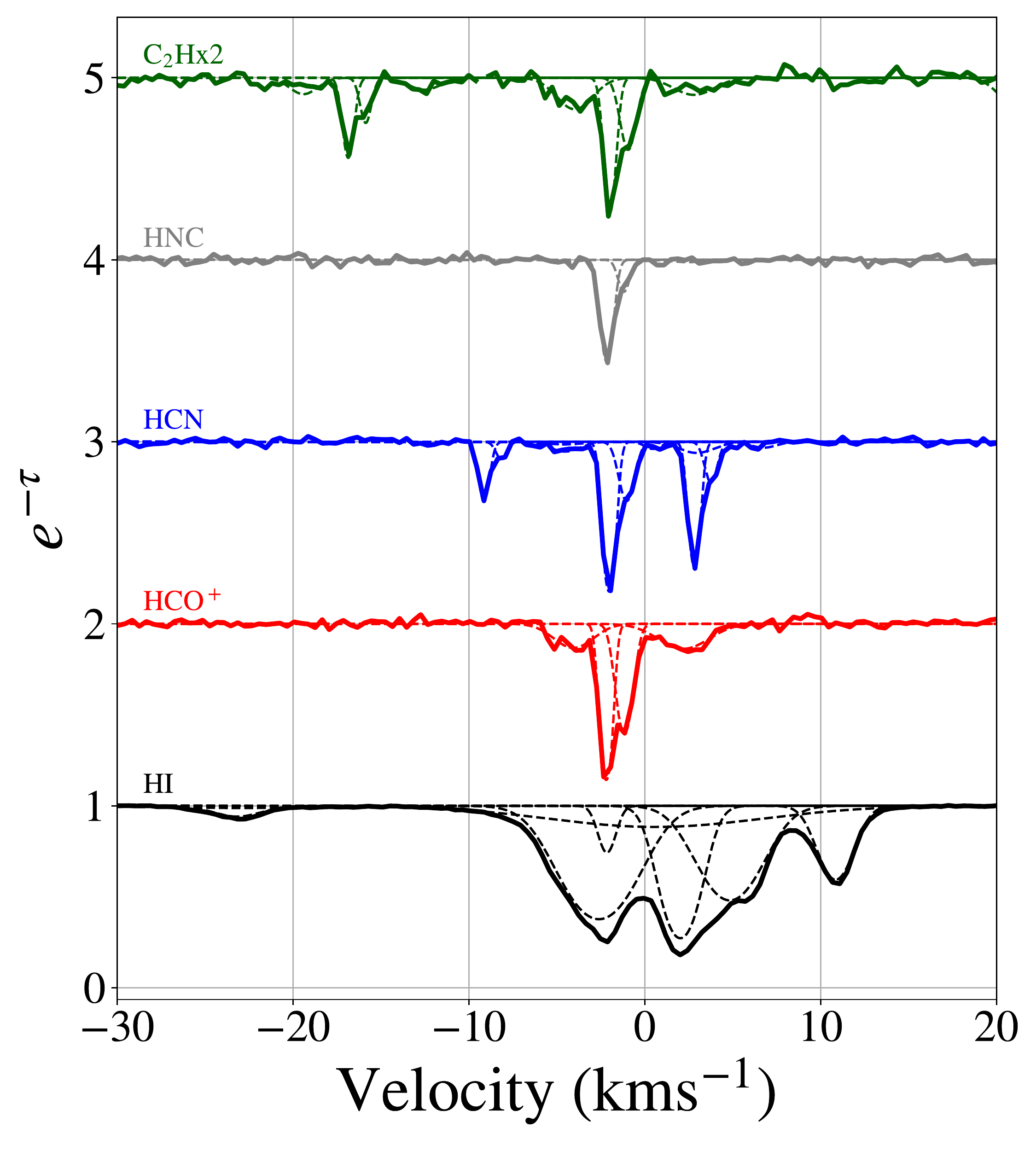}{0.3\textwidth}{3C154}
\fig{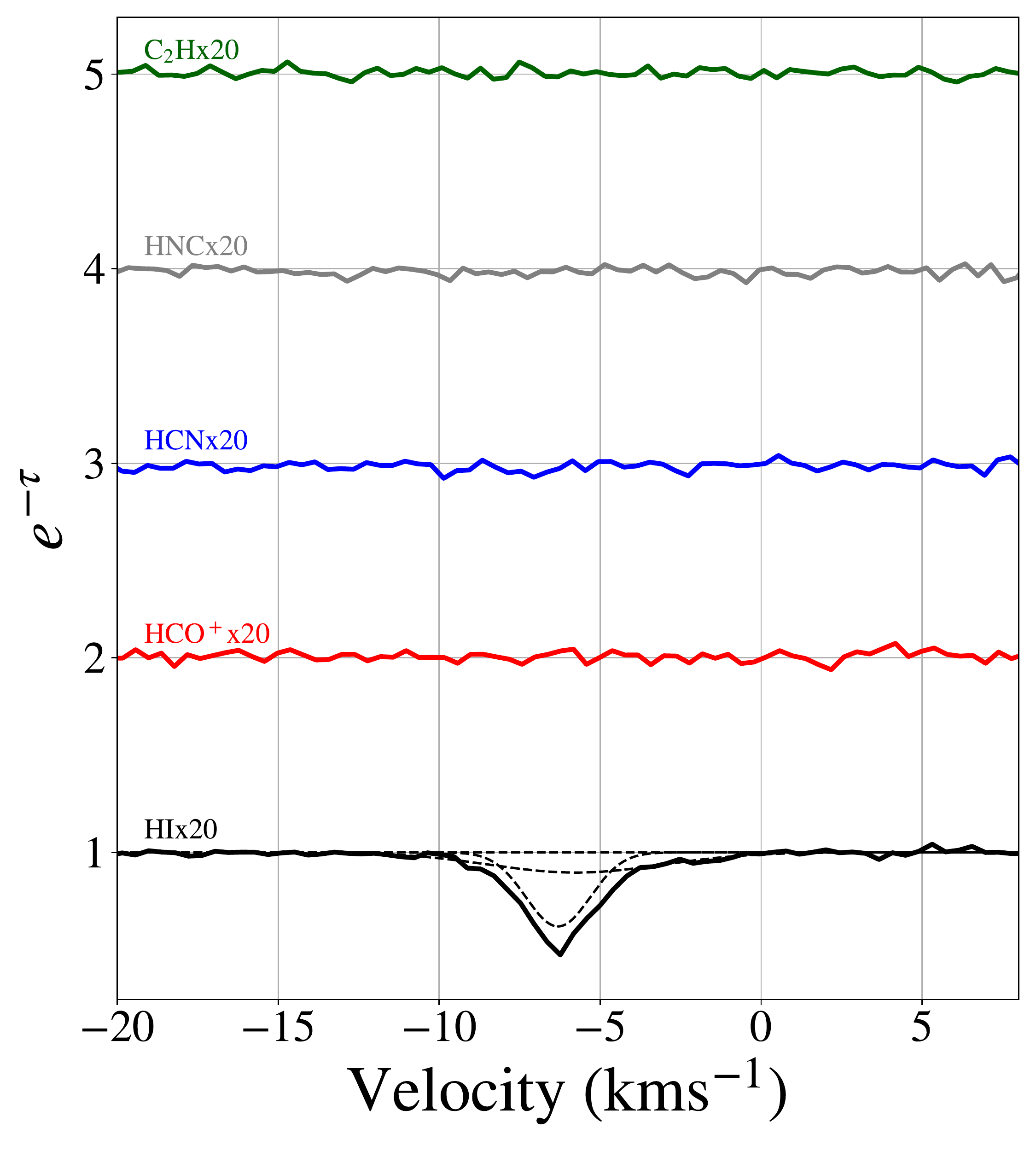}{0.3\textwidth}{3C273}}
\caption{}
\end{figure*}
\begin{figure*}\ContinuedFloat
\gridline{\fig{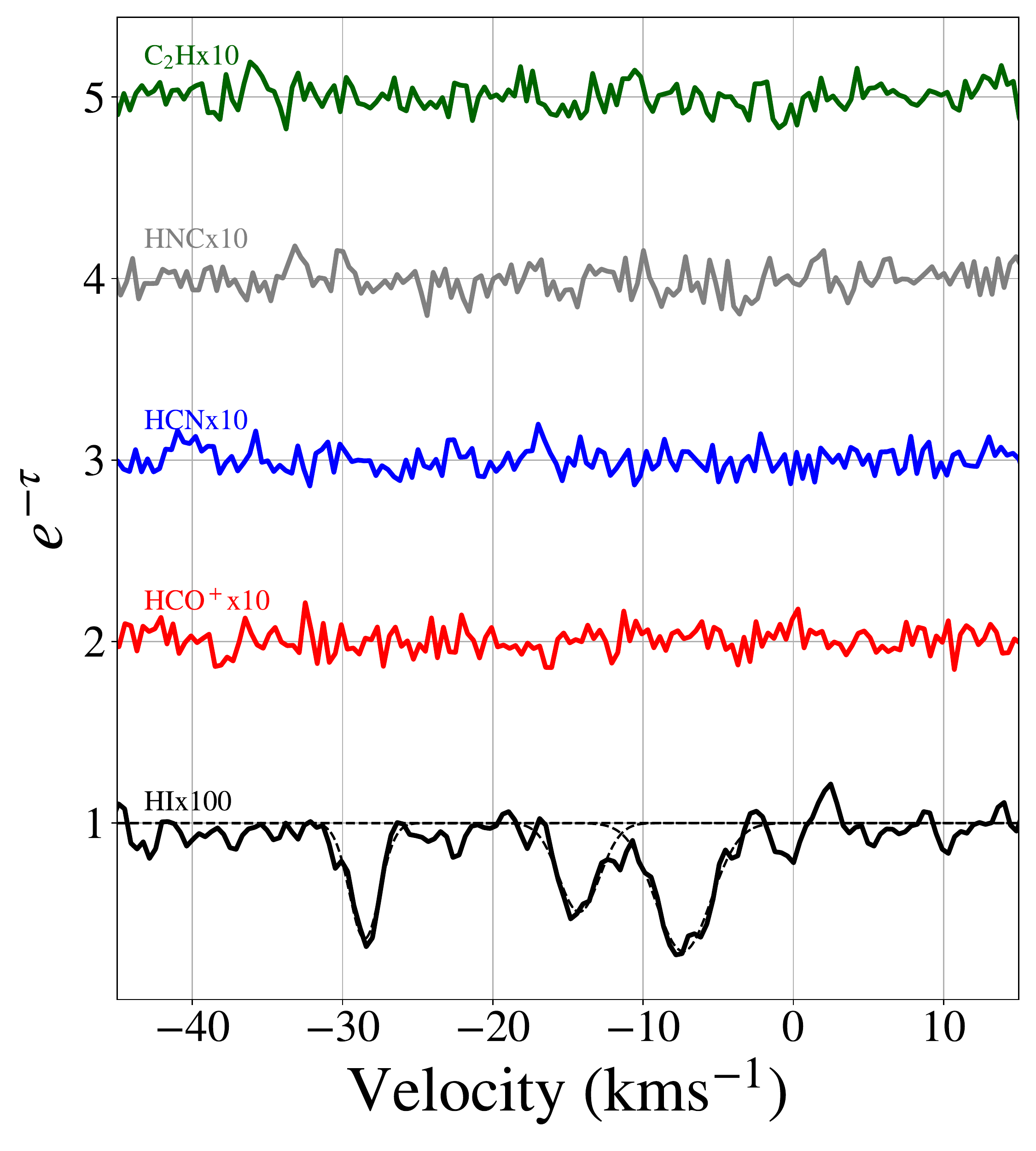}{0.3\textwidth}{3C286}
\fig{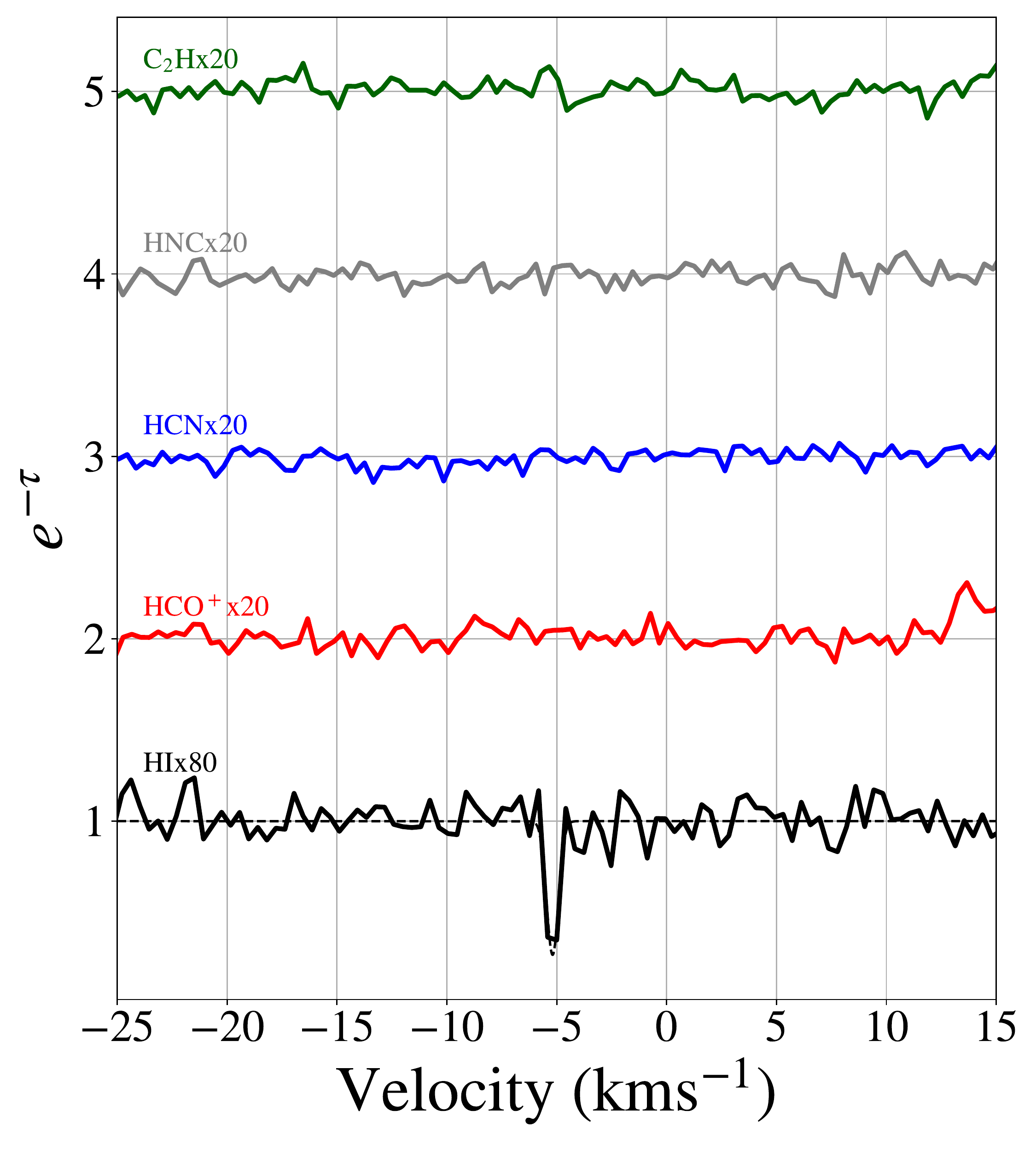}{0.3\textwidth}{3C345}
\fig{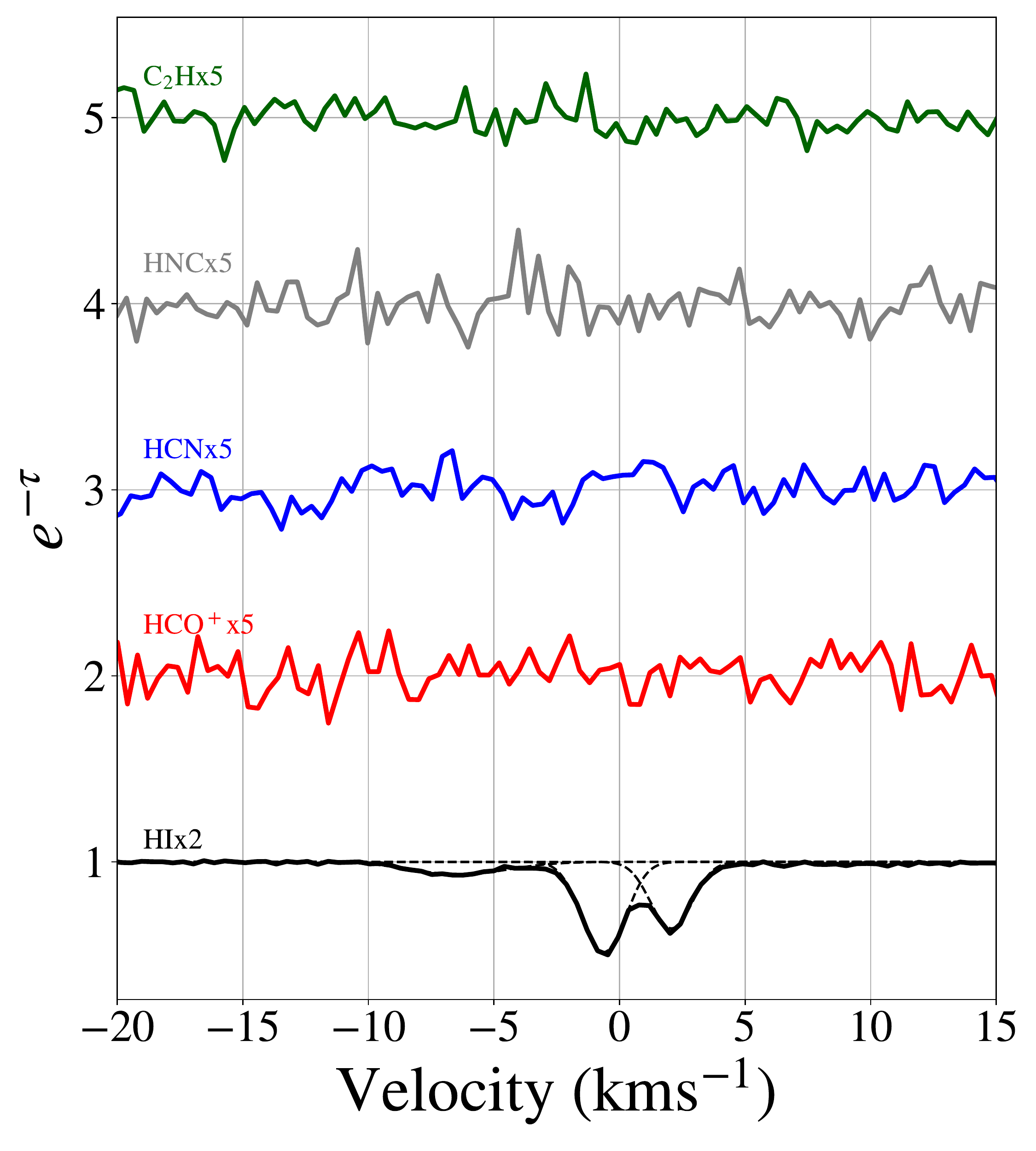}{0.3\textwidth}{3C346}}
\gridline{\fig{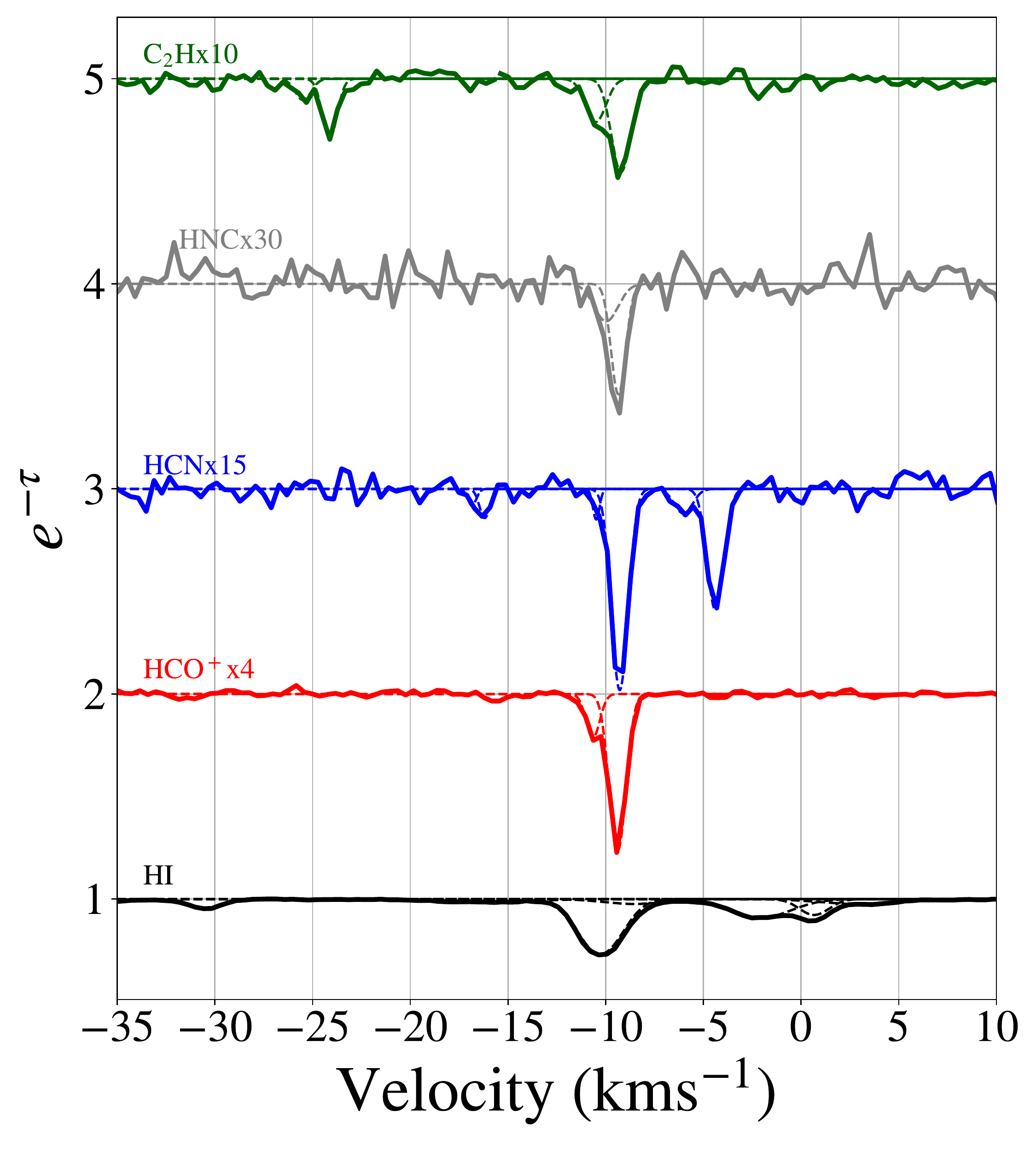}{0.3\textwidth}{3C454.3}
\fig{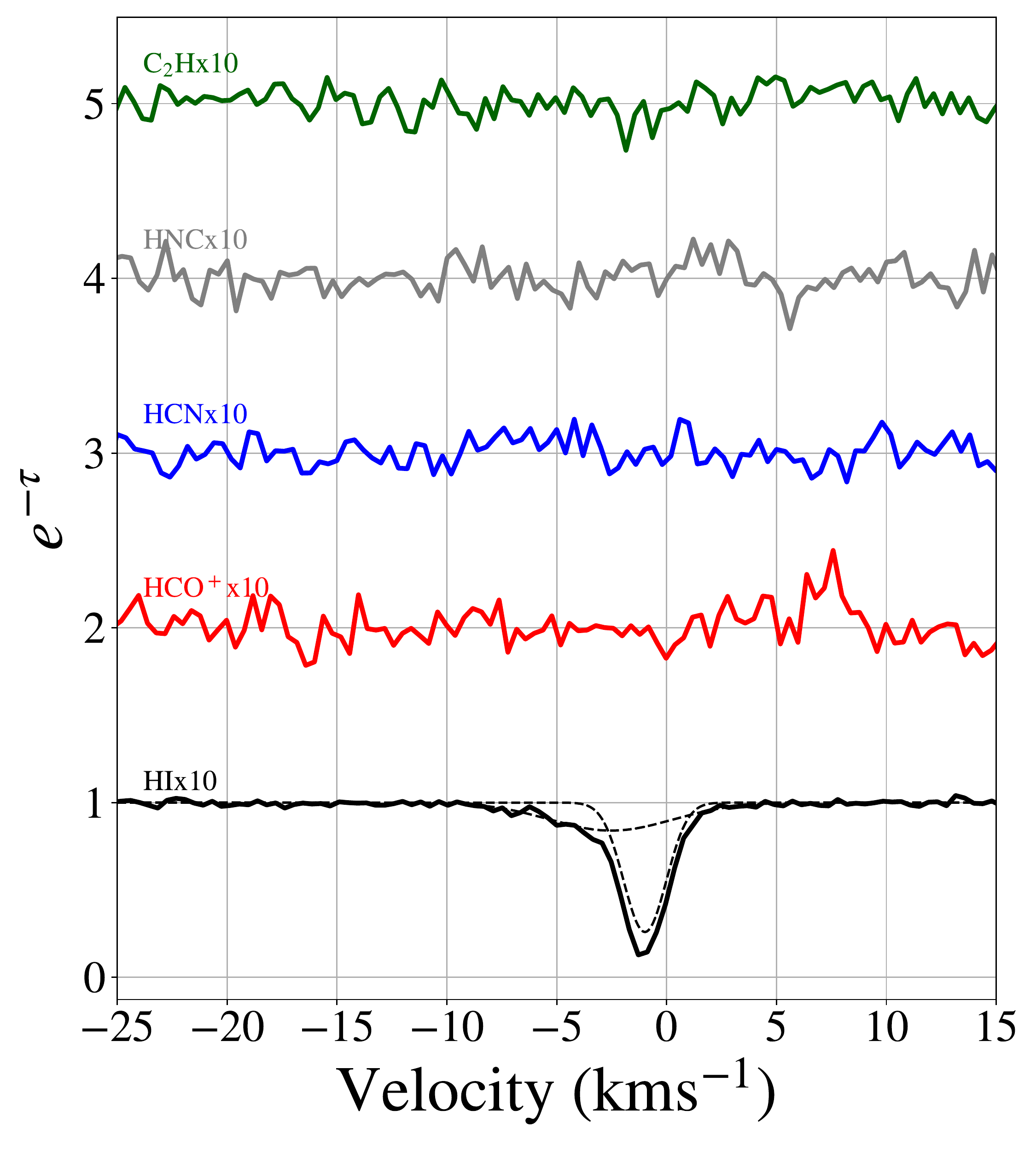}{0.3\textwidth}{4C12.50}
\fig{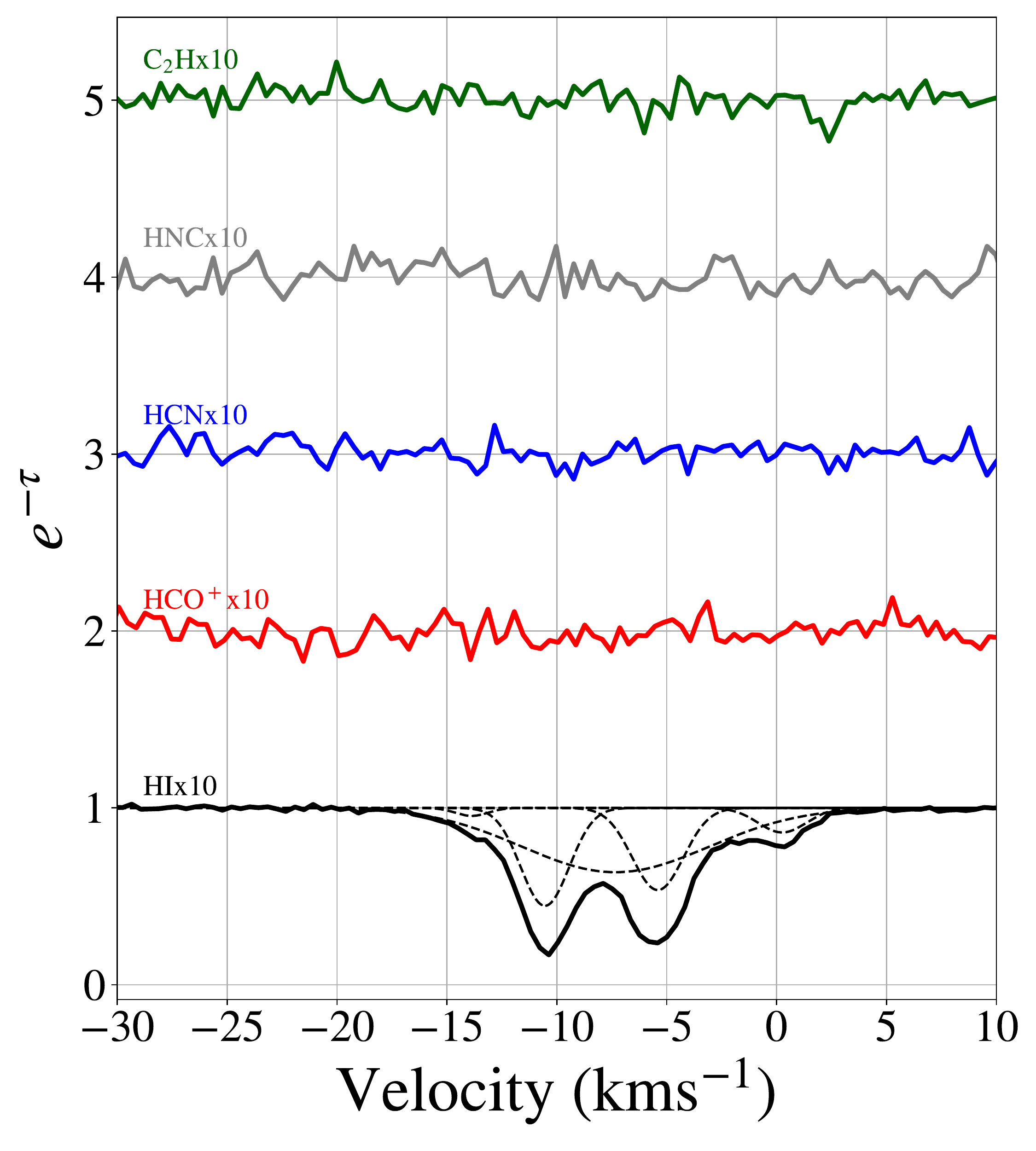}{0.3\textwidth}{4C15.05}}
\gridline{\fig{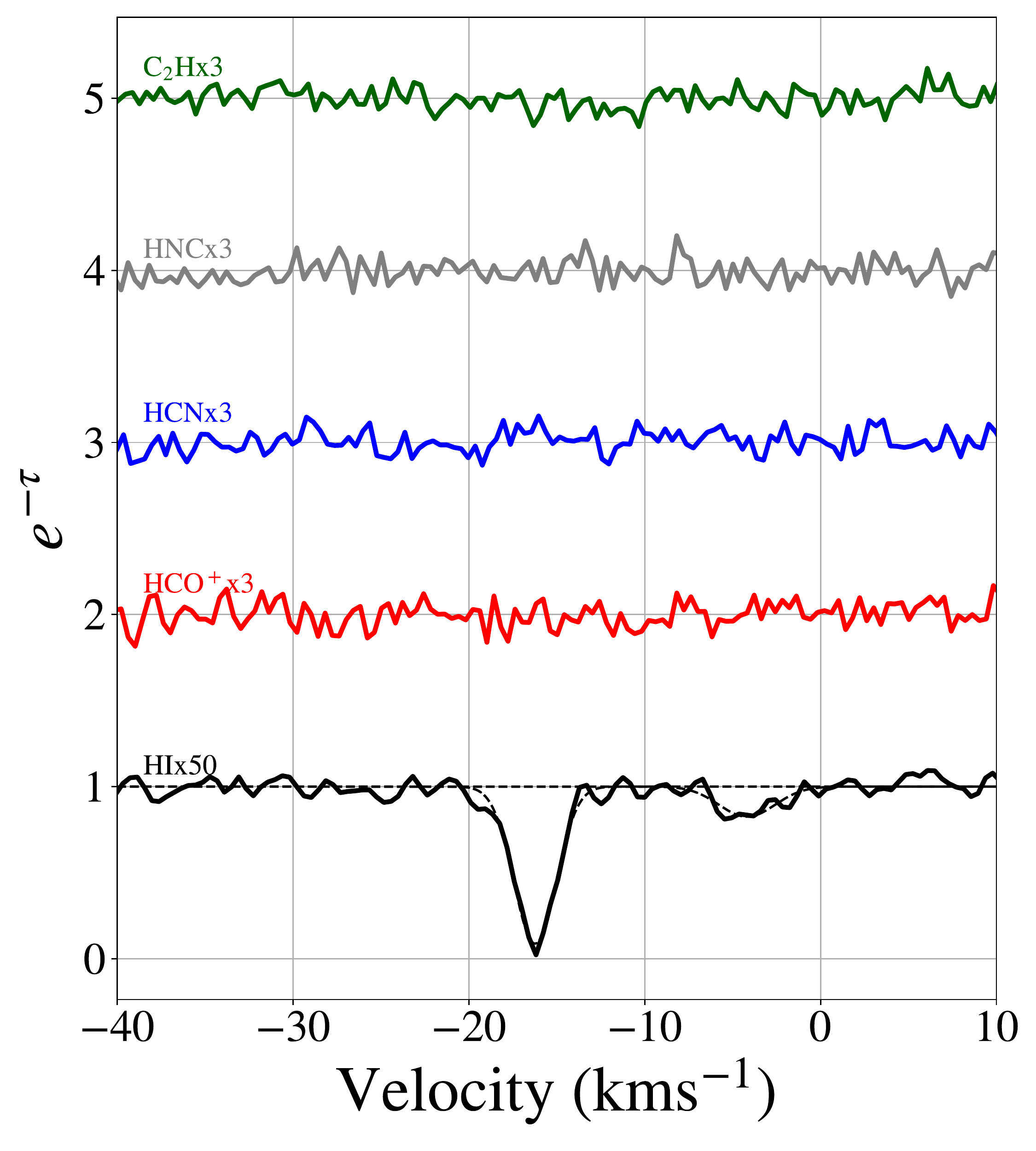}{0.3\textwidth}{4C32.44}
\fig{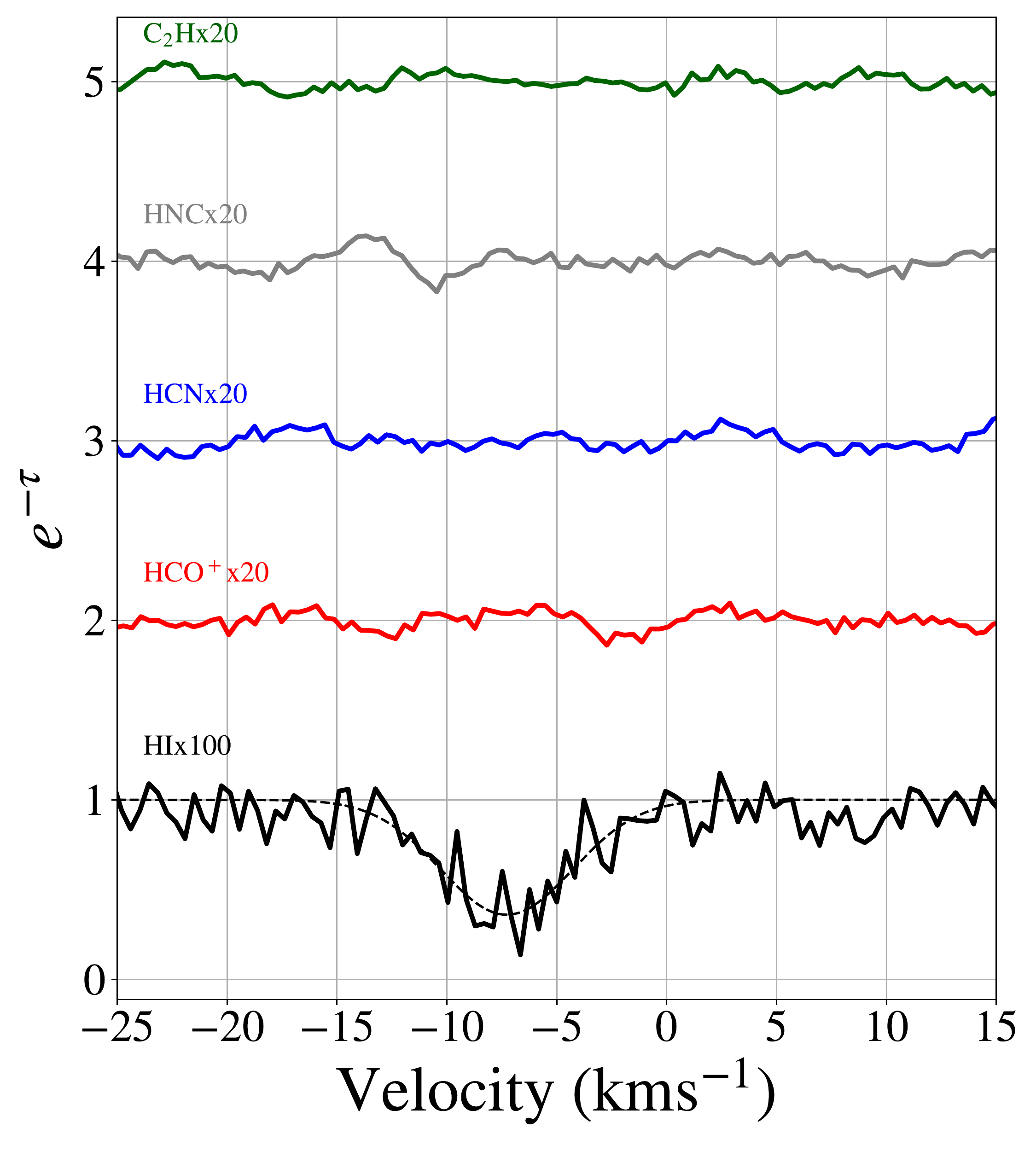}{0.3\textwidth}{1055+018}
\fig{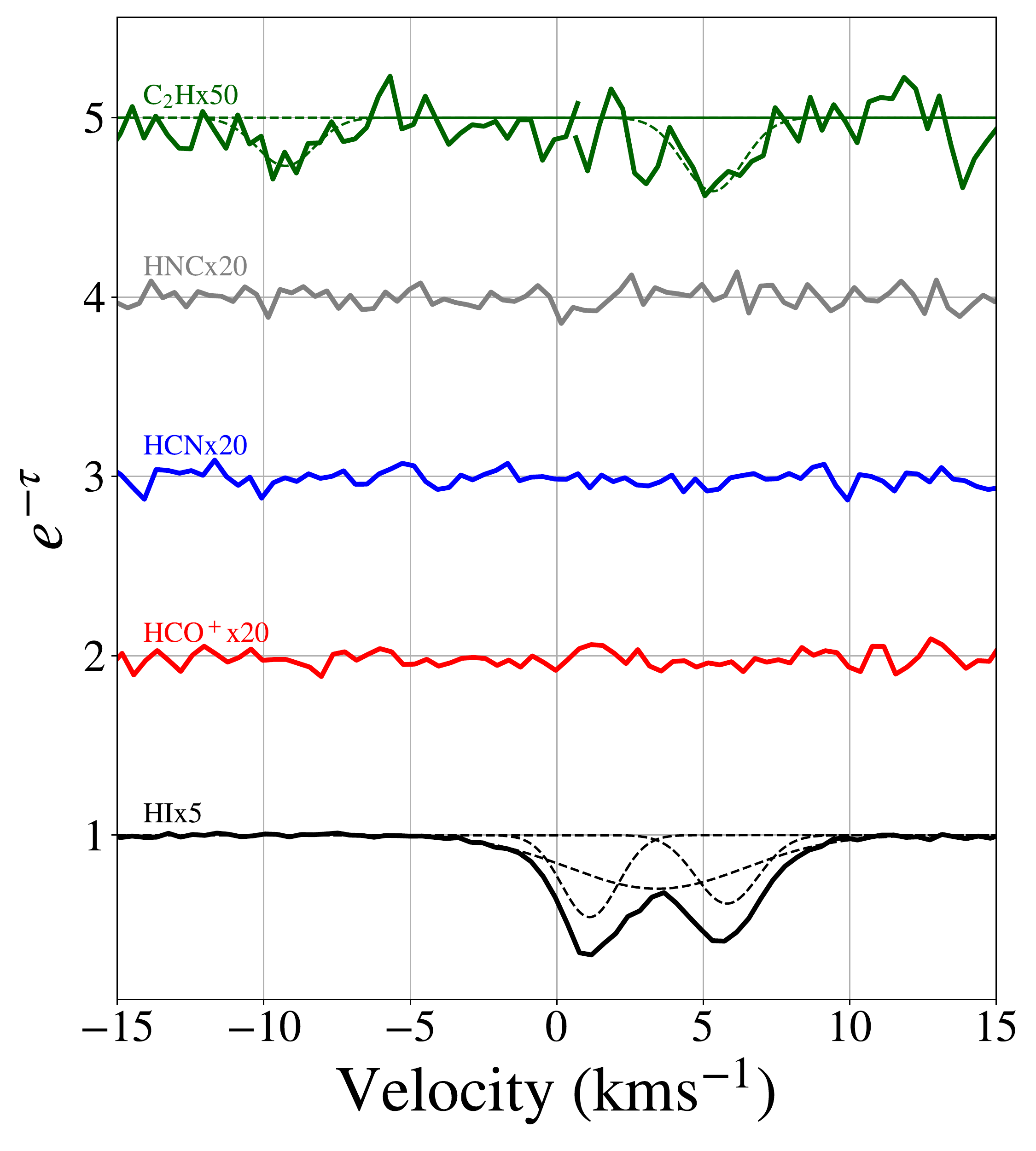}{0.3\textwidth}{J2136}}
\caption{}
\end{figure*}
\begin{figure*}\ContinuedFloat
\gridline{\fig{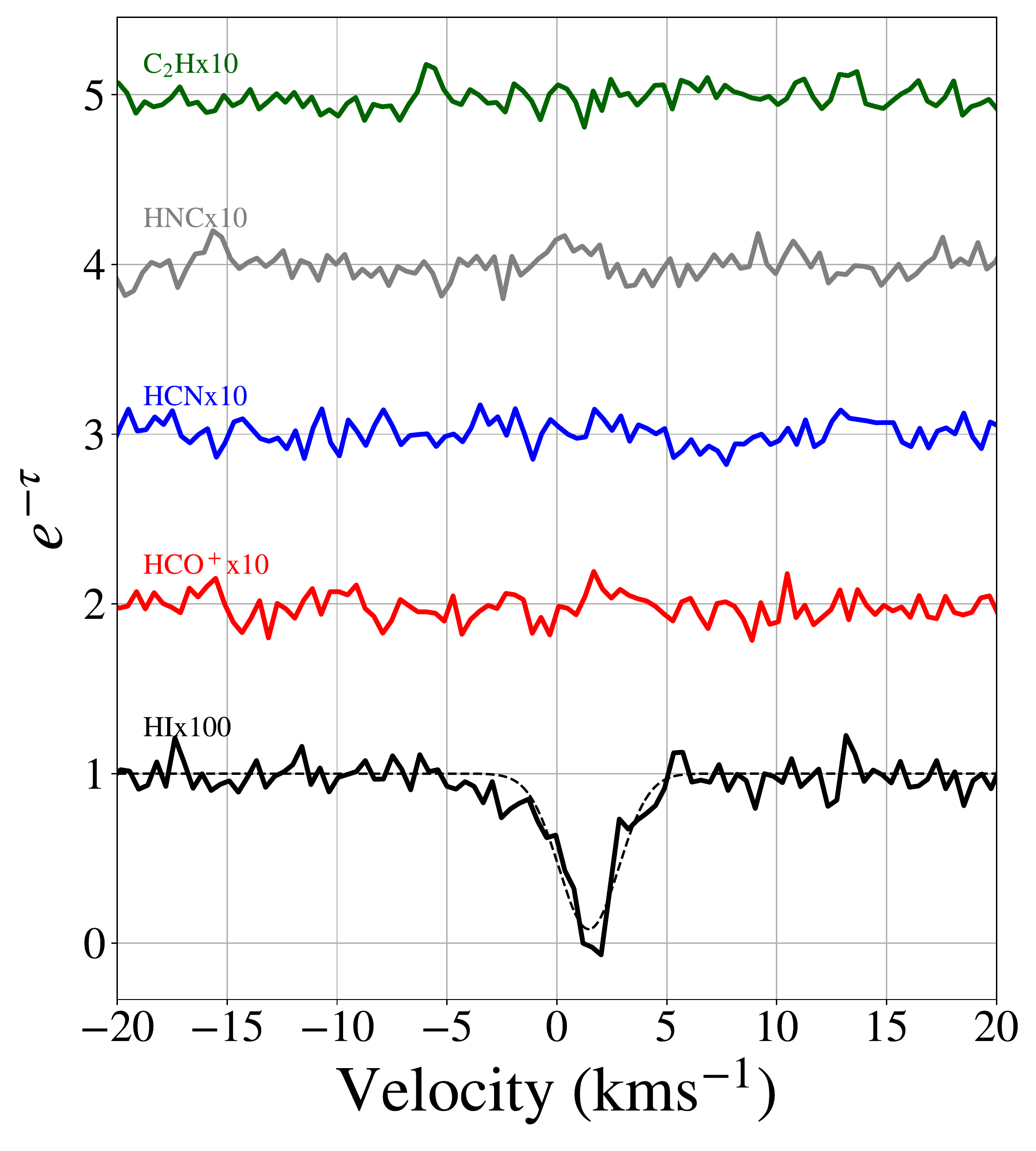}{0.3\textwidth}{PKS0742}
\fig{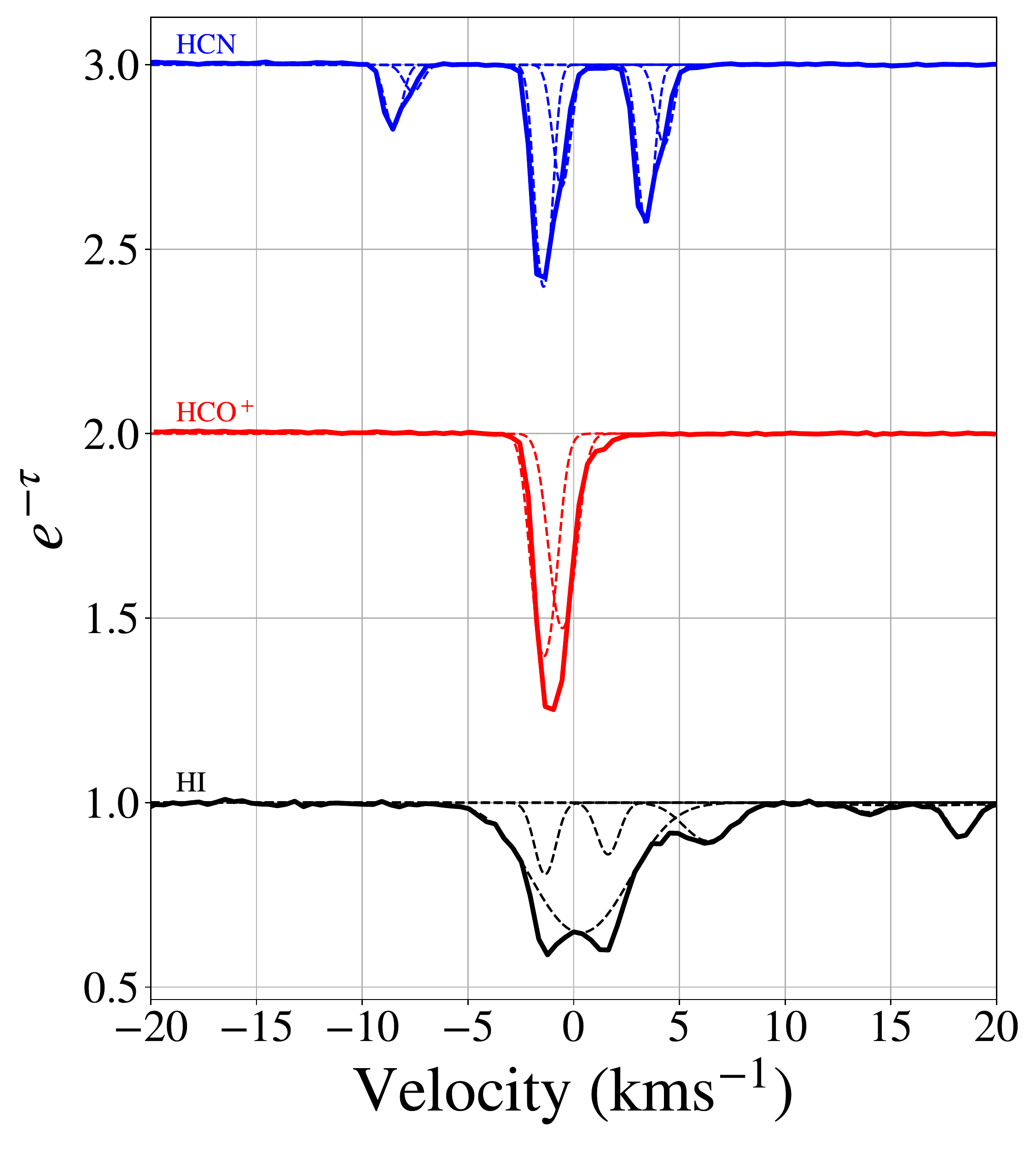}{0.3\textwidth}{BL Lac}}
\caption{The optical depth spectra, $e^{-\tau}$, versus LSR velocity for \hi{} \citep[bottom, black;][]{2018ApJS..238...14M}, \hcop{} (second from bottom, red), \hcn{} (middle, blue), \hnc{} (second from top, gray), and \cch{} (top, green). \hcn{} velocities correspond to the $F=2-1$ transition. \cch{} velocities correspond to the $F=2-1$ transition. The velocity of the $F=1-0$ \cch{} transition has been shifted by $+25$ \kms{} for viewing convenience. Gaussian fits to the \hi{} spectra from 21-SPONGE  \citep{2018ApJS..238...14M} and Gaussian fits to the molecular absorption spectra from this work (see Section \ref{subsec:gauss_decomp}) are shown as dashed lines. Absorption spectra that have been amplified are labeled with the multiplication factor.}
\label{fig:allspectra}
\end{figure*}

\section{Molecular gas in the diffuse ISM} \label{sec:thresholds}
The distribution of these background sources across the sky is shown in Figure \ref{fig:FullSkyCO}, overlaid on the CO integrated intensity image from \citet{2014A&A...571A..13P}.
We show our ALMA/NOEMA observations of \hcn{}, \cch{}, \hcop{}, and \hnc{}, as well as \hi{} absorption from 21-SPONGE, in Figure \ref{fig:allspectra}. 
Table \ref{tab:N_meas} summarizes the observational results in these directions, including the \hi{} column density \citep{2018ApJS..238...14M}, reddening \citep{2014A&A...571A..11P}, and molecular optical depth integrals, calculated for channels with an optical depth $\tau(v)\geq3\sigma_{\tau}$. Sources are listed in order of increasing Galactic latitude.

For most sources at latitudes $\lesssim40^\circ$ degrees, we detect at least one of four transitions. For sources at latitude $>40^\circ$, we detect marginal \hcop{} absorption along only one sightline, and no molecular absorption in any other direction. This is evident in Figure \ref{fig:FullSkyCO}, where sources with molecular absorption detections are indicated using stars, while sources with no molecular detections are indicated using circles. There is no significant CO emission observed in the \citet{2014A&A...571A..11P} or \citet{2001ApJ...547..792D} data in the direction of three sightlines with molecular absorption detections---3C120, 3C78, and J2136. These sources have $|b|=27^\circ$--$44^\circ$. We discuss this further in Section \ref{sec:broad_hcop}.

The integrated optical depths---and therefore column densities (Equation \ref{eq:N})---span over two orders of magnitude across our sample for each of the molecular species, indicating that we are probing diverse interstellar environments. In Section \ref{sec:temporal_stability}, we compare the results in Table \ref{tab:N_meas} to previous observations of \hcn{}, \cch{}, \hcop{} and \hnc{} in the same directions, taken between 3 and 25 years ago.
All variations are at a level $\lesssim3\sigma$, demonstrating that our observing and processing strategies are sound.

\begin{figure*}[]
\gridline{\fig{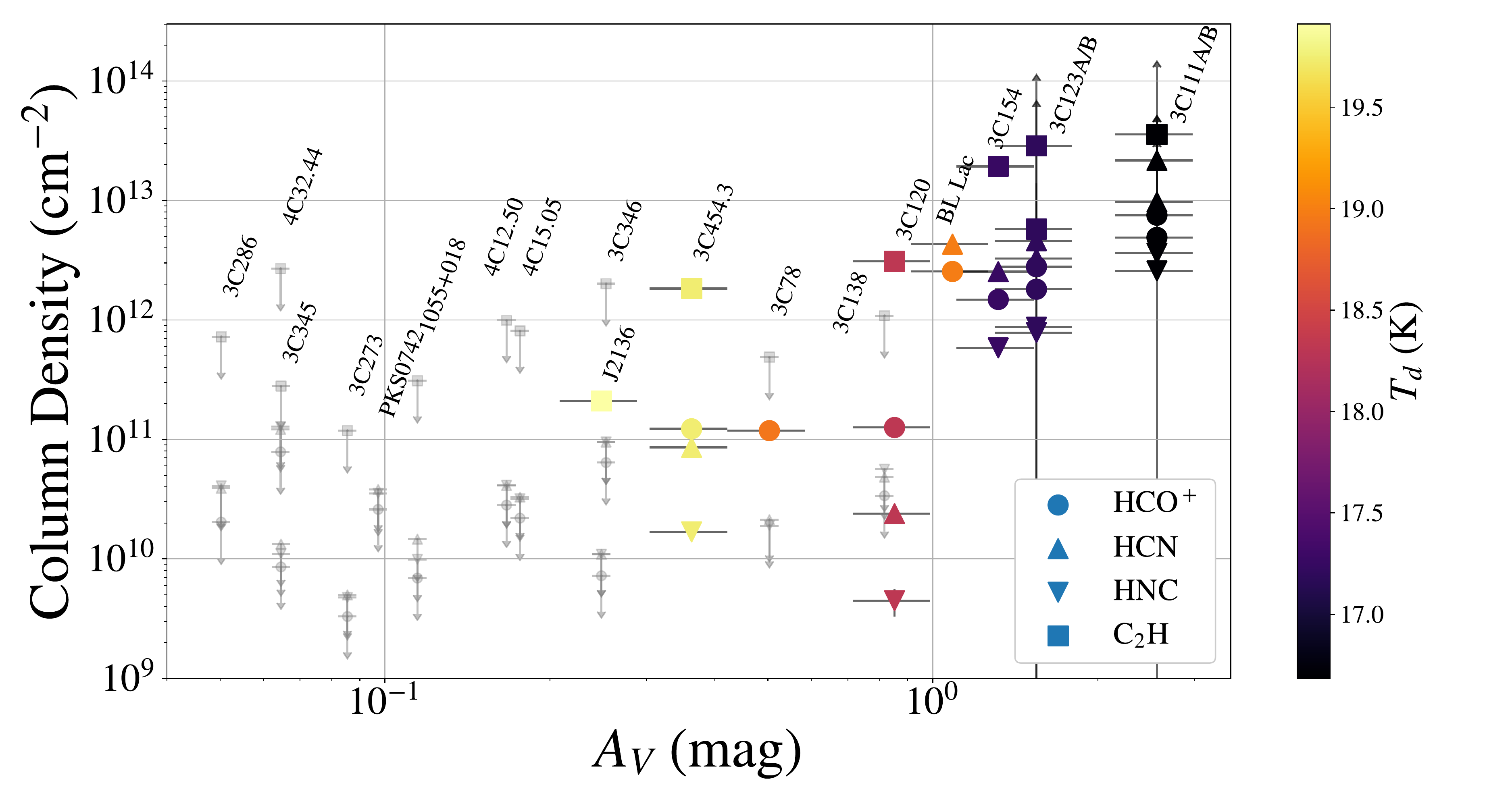}{0.52\textwidth}{}
\fig{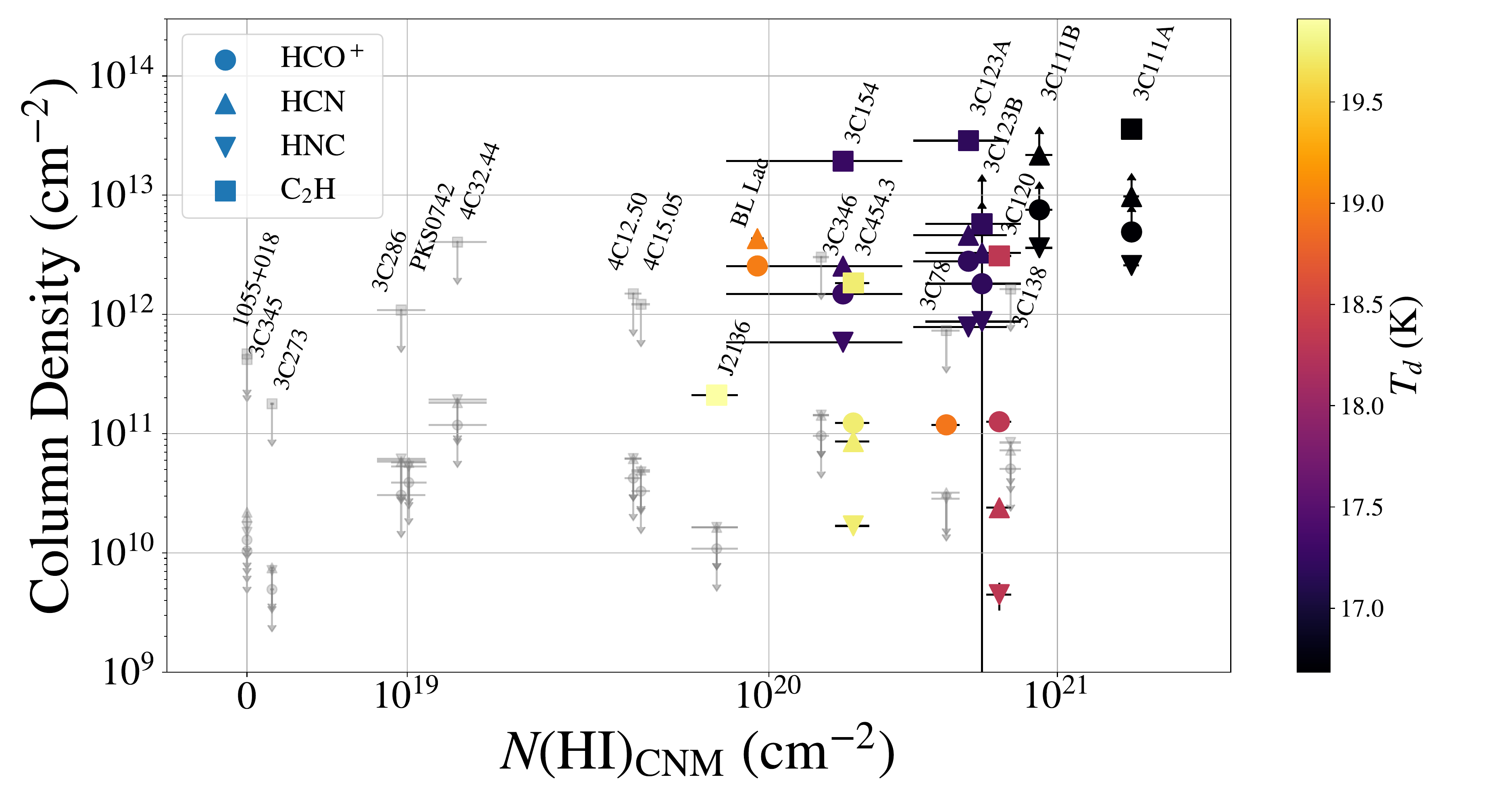}{0.52\textwidth}{}}
\gridline{\fig{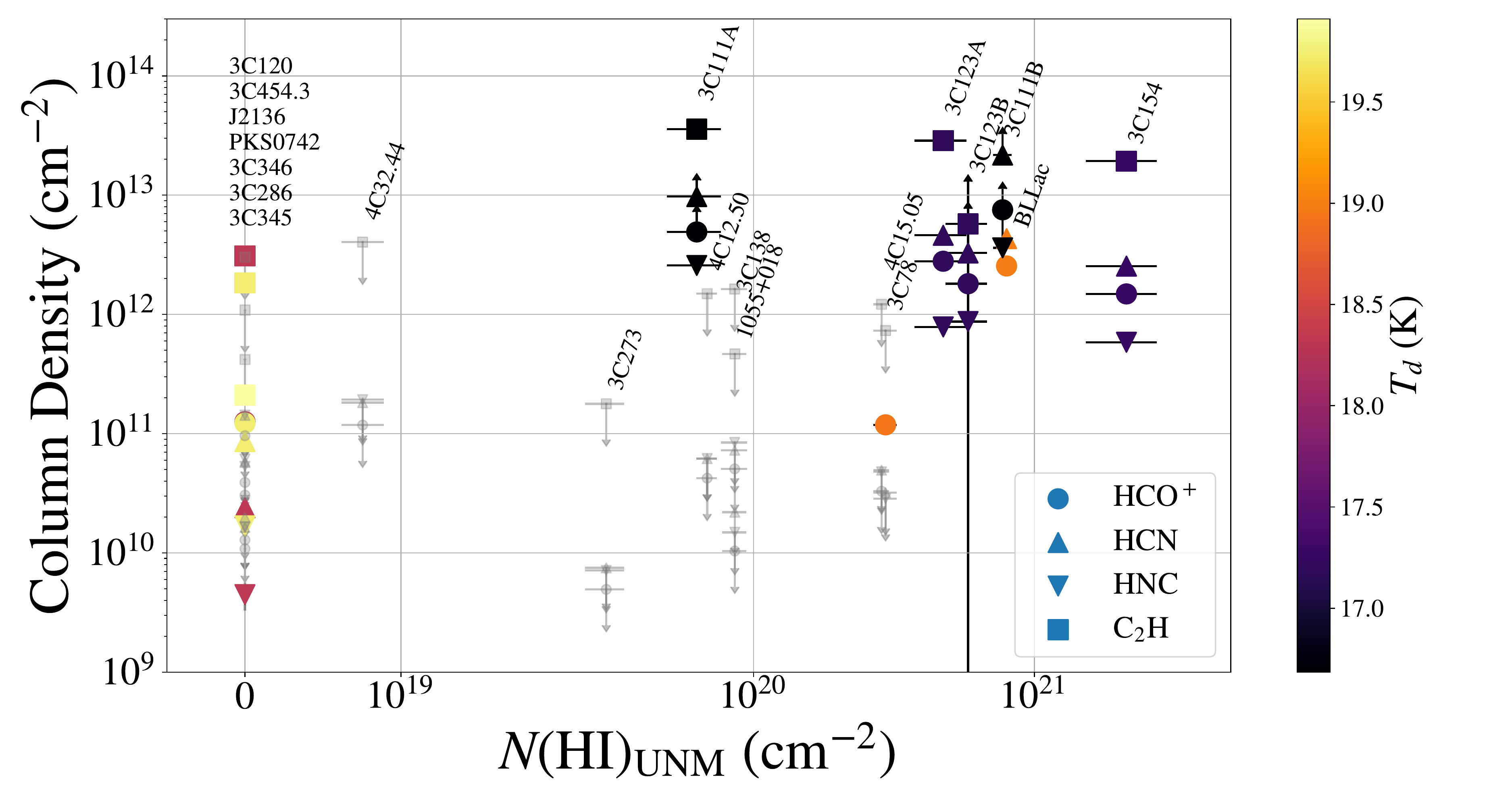}{0.52\textwidth}{}
\fig{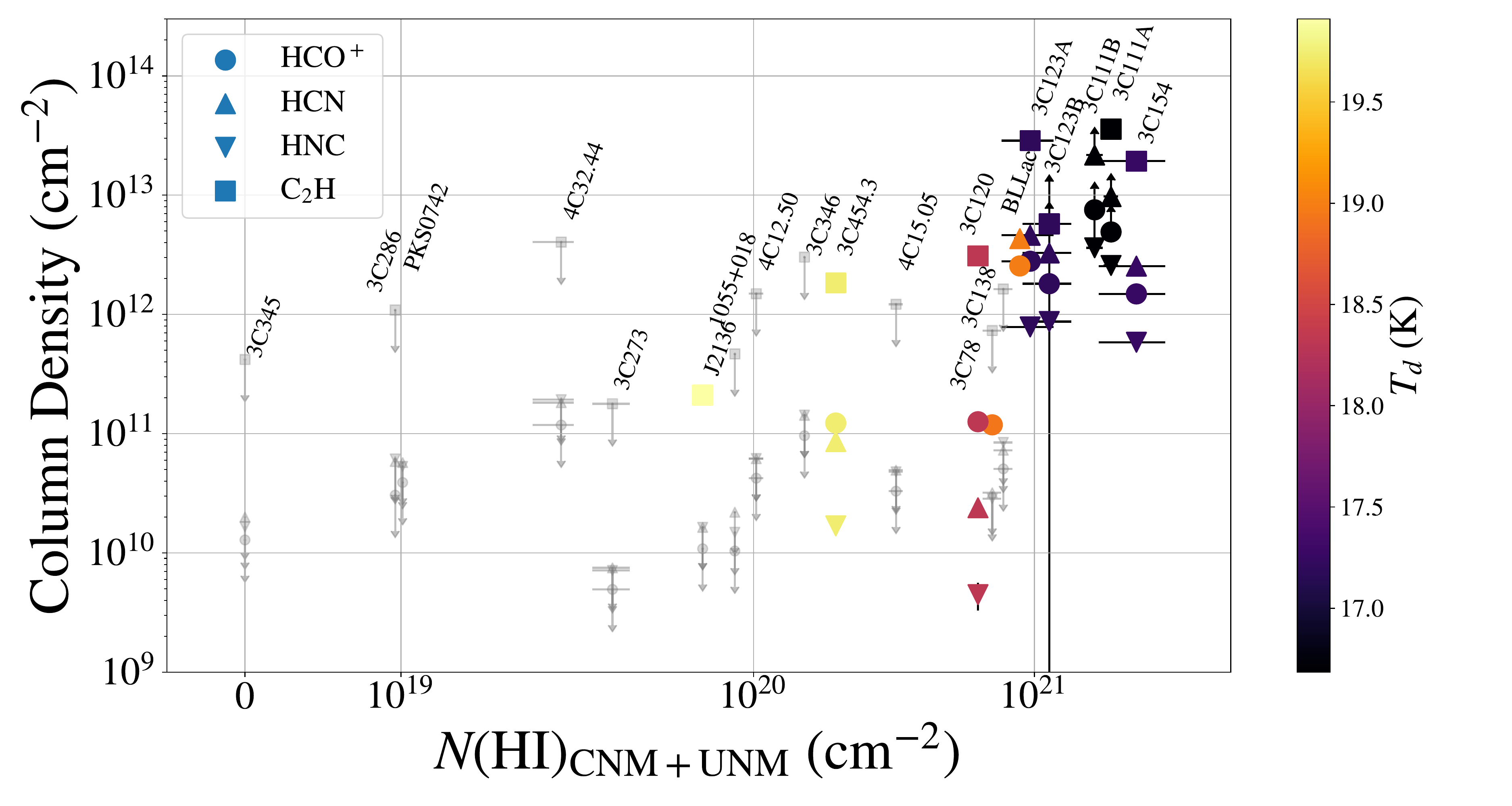}{0.52\textwidth}{}}
\caption{Molecular column densities versus visual extinction ($A_V$; upper left), \hi{} CNM column density (upper right), \hi{} UNM column density (lower left), and the combined column density of the CNM and UNM (lower right). Colors indicate the dust temperature derived by \citet{2016A&A...596A.109P}. Because the resolution of the \textit{Planck} extinction maps is $5\arcmin$, the A and B components of 3C111 and 3C123 have the same $x$ coordinate in the upper left panel, so are not labeled separately. \label{fig:NX_v_var}}
\end{figure*}

\subsection{Thresholds for molecule formation}

In Figure \ref{fig:NX_v_var}, we compare the observed molecular column densities to the neutral and total gas column densities. The total gas column density is traced via the visual extinction ($A_V=3.1E(B-V)$; upper left). For the \hi{} gas we investigate the CNM column density (the \hi{} column density occupied by gas with $T_S<250$ K; upper right), the UNM column density, (the \hi{} column density occupied by gas with $250 \text{ K}<T_S<1000$ K; lower left), and the sum of the CNM and UNM column densities (lower right), for each line of sight. For non-detections, we show $3\sigma$ upper limits to the molecular column densities assuming a FWHM of 3 channels or 1.2 \kms{}. The data points are color-coded based on the dust temperature provided by \textit{Planck}, which can be used as a proxy for the ISRF.

The upper left panel of Figure \ref{fig:NX_v_var} suggests a threshold visual extinction $A_V\approx0.25$ below which no molecular absorption is detected and above which molecular abundances tend to increase with increasing $A_V$. A visual extinction $A_V\gtrsim0.25$ corresponds to a total hydrogen column density $N_\mathrm{H}\gtrsim 5\times10^{20}$ \persc{} \citep[e.g.,][]{2009MNRAS.400.2050G,2017MNRAS.471.3494Z}, which is roughly the threshold column density for the \hi{}-to-H$_2$ transition at solar metallicity \citep[e.g.,][]{1973ApJ...181L.116S,1977ApJ...216..291S,2009ApJ...693..216K,  Gong2018}. \citet{1996A&A...307..237L} identified a similar threshold from observations of \hcop{} (as well as OH), and \citet{2000A&A...358.1069L} showed that \cch{} was as widespread as \hcop{}. These results suggest that a similar column density threshold applies in the case of \hcop{}, HCN, HNC and \cch{}. Essentially, as soon as conditions are suitable for \htwo{} survival, these other species follow quickly.
The sightline toward 3C138, is an exception---it has a visual extinction of 0.815 but no detectable molecular absorption. This sightline is explored in greater depth in Section \ref{subsec:tsas_discussion}. The $A_V$ trend shown in this figure also follows the dust temperature distribution, again highlighting the importance of shielding against the radiation field. The five sources with the highest molecular column densities ($\gtrsim10^{12}$ \persc{} for all four species) have the lowest dust temperature, $T_d\sim17$ K, and FUV radiation field $G'\sim0.4$, where $G'$ is the radiation field strength relative to the standard \citet{Draine1978} field, calculated using $T_d$ and $\beta$ from \textit{Planck} (see Section \ref{subsec:planck}). The rest of the sources have molecular column densities $\la 10^{12}$ cm$^{-2}$, $T_d\sim19$ K, and have $G' \sim1$.

The lower left panel of Figure \ref{fig:NX_v_var} shows the molecular column densities as a function of the column density of the thermally unstable \hi{}. While this graph lacks a correlation, there is a clear dichotomy of sources: the sources with the highest molecular column densities ($>10^{12}$ cm$^{-2}$ for all species) all have  $N(\hi{})_{\mathrm{UNM}}\sim 10^{20-21}$ cm$^{-2}$, and the fraction of \hi{} in the UNM is $\sim30$--$70\%$ for these sightlines); all other sources have essentially no UNM gas.
The only exception is 3C78, which has a high UNM column density and low molecular column densities. This suggests that UNM gas likely plays an important role for molecule formation and survival.

Meanwhile, in the two right panels of Figure \ref{fig:NX_v_var}, we see that there is a threshold of $\gtrsim10^{20}$ \persc{} for the column density of cold \hi{} below which no molecular absorption is detected (J2136---the only exception---is the lowest column density source to show molecular absorption, with $N(\hi{})\approx4\times10^{19}$ \persc{} and very weak \cch{} absorption).  Again, this threshold is similar to the column density threshold for the \hi{}-to-\htwo{} transition  \citep[$\sim5\times10^{20}$ \persc{}][]{1977ApJ...216..291S}. The correlation seen in the lower right plot, where the $x$-axis traces the total amount of cold \hi{}, is qualitatively similar to that seen in the upper left plot, and again reflects the importance of shielding.

We do not see any difference here across the different chemical species probed in this study. For example, Goddard et al. (2014) suggested that \hcop{} and \cch{} are more sensitive to turbulent dissipation than HCN or HNC. We note, though, that our sample size is modest, and very few structures do not show absorption from all four species.

\begin{figure*}[]
\gridline{\fig{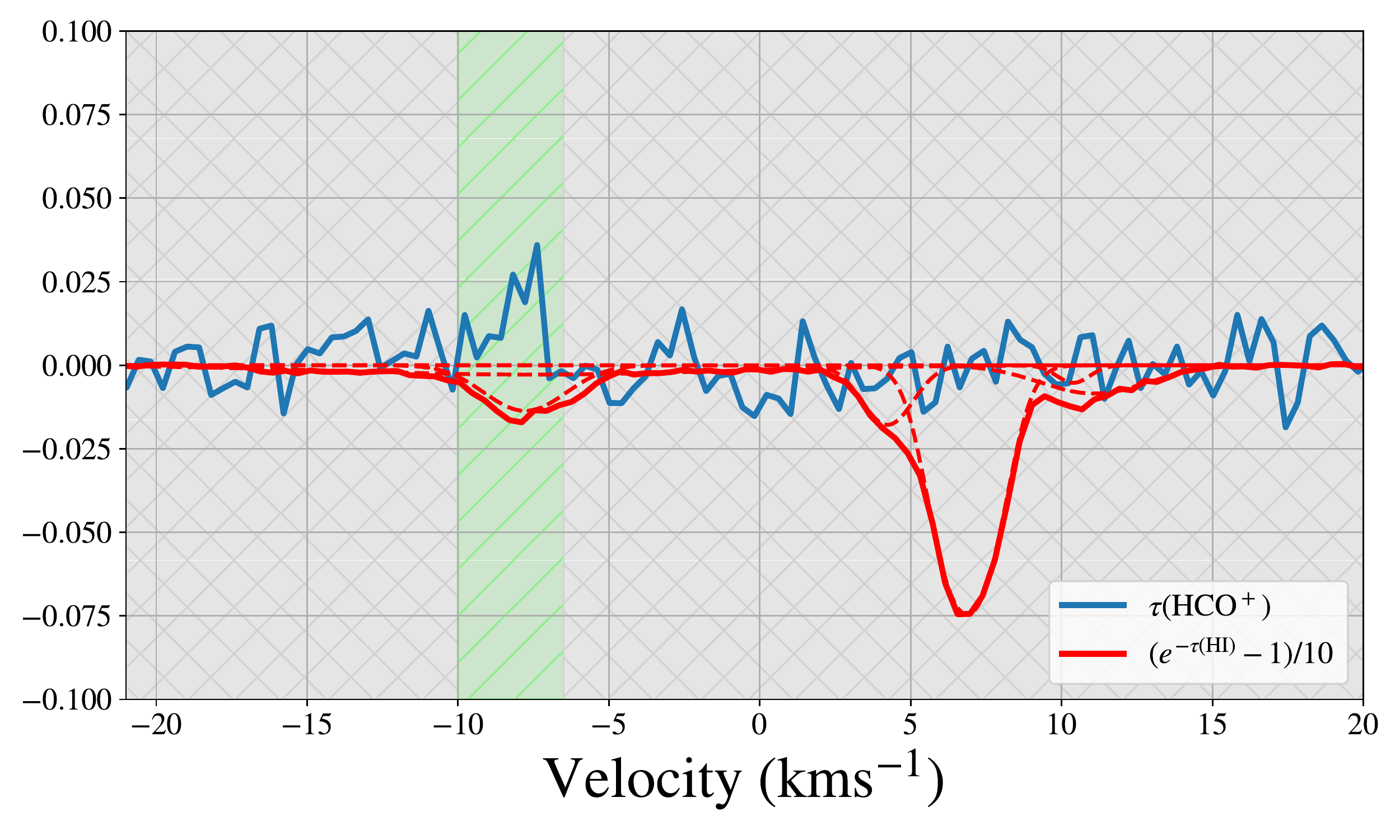}{0.4\textwidth}{3C78}
\fig{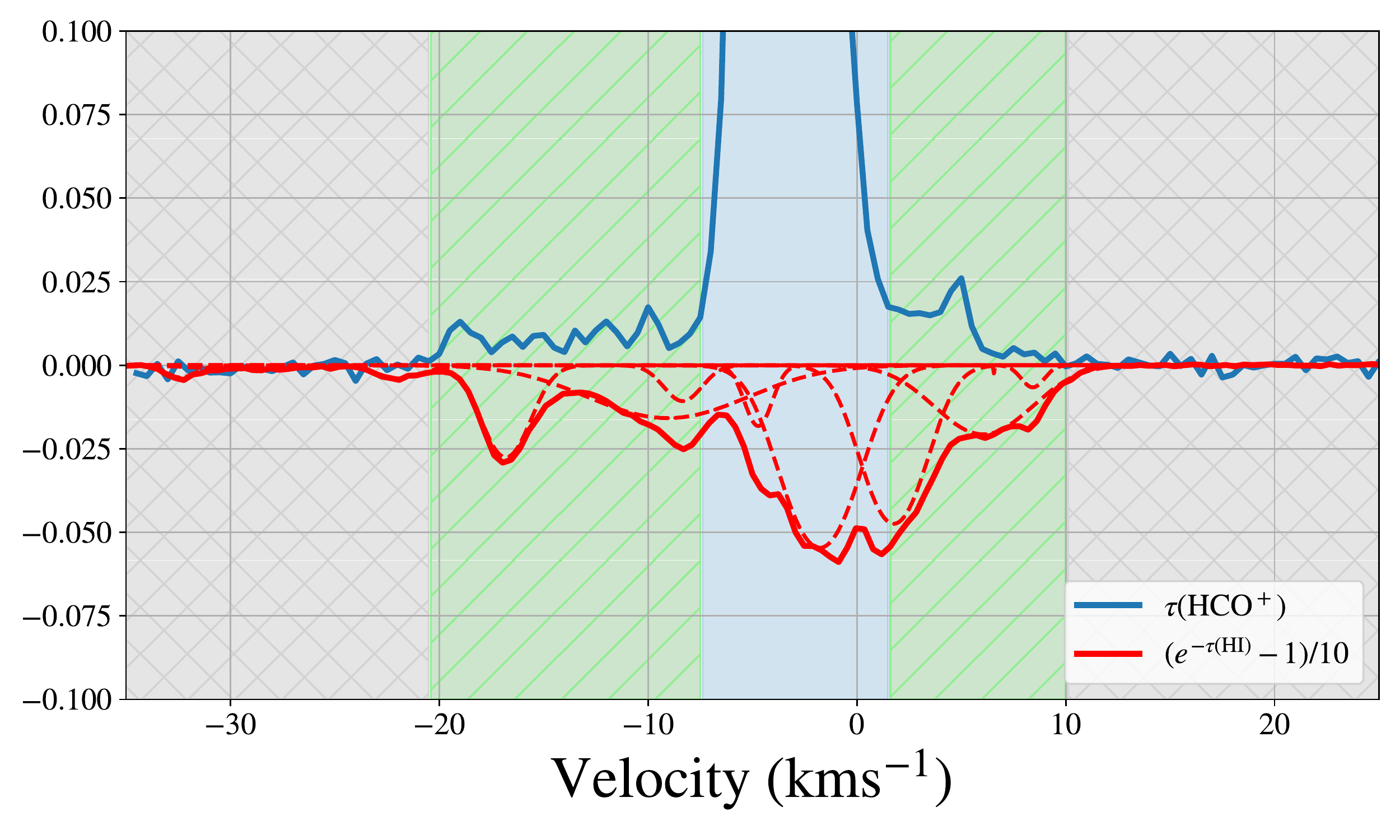}{0.4\textwidth}{3C111A}}
\gridline{\fig{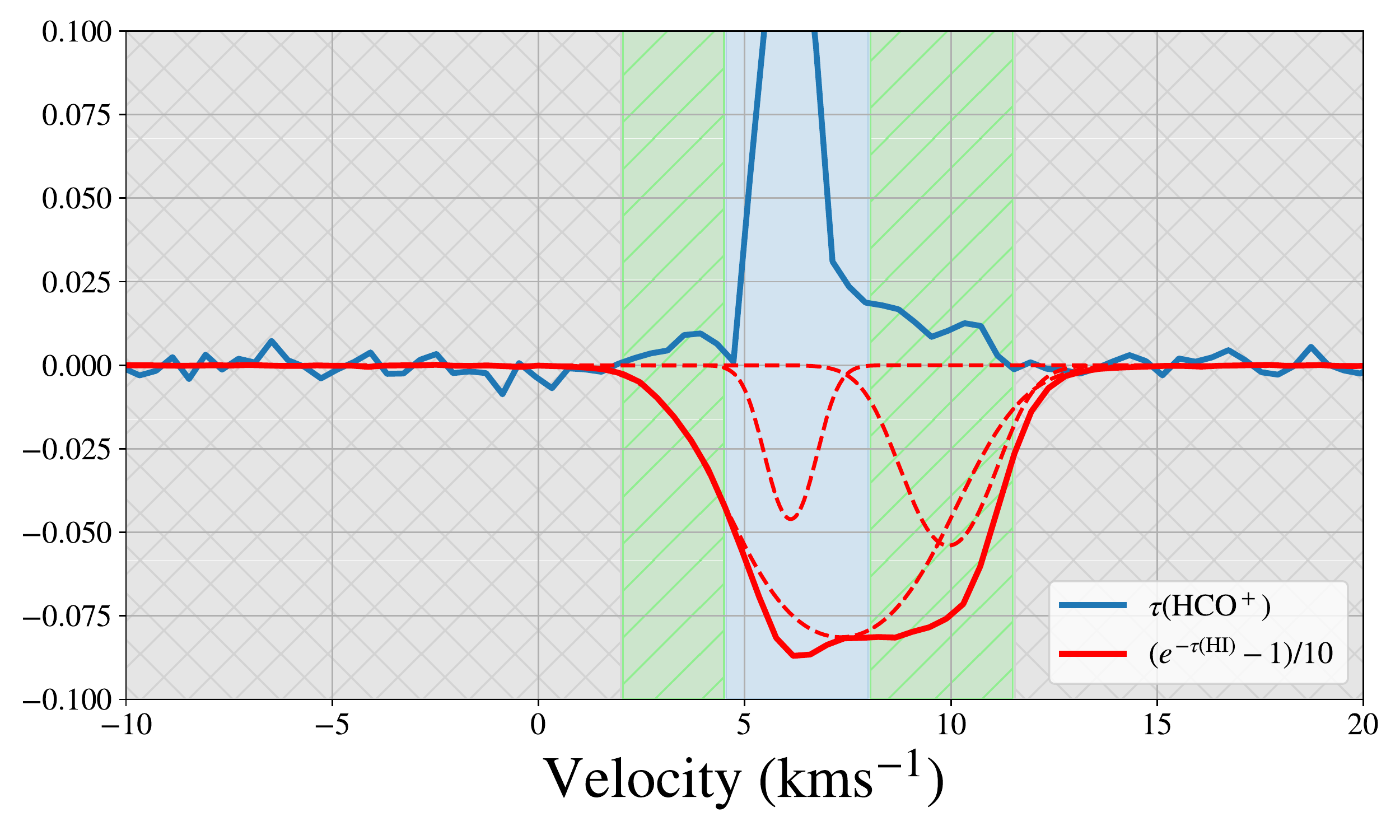}{0.4\textwidth}{3C120}
\fig{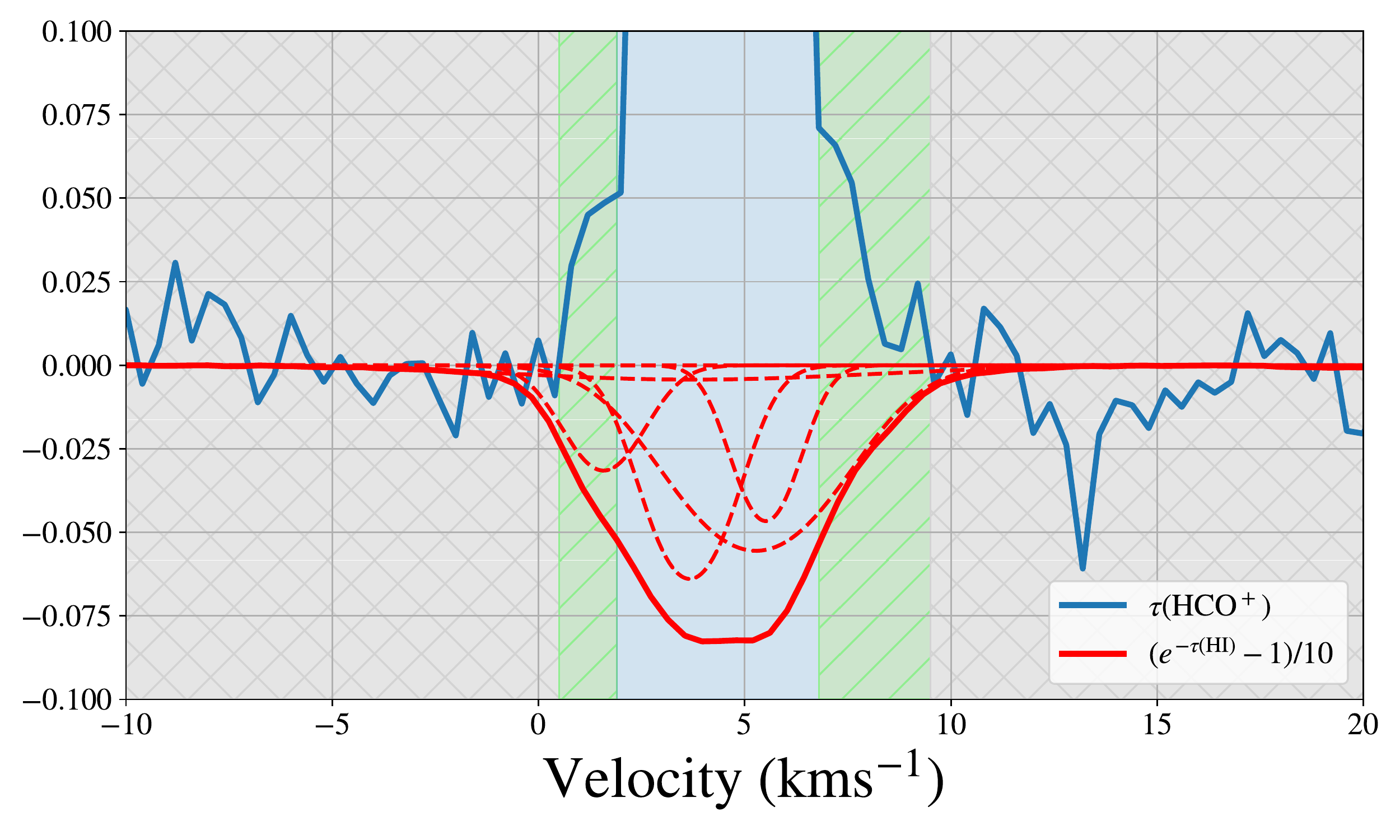}{0.4\textwidth}{3C123A}}
\gridline{\fig{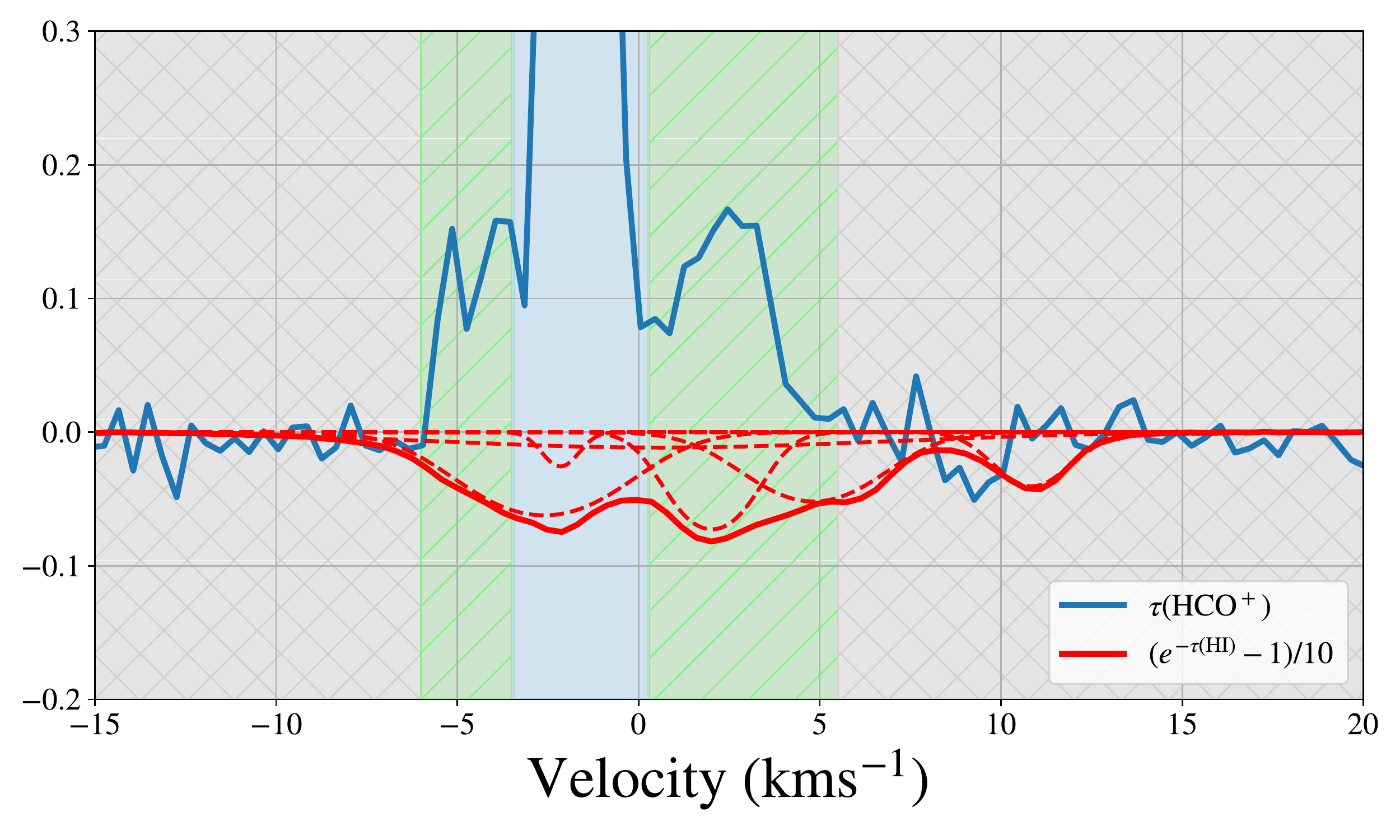}{0.4\textwidth}{3C154}
\fig{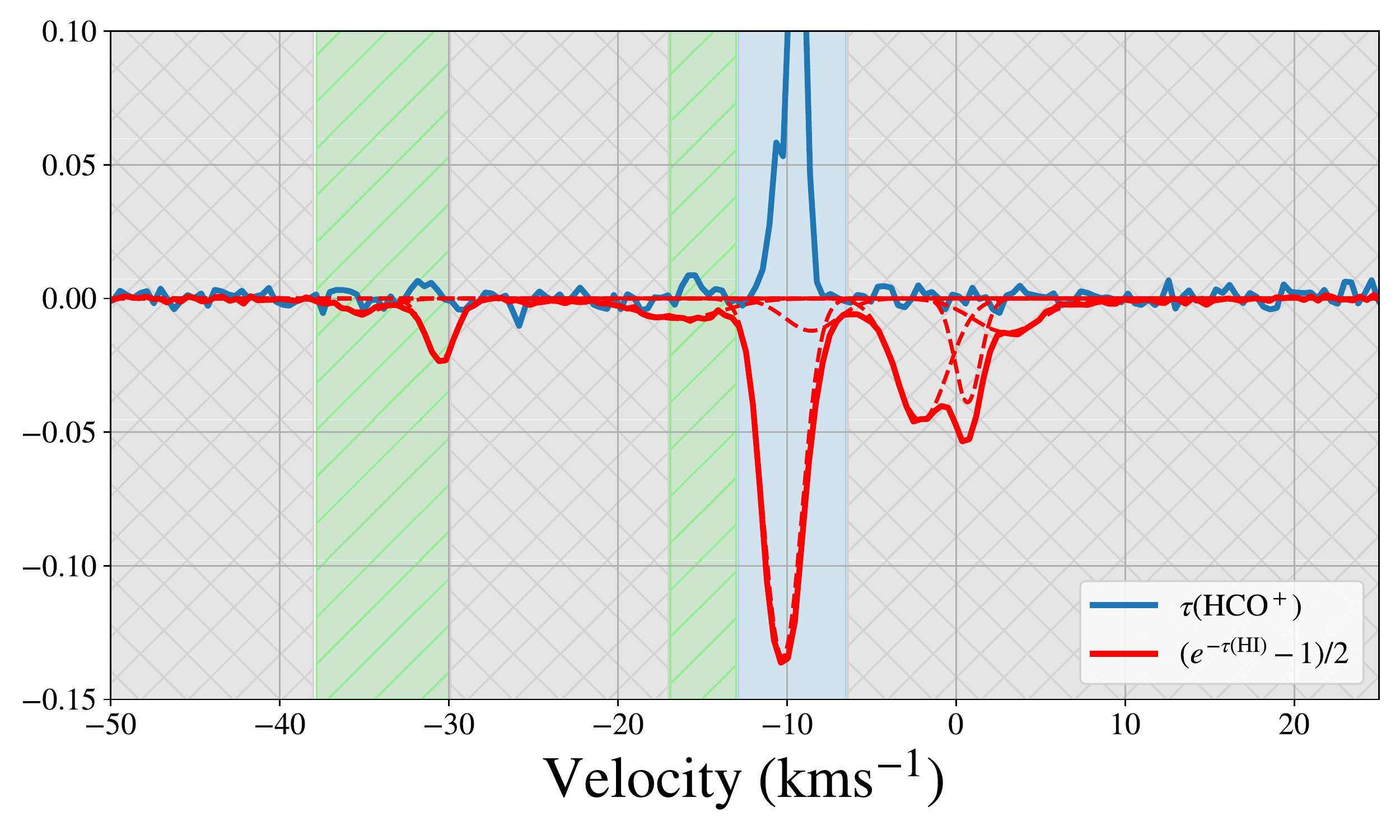}{0.4\textwidth}{3C454.3}}
\gridline{\fig{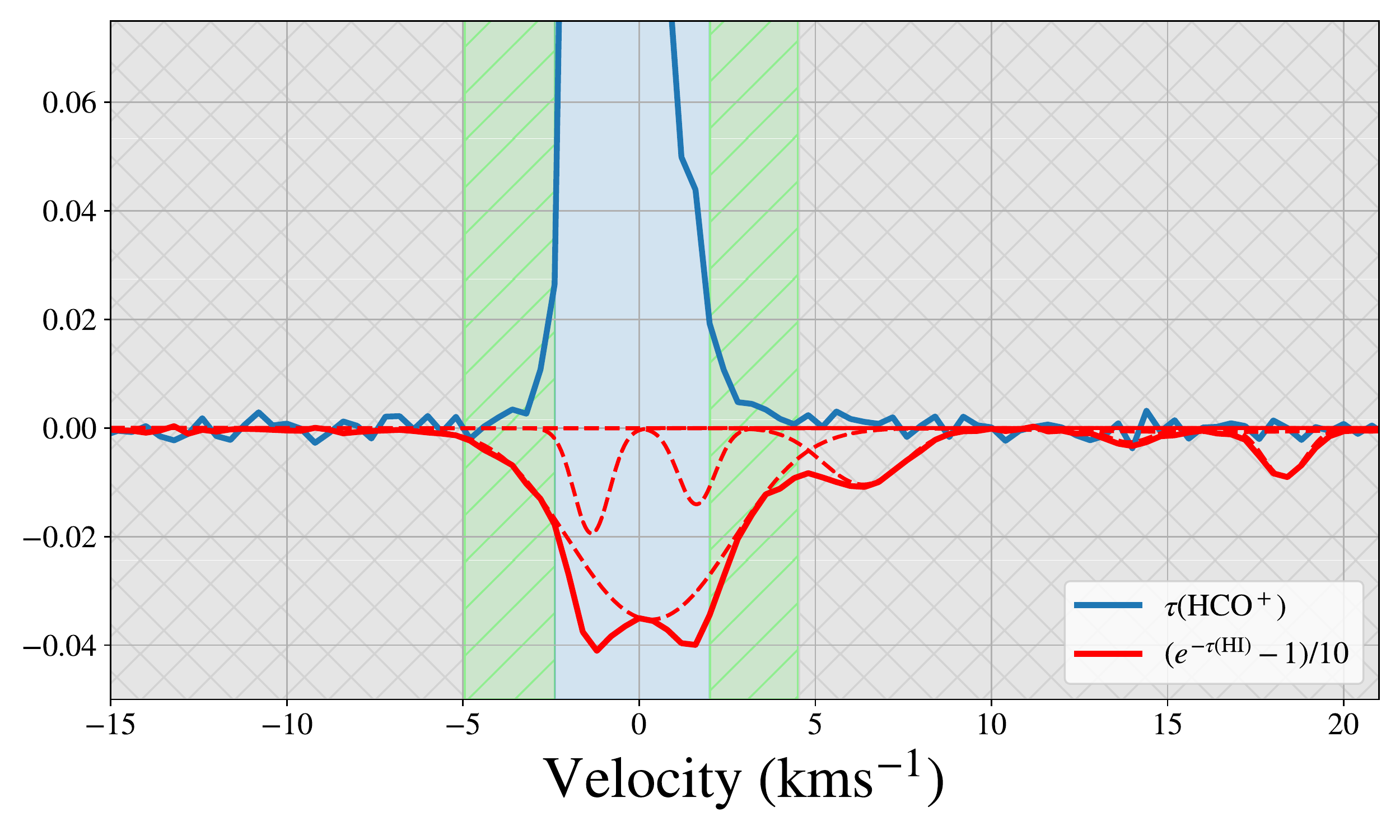}{0.4\textwidth}{BLLac}}
\caption{A comparison of the \hcop{} optical depth (blue, positive) with \hi{} absorption \citep[red, negative, with Gaussian components shown as dahsed lines;][]{2018ApJS..238...14M}. Regions without \hcop{} absorption are shaded in cross-hatched gray. Regions with weak, broad \hcop{} absorption are shaded in diagonally-hatched green. Regions of strong, narrow \hcop{} absorption are shaded in unhatched blue.}
\label{fig:broadHCOp}
\end{figure*}

\begin{figure}
    \centering
    \includegraphics[width=1.05\columnwidth]{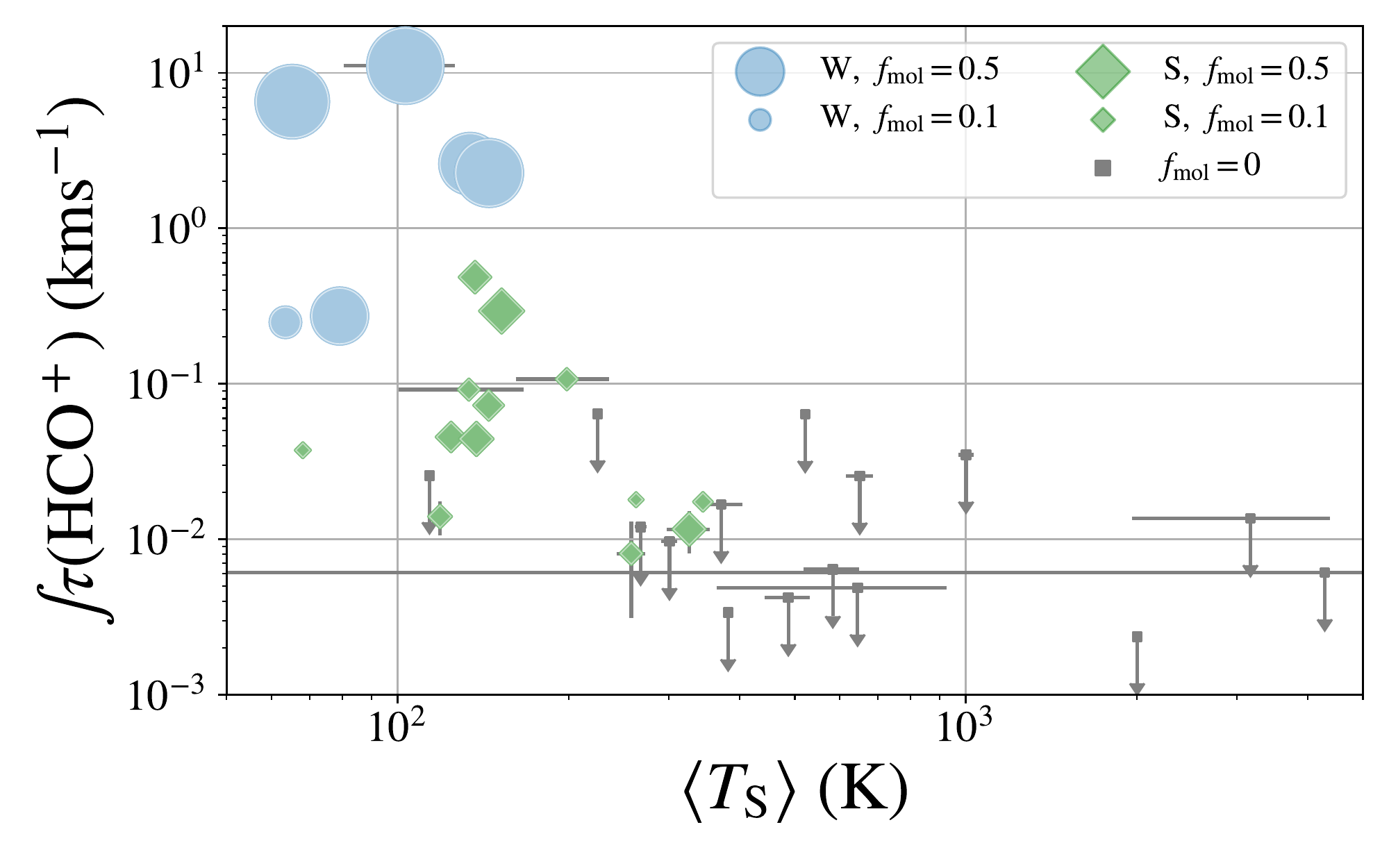}
    \caption{The integrated \hcop{} optical depth versus the \hi{} optical depth-weighted mean spin temperature in regions with strong \hcop{} absorption (blue circles), weak \hcop{} absorption (green diamonds), and no \hcop{} absorption (gray points, shown as upper limits), as illustrated in Figure \ref{fig:broadHCOp}. The sizes of the plotted points correspond to the molecular fraction, $f_\mathrm{mol}=2N(\htwo{})/N_\mathrm{H}$, and the colors indicate the \hi{} column densities.}
    \label{fig:int_hcop_v_Ts}
\end{figure}

\subsection{A broad, faint component of \hcop{} absorption}
\label{sec:broad_hcop}
In addition to the narrow, strong absorption features easily identifiable in the spectra shown in Figure \ref{fig:allspectra}, we also detect a broad, faint component of \hcop{} absorption in a majority of sightlines. Typical \hcop{} optical depths are between 0.01 and 0.1 for the broad component, while narrow absorption lines have typical optical depths $>0.1$. In the direction of 3C111A, 3C120, and 3C123A, we detect \hcop{} absorption across nearly all velocities where 21-SPONGE detected \hi{} absorption (Figure \ref{fig:broadHCOp}). The \hcop{} absorption in the direction 3C154 and BL Lac spans $\gtrsim10$ \kms{}, covering more than half of the velocity range where \hi{} absorption is observed. In the direction of 3C454.3, the \hcop{} absorption is localized mostly to a narrow range of velocities around $-10$ \kms{}, but there are marginal detections at $v\approx-30$ \kms{} and $v\approx-15$ \kms{}, where \hi{} absorption is also observed. In the direction of 3C78, \hcop{} absorption is weak, and confined only to a narrow velocity range compared to the \hi{} absorption. Some of these broad features are fit with Gaussians, while some features are only apparent in the residuals (since they have peak optical depths $\lesssim0.05$ and widths $\gtrsim3\kms{}$). We do not consider 3C111B or 3C123B here because of the relatively poor optical depth sensitivity in these directions. 

\subsubsection{Physical properties of the broad HCO$^+$ absorption}

Most \hcop{} absorption features detected in the literature have narrow line widths, except in a handful of more sensitive, recent studies, e.g., \citet{2000A&A...355..333L}, \citet{2018A&A...610A..49L}. To characterize the physical properties of the broad absorption features, we divide the spectra into velocity segments of no \hcop{} absorption ($\tau\lesssim0.01$), weak \hcop{} absorption ($0.01\lesssim\tau\lesssim0.05$), and strong \hcop{} absorption ($\tau\gtrsim0.05$).
These spectral regions are shown in cross-hatched gray, diagonally hatched green, and unhatched blue, respectively, in Figure \ref{fig:broadHCOp}. 

For each velocity region, we calculate the \hi{} optical depth-weighted mean spin temperature, 
\begin{equation}
\langle T_s \rangle=   \frac{ \int \tau(v)\, T_{\mathrm{B}}(v)/(1-e^{-\tau(v)})dv}{\int \tau(v) dv} .
\end{equation}
While the definition of $\langle T_s \rangle$ for selected velocity intervals is subjective, as it depends on where the velocity limits are set, it is useful for comparing atomic and molecular properties within the same velocity intervals \citep[see discussion below and ][]{2000A&A...355..333L}.

We also integrate the \hcop{} optical depth and estimate the \hi{} and \htwo{} column densities for each region.
We use the isothermal approximation to \hi{} column density, $N(\hi{})=1.823\times10^{18} \persc{}/(\kms{}~\rm{K}) \int \tau(v)\, T_{\mathrm{B}}/(1-e^{-\tau(v)})dv$. For the \htwo{} column density, we assume that $N(\hcop{})=2\times10^{-9}N(\htwo{})$ for all sightlines \citep{2000A&A...355..333L}\footnote{More recent work \citep[e.g.,][]{2010A&A...518A..45L} suggests  $N(\hcop{})/N(\htwo{})=3\times10^{-9}$. We use  $N(\hcop{})/N(\htwo{})=2\times10^{-9}$ to be consistent with \citet{2000A&A...355..333L}, who performed a similar analysis. Using $3\times10^{-9}$ produces results that are qualitatively the same.}. 

Figure \ref{fig:int_hcop_v_Ts} shows the integrated \hcop{} optical depth versus $\langle T_s \rangle$ in each spectral region. For regions with strong, narrow \hcop{} absorption, we use blue circles. For regions with broad, weak \hcop{} absorption, we use green diamonds. Regions with no \hcop{} absorption are shown as upper limits. Points are sized according to the molecular fraction, $f_\mathrm{mol}=2N(\htwo{})/N_\mathrm{H}$.

Figure \ref{fig:int_hcop_v_Ts} shows an anti-correlation between the integrated optical depth of \hcop{} and the optical depth-weighted mean spin temperature of \hi{}. As shown in numerical simulations by \citet{Kim2014}, the CNM fraction $f_{\rm {CNM}} \propto T_c/\langle T_s \rangle$, where $T_c$ is the intrinsic CNM temperature. \citet{2015ApJ...804...89M} showed that most 21-SPONGE observations are consistent with $f_{\rm{CNM}} \propto 50~\rm{K}/\langle T_s \rangle$ although there was a reasonable scatter. The strong \hcop{} absorption is associated with \hi{} that has $\langle T_s \rangle <140$ K, while weak and broad \hcop{} absorption is primarily associated with $\langle T_s \rangle$ in the range of 140--400 K (Figure \ref{fig:int_hcop_v_Ts}).
Thus, based on the relationship between $f_{\rm{CNM}}$ and $\langle T_s \rangle$ established by simulations and observations, the strong \hcop{} absorption traces gas with a high CNM fraction while the weak \hcop{} absorption traces gas with a lower CNM fraction. This is in agreement with the H$_2$ fraction shown in the same figure: a high H$_2$ fraction is associated with high CNM fraction and strong \hcop{} absorption, while lower H$_2$ and CNM fractions are associated with weak and broad \hcop{} absorption. 

Our observations suggest that broad HCO$^+$ absorption is common and associated with gas that has low CNM and H$_2$ fractions.
This diffuse molecular gas may originate in the outer layers of molecular structures. Such regions have been proposed to have significant turbulent motions and mixing between the CNM and WNM \citep{Valdivia2016H2,Lesaffre2020,Hennebelle2006}. 
Alternatively, this broad \hcop{} may simply trace lower density environments. Molecular gas is preferentially formed out of the CNM, so the low \hcop{} column densities may simply be a result of the relatively small quantity of CNM gas at these velocities.

\subsubsection{Broad HCO$^{+}$ absorption as  a tracer of the CO-dark molecular gas?}

\begin{figure}
    \centering
    \includegraphics[width=\columnwidth]{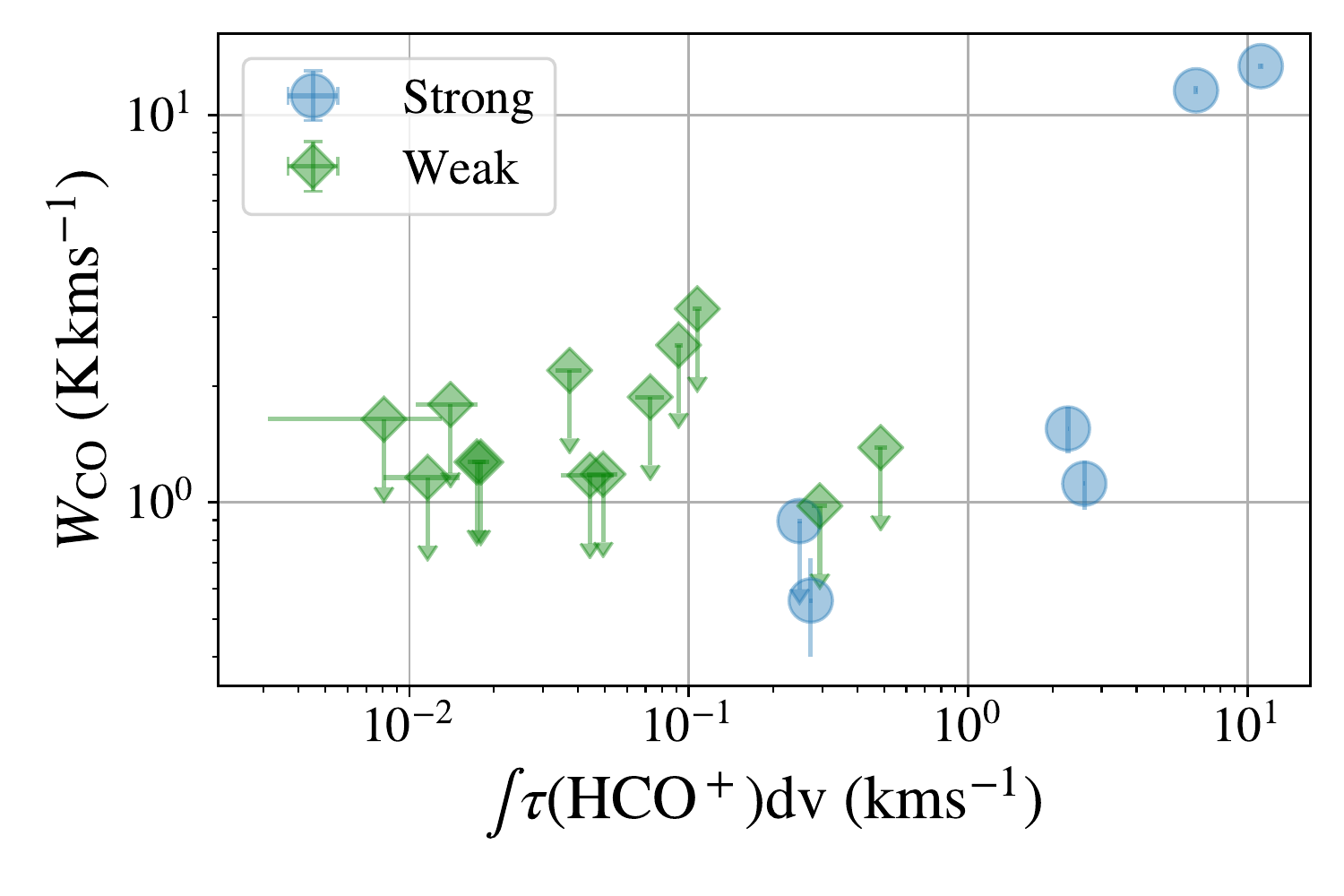}
    \caption{The integrated CO intensity ($W_{\rm{CO}}=\int T_b dv$) versus the integrated \hcop{} optical depth for velocity ranges with strong, narrow absorption (blue circles) and weak, broad absorption (green diamonds) outlined in Figure \ref{fig:broadHCOp}. The CO spectra are from Dame et al. (in preparation). Upper limits are indicated with downward arrows.}
    \label{fig:WCO_strongweak}
\end{figure}

It is intriguing to investigate how this broad component of \hcop{} absorption might contribute to CO dark molecular gas. \citet{2018A&A...617A..54L} have shown that \hcop{} is a CO dark molecular gas tracer in some environments and that CO line profiles are systematically narrower than \hcop{} line profiles. In our sample, CO emission spectra from \citet{2001ApJ...547..792D} do not show emission at velocities where broad, faint \hcop{} absorption is observed, as shown in Figure \ref{fig:WCO_strongweak}. The integrated CO intensity is plotted against the integrated \hcop{} optical depth for the regions outlined in Figure \ref{fig:broadHCOp}, with regions of strong, narrow absorption again shown in blue circles and regions with weak, broad absorption shown in green diamonds. 

At a level of $3\sigma$, we do not find any CO emission at the velocities where weak, broad \hcop{} absorption is observed. On the contrary, we detect CO emission in five of the six velocity ragnes where strong, narrow \hcop{} absorption is observed.
However, we note that the typical uncertainty in CO brightness temperature is $\sim0.15$--$0.3$ K per channel so our sensitivity may not be adequate to detect any potential low column density CO. 
Moreover, due to low excitation temperature, CO may not be detectable in 
emission  \citep[e.g.,][]{1996A&A...307..237L,2000A&A...355..333L}. The broad component of \hcop{} absorption is seen to be CO dark in at least one experiment with sensitive observations of \hcop{} and CO in absorption, though---see Figure 1 of \citet{2020ApJ...889L...4L}. 
Future ALMA and NOEMA observations of CO absorption in the direction of our sources are important to determine how much this broad, faint component to the \hcop{} absorption traces the total budget of the CO-dark molecular gas. Due to the low molecular fraction at these velocities (Figure \ref{fig:int_hcop_v_Ts}), though, this may trace only a small fraction of the CO-dark molecular gas.

\section{Comparison to atomic gas properties}
\label{sec:hi_molecules}
To investigate the connection between atomic gas properties and molecule formation, we compare the Gaussian-fitted components in the \hi{} absorption spectra to those in the molecular absorption spectra (see Figure \ref{fig:allspectra}). For each molecular absorption feature, we find the \hi{} absorption feature closest in velocity space and assume that they are associated with the same interstellar absorbing structure\footnote{Hereafter ``structure'' refers to the individual 
physical entities which correspond to velocity components 
identified in the 21-SPONGE Gaussian decomposition. }.
We make exceptions for low latitude sources, where velocity crowding introduces significant ambiguity. In these cases, if two \hi{} absorption features are within 1 \kms{} but one is a broad, weak feature ($\Delta v > 10$ \kms{}, $\tau < 0.1$), we take the narrower, stronger feature to be the most probable match, even if it is not the closest in velocity space. In cases where multiple molecular absorption features are matched to the same atomic absorption feature, we consider the sum of the column densities of all matched molecular components in the following analysis.

\begin{figure}
    \centering
    \gridline{\fig{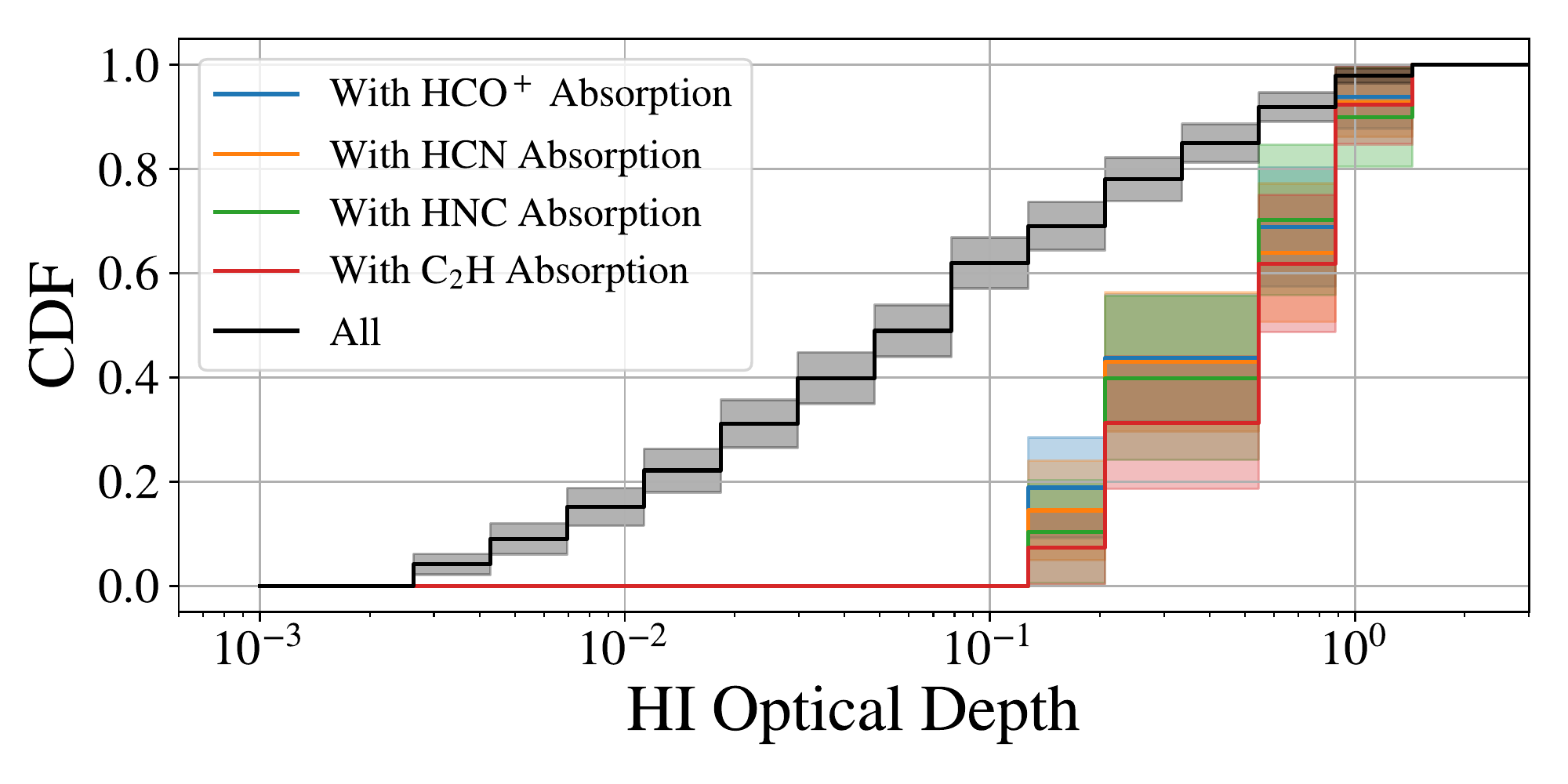}{\columnwidth}{}}
    \gridline{\fig{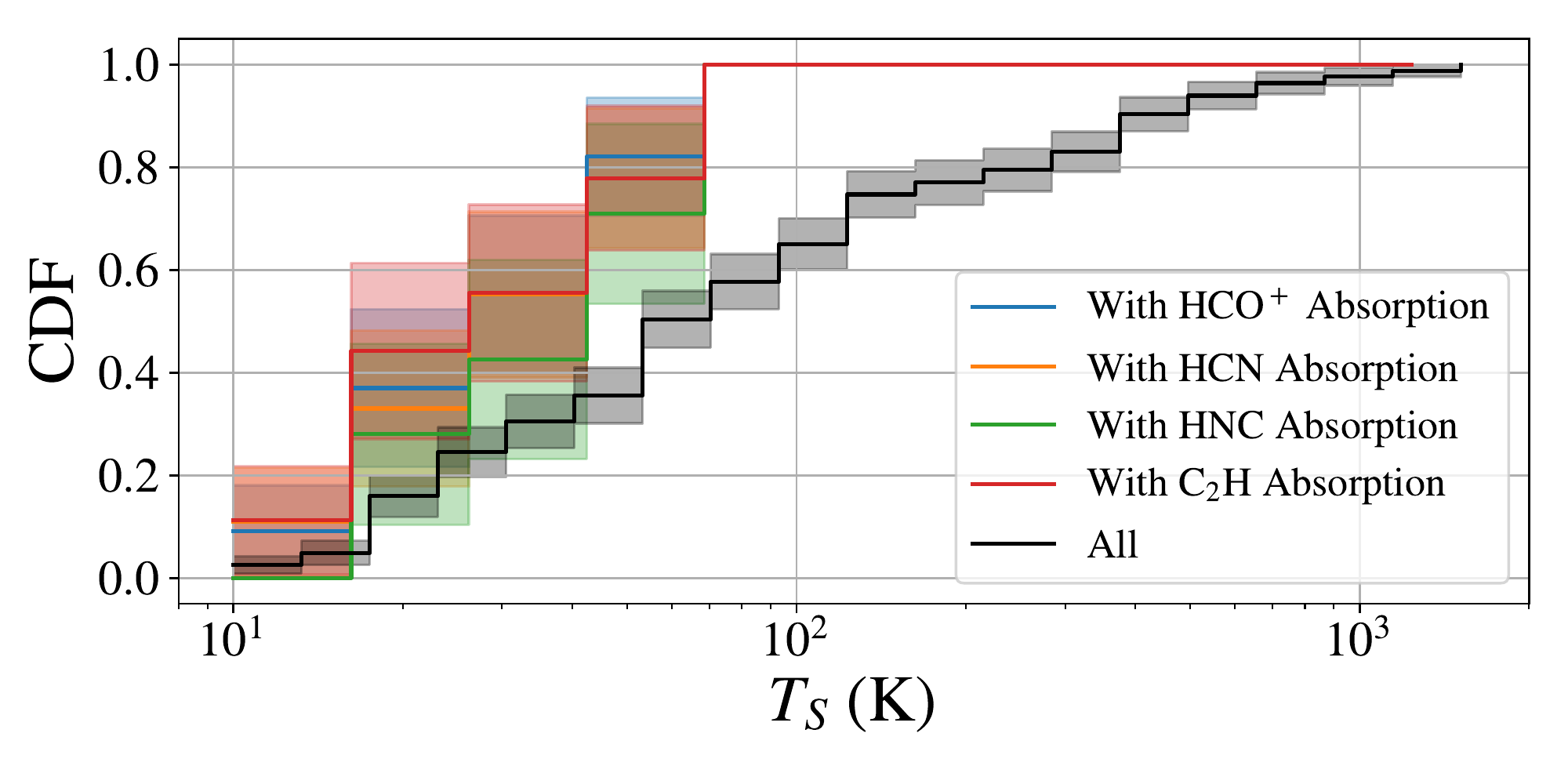}{\columnwidth}{}}
    \gridline{\fig{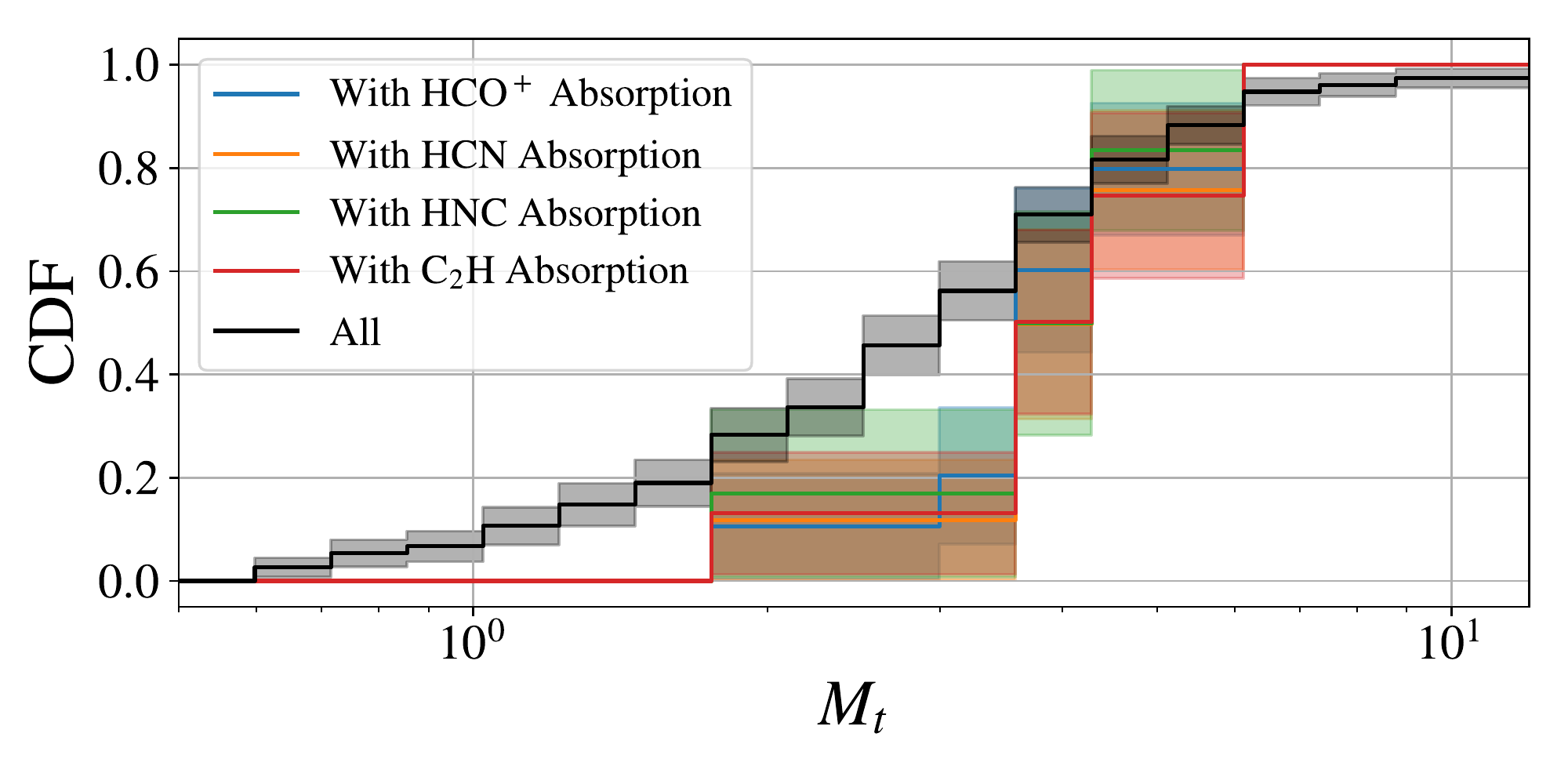}{\columnwidth}{}}
    \caption{CDFs of the \hi{} optical depth (top), the \hi{} spin temperature (middle) and the \hi{} turbulent Mach number (bottom) measured by \citet{2018ApJS..238...14M} in the direction of the background sources in this work. The CDF for all components is shown in black. The CDFs for \hi{} features associated with \hcop{}, \hcn{}, \hnc{}, and \cch{} absorption are shown in blue, orange, green, respectively. Error estimates derived from bootstrapping are shown.}
    \label{fig:hi_cdfs}
\end{figure}

In Figure \ref{fig:hi_cdfs}, we show the cumulative distribution functions (CDFs) of the spin temperature, optical depth, and turbulent Mach number of \hi{} structures seen in absorption towards our background sources (all quantities were constrained by 21-SPONGE). The turbulent Mach number---the ratio of the three dimensional turbulent velocity to the sound speed---is given by 
\begin{equation}
    M_t = 2.05 \Bigg(\frac{21.866 \Delta v_{\hi{}}^2}{T_s} - 1 \Bigg)^{1/2},
\end{equation}
where $\Delta v_{\hi{}}$ is the FWHM of the Gaussian \hi{} component.
The black CDF is for all \hi{} structures, while the blue, orange, green, and red CDFs are for the \hi{} structures that also show \hcop{}, \hcn{}, \hnc{}, and \cch{} absorption, respectively. 

From these CDFs, it is immediately clear that \hi{} structures with a molecular component have higher optical depths and lower spin temperatures (so lower kinetic temperatures) than the general population of cold \hi{} structures in the directions of our background sources. In particular, whereas the total population of \hi{} structures exhibits a large range in \hi{} optical depths and spin temperatures, the \hi{} structures with a molecular component all have optical depths $>0.1$ and spin temperatures $\lesssim80$ K. There is not such a marked difference in the turbulent Mach number of the different distributions, but the turbulent Mach numbers do appear higher for structures with a molecular component, $M_t\gtrsim2$. There is no apparent difference between different molecular species, but this analysis does not necessarily include the broad component \hcop{} absorption that is not always fit in our Gaussian decomposition (Section \ref{sec:broad_hcop}). Moreover, our sample size is modest, and few structures do not show absorption from all four species.

These results suggest that regions of the CNM with higher optical depths (more CNM increases the production of \htwo{} and other molecules) and lower temperatures (meaning more shielded against the ISRF) are more conducive to forming molecules. Such regions also tend to have a higher turbulent Mach number. However, in most cases the 1D turbulent velocities are not systematically higher for features with a molecular component, suggesting that the higher turbulent Mach number is largely a reflection of the lower $T_s$.

\begin{figure}
    \centering
    \gridline{\fig{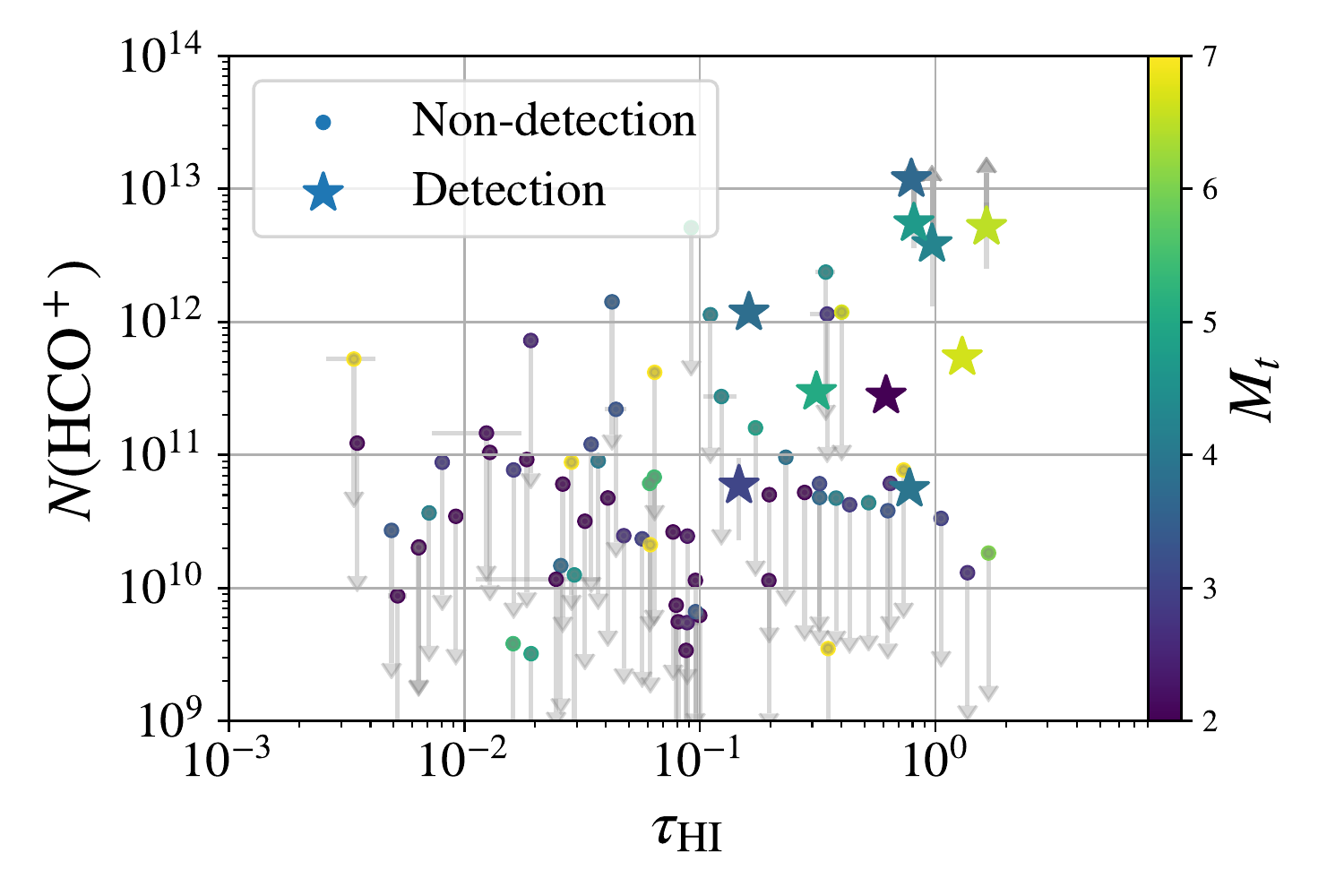}{\columnwidth}{}}
    \gridline{\fig{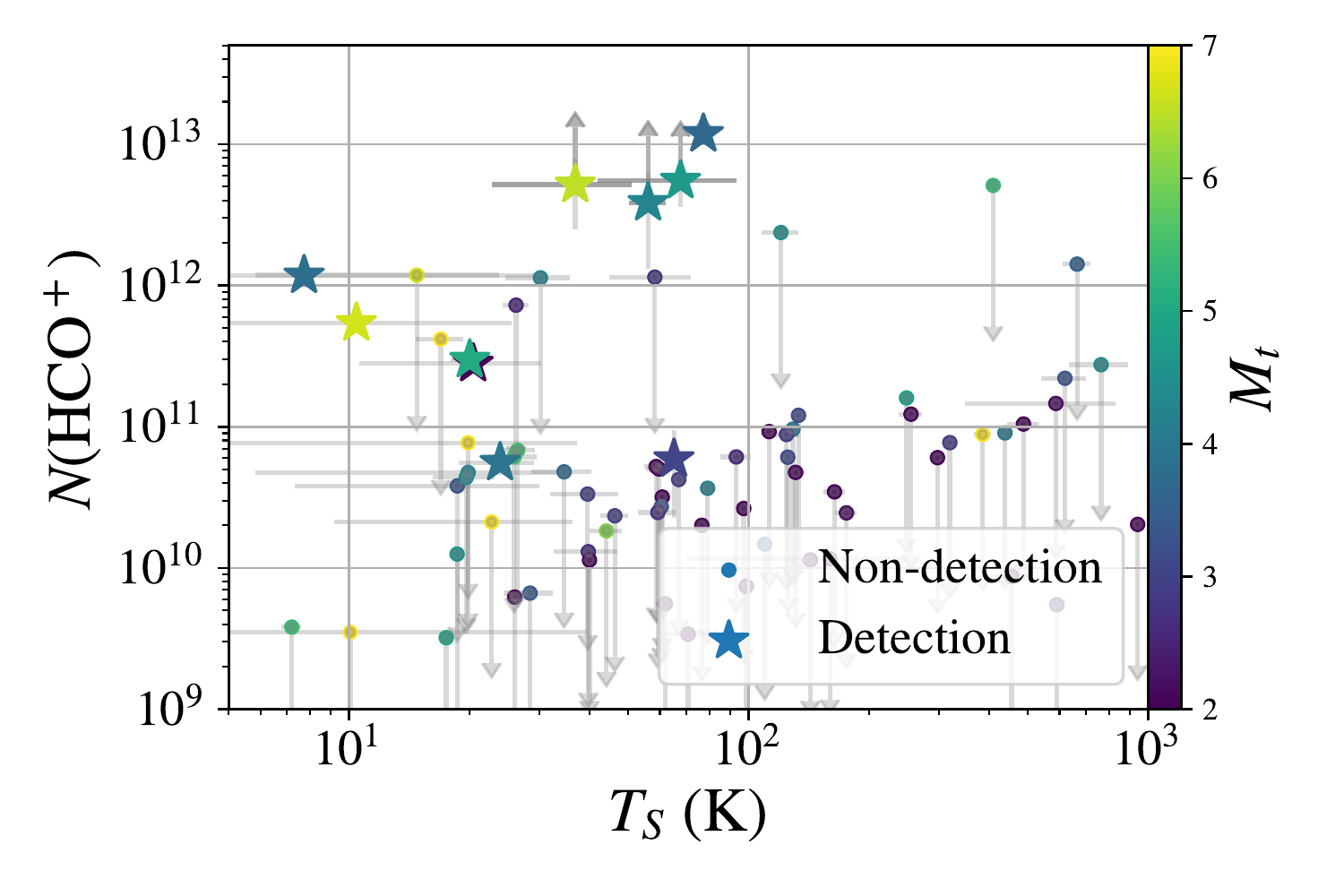}{\columnwidth}{}}
    \caption{The \hcop{} column density versus the \hi{} optical depth (top) and spin temperature (bottom) for Gaussian components identified in the \hi{} and \hcop{} absorption spectra. \hcop{} detections are shown as filled stars. The $3\sigma$ upper limits for non-detections are shown as filled circles. All points are colored according to the \hi{} turbulent Mach number, $M_t$. Structures for which 21-SPONGE could not determine the spin temperature are not included in these plots.}
    \label{fig:NHCOp_v_tauTs_nofits}
\end{figure}

These results are also summarized in Figure \ref{fig:NHCOp_v_tauTs_nofits}, where we compare \hcop{} column densities with \hi{} optical depths and spin temperatures. Points are colored according to turbulent Mach number. \hcop{} absorption detections are shown as stars, while $3\sigma$ upper limits for non-detections are shown as circles. We do not show here features for which 21-SPONGE was unable to determine the \hi{} spin temperature. Again, we see that molecular gas is associated with \hi{} that is colder, optically thicker, and has a higher turbulent Mach  number than the mean of the \hi{} cloud sample in this study. For all \hi{} structures along these sightlines, the mean values $(\tau, T_S, M_t)$ are $(0.2,171~\rm{K},3.6)$, while for \hi{} structures with a molecular component, the mean values are $(0.7,42~\rm{K},4.4)$.

Figure \ref{fig:NHCOp_v_tauTs_nofits} also shows that many components with $T_s<80$ K, as well as many components with $\tau>0.1$, do not have detected molecular column densities. This means that low $T_s$ and high \hi{} optical depth are necessary but not sufficient conditions for molecule production.

\begin{figure}
    \centering
    \gridline{\fig{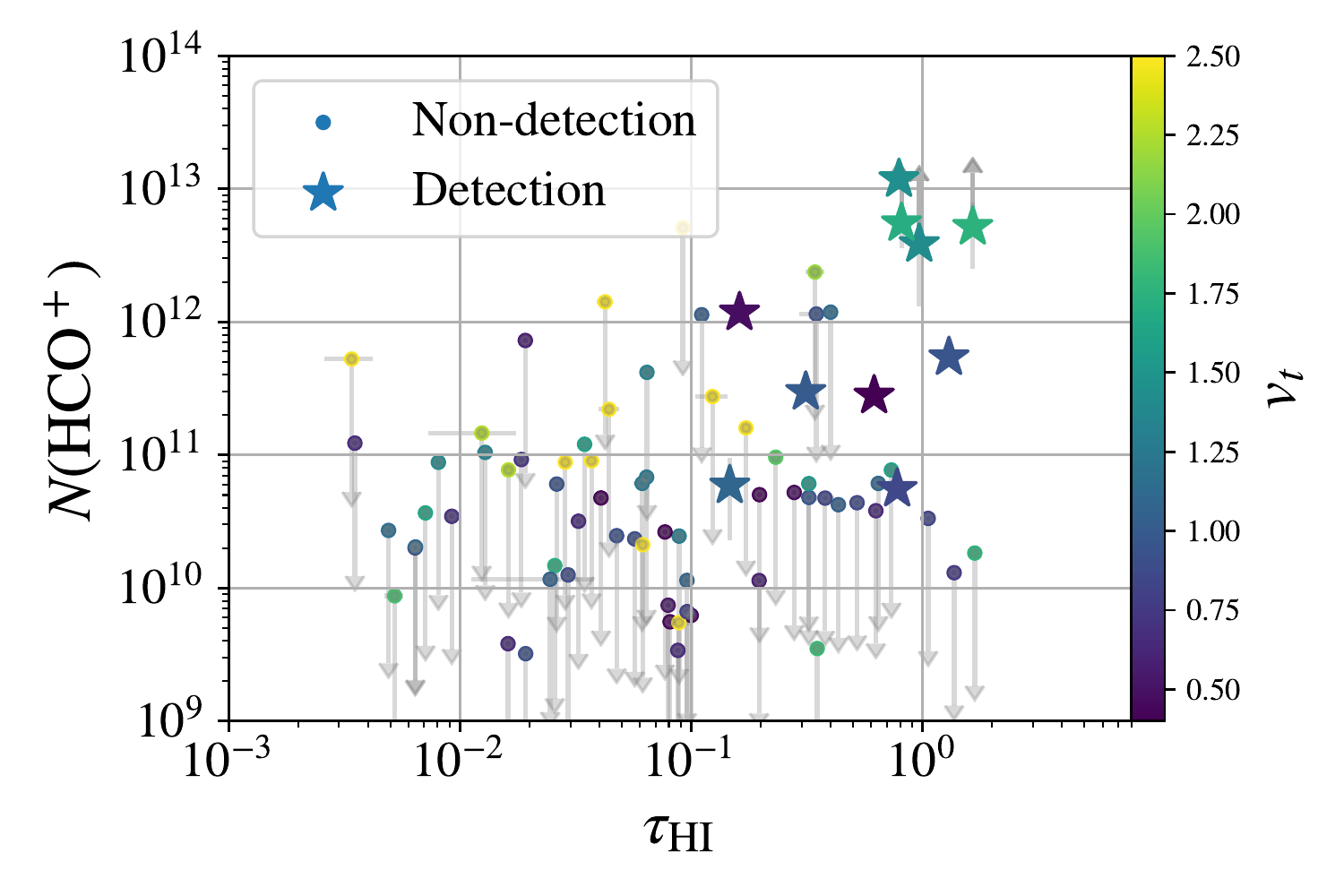}{\columnwidth}{}}
    \gridline{\fig{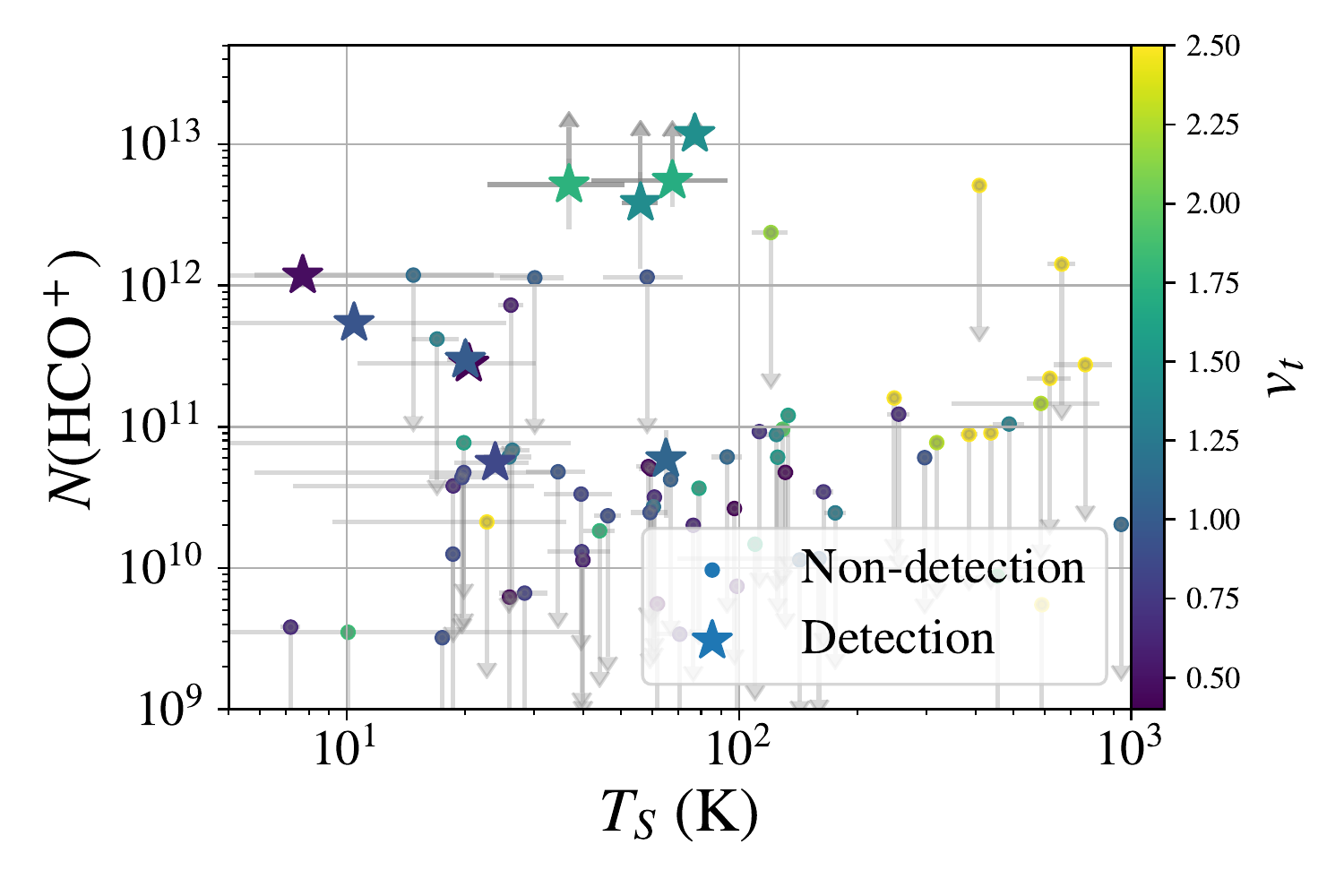}{\columnwidth}{}}
    \caption{Same as Figure \ref{fig:NHCOp_v_tauTs_nofits}, but points are colored according to the 1D turbulent velocity, $v_t$, of the atomic gas structures.}
    \label{fig:NHCOp_v_tauTs_nofits_vt}
\end{figure}

Previously, \citet{2010A&A...520A..20G} showed that it is hard to reproduce \hcop{} column densities larger than few $\times 10^{12}$ \persc{} in the diffuse ISM with PDR models. In Figure \ref{fig:NHCOp_v_tauTs_nofits} we see several detections with such high column densities. 
We also see that these components have \hi{} temperature in the range 40--80 K \citep[the peak of the CNM distribution was found to be around 50--60 K;][]{2003ApJ...586.1067H,2018ApJS..238...14M}, peak \hi{} optical depth $\sim1$, and $M_t>3$ (colors indicate turbulent Mach number). These four detections also have higher 1D turbulent velocities than detections that show only weak molecular absorption, $v_t\gtrsim1$ \kms{} (see Figure \ref{fig:NHCOp_v_tauTs_nofits_vt}, where points are colored according to the 1D turbulent velocity), suggesting that the higher turbulent Mach numbers are a product of both the colder temperatures and the turbulent velocities. We discuss these relatively warmer, more turbulent features further in \papertwo{}.

\section{Comparison of different molecular species}
\label{sec:species_comparison}

A comparison of different molecular species contains valuable information about formation and destruction processes. We therefore compare the column densities and absorption line properties of \hcn{}, \cch{}, \hcop{}, and \hnc{}, and place our results in the context of previous studies.

\subsection{Line of sight abundances} \label{subsec:los_abundances}

\begin{deluxetable}{ccc}
\tablenum{7}
\tablecaption{Column density ratios for different sightlines observed in this work. Column 1: name of background radio continuum source; Column 2: ratio of \hcn{} column density to \hnc{} column density; Column 3: ratio of \cch{} column density to \hcop{} column density. A dagger ($\dagger$) indicates that one or both of the measured column densities is a lower limit. \label{tab:line_ratios}}
\tablehead{\colhead{Source} & \colhead{$N({\hnc{}})/N({\hcn{}})$} & \colhead{$N({\cch{}})/N({\hcop{}})$}}
\startdata
3C111A     & $0.26 \pm 0.02^\dagger$ & $4.54 \pm 0.43^\dagger$ \\
3C120      & $0.19 \pm 0.05$ & $15.36 \pm 0.60$ \\
3C123A     & $0.17 \pm 0.01$ & $6.41 \pm 0.29$ \\
3C123B     & $0.27 \pm 0.05^\dagger$ & $1.99 \pm 0.62^\dagger$ \\
3C154      & $0.23 \pm 0.01$ & $13.79 \pm 0.58$ \\
3C454.3    & $0.20 \pm 0.02$ & $9.33 \pm 0.36$ \\
\enddata
\end{deluxetable}

The abundance of molecular species studied here is determined by the rates of formation and destruction processes and may depend on different interstellar environments. For example, in cold and dense interstellar clouds, where the gas is largely shielded from external ultraviolet radiation, the HNC/HCN abundance ratio is expected to be close to 1 based on equilibrium chemical models  \citep{2017ApJ...838...33A}. However, in regions illuminated by ultraviolet photons, HNC is photodissociated faster than HCN resulting in HCN being significantly more abundant than HNC. This could happen in both diffuse interstellar clouds \citep{2001A&A...370..576L,2010A&A...520A..20G} and in photon-dominated regions \citep{2017ApJ...838...33A}. 
Similarly, considering that both \hcop{} and \cch{} are products of CH$_3^+$, the abundances of these molecules are expected to be correlated, e.g. \citet{2009A&A...495..847G}.

For the four lines of sight where both \hnc{} and \hcn{} are detected (3C120, 3C123A, 3C154, and 3C454.3, excluding lower limits; see Table \ref{tab:line_ratios}) we find $\langle N(\hnc{})/N(\hcn{})\rangle=0.20\pm0.02$. Similarly, for the four lines of sight where both \cch{} and \hcop{} are detected (3C120, 3C123A, 3C154, and 3C454.3, excluding lower limits; see Table \ref{tab:line_ratios}), we find  $\langle N(\cch{})/N(\hcop{})\rangle=13.3\pm6.9$. This ratio decreases with increasing $N(\hcop{})$ (see Section \ref{sec:gauss_results} for a further discussion of this observation). Our results are consistent with previous measurements from \citet{2001A&A...370..576L}, who found $\langle N(\hnc{})/N(\hcn{})\rangle=0.21\pm0.05$, and   \citet{2000A&A...358.1069L}, who found $\langle N(\cch{})/N(\hcop{})\rangle=14.5\pm6.7$.

Our results for HCN and HNC are generally not consistent with predictions from PDR models from the literature. For example, \citep{2017ApJ...838...33A} used the Meudon PDR code \citep{2006ApJS..164..506L} to model the chemistry of a typical diffuse cloud. They considered a plane-parallel cloud with a total visual extinction of $A_V=1$ mag illuminated at both sides by the ISRF of 1--3 Draine fields, and considered a range of densities of H nuclei of $10^2$--$10^4$ \percc{}. They found $N(\hnc{})/N(\hcn{})\sim0.5$--1.  While these results are only slightly higher than the observed values, they propagate to CN/HCN ratios one or two orders of magnitude above the observed value.

\subsection{Gaussian components} \label{sec:gauss_results}

\subsubsection{Column density ratios} \label{subsubsec:gauss_column}
\begin{figure}
    \centering
    \gridline{\fig{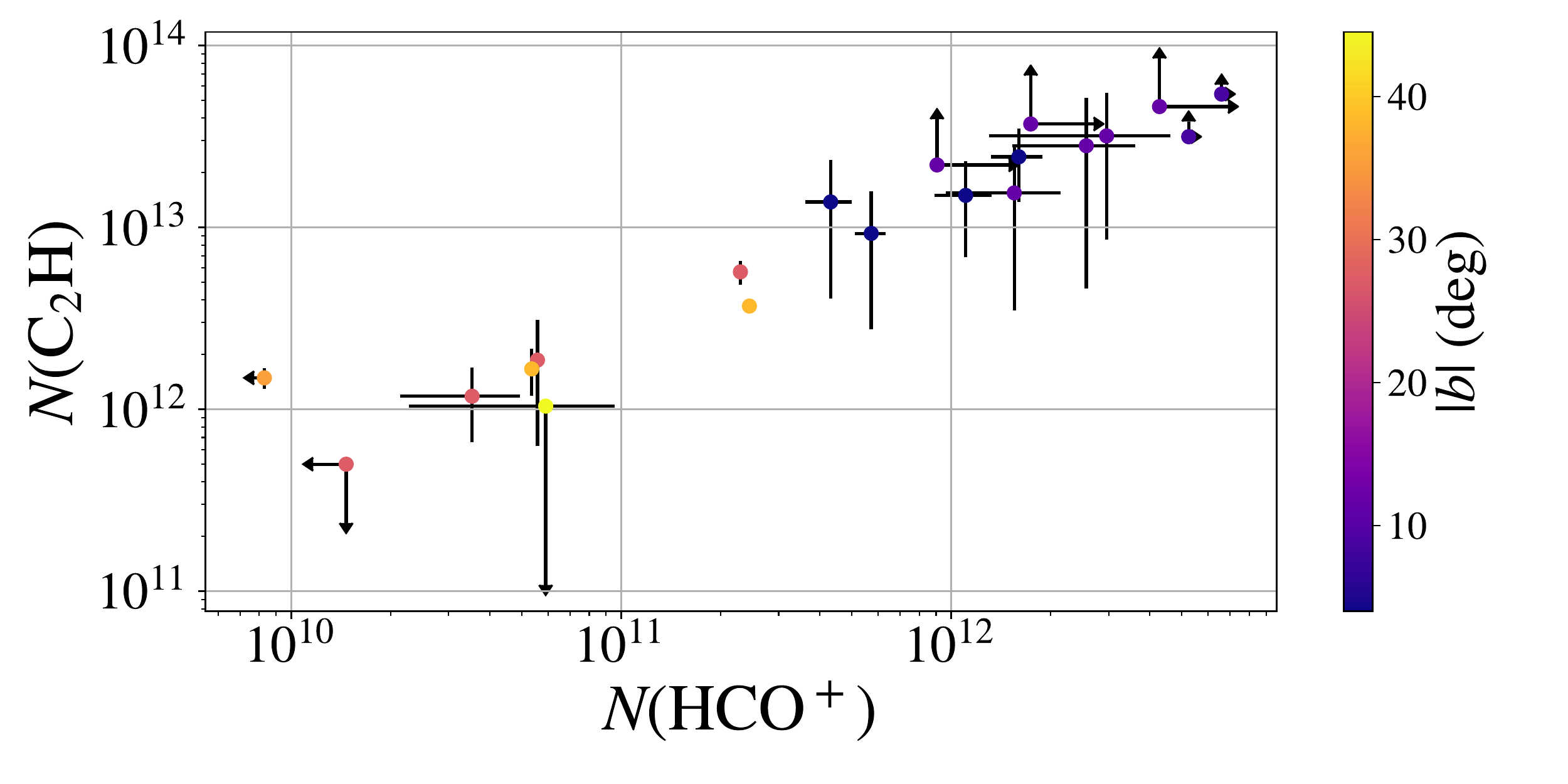}{1.1\columnwidth}{}}
    \gridline{\fig{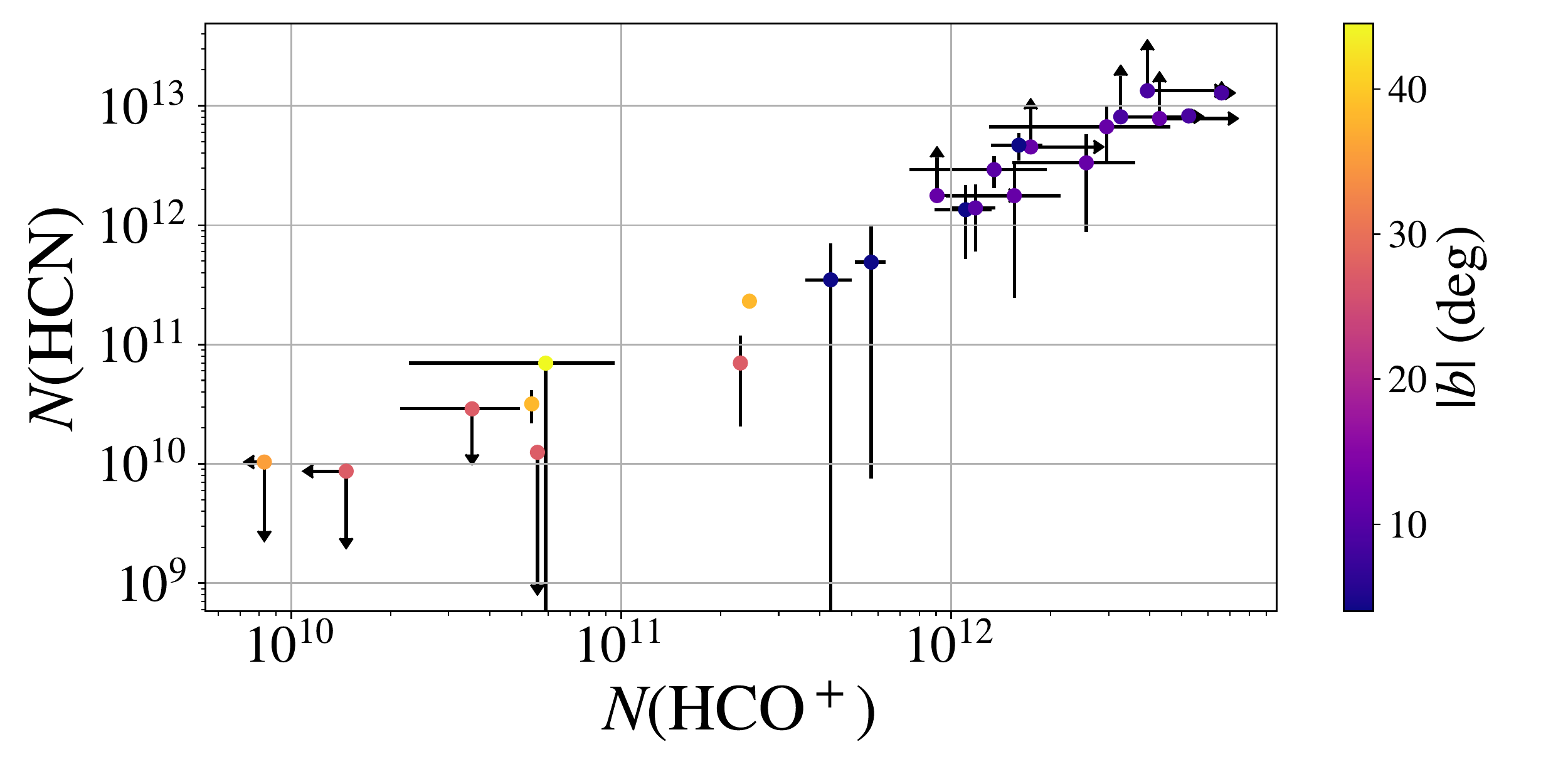}{1.1\columnwidth}{}}
    \gridline{\fig{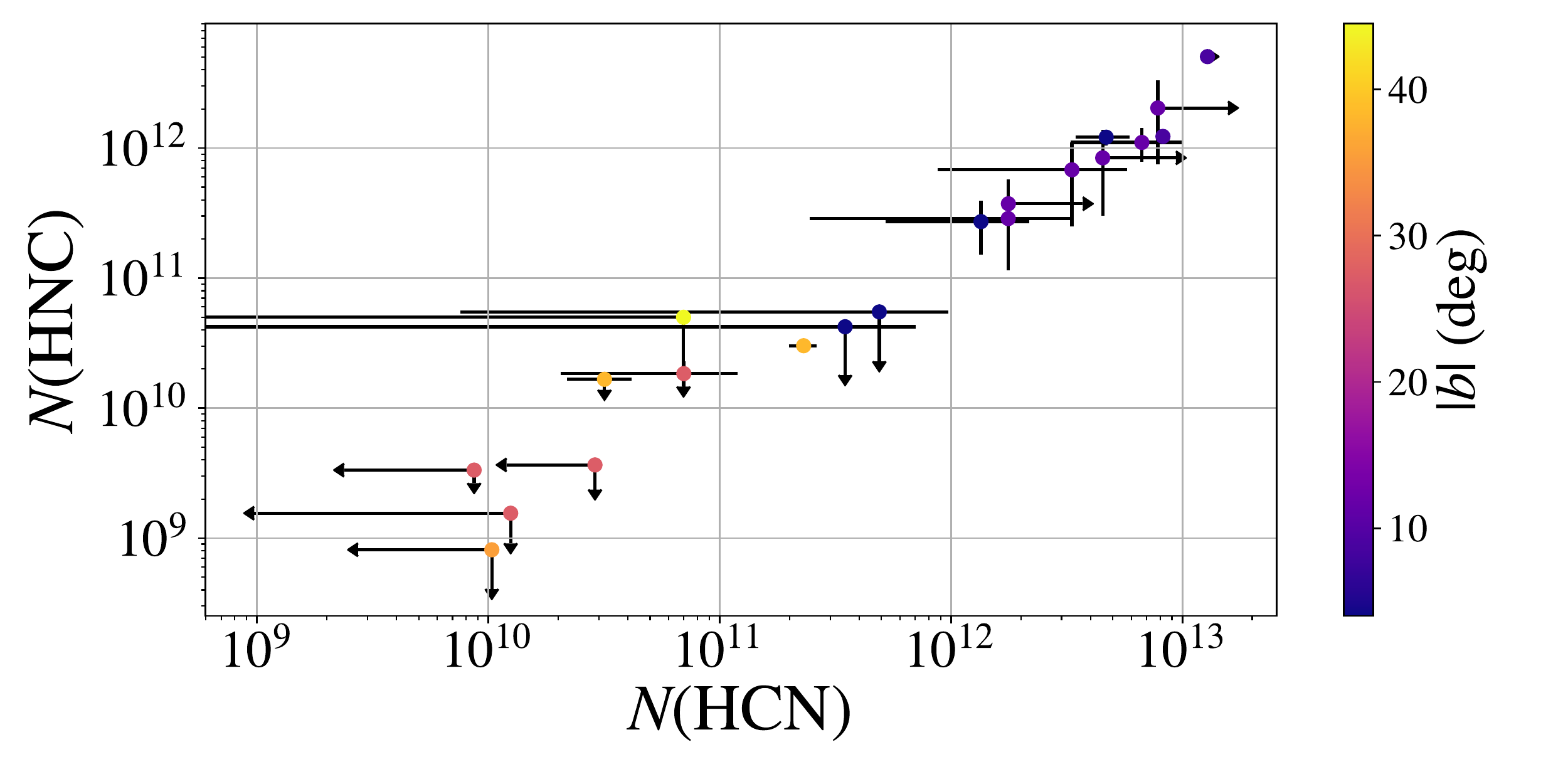}{1.1\columnwidth}{}}
    \caption{Comparisons of the \cch{} column densities to the \hcop{} column densities (top), the \hcn{} column densities to the \hcop{} column densities (middle), and the \hnc{} column densities to the \hcn{} column densities (bottom), for features identified in our Gaussian decomposition.}
    \label{fig:column_ratios_g}
\end{figure}

We measure the column density of each Gaussian component, where $\int\tau dv = 1.064 \, \tau_0 \, \Delta v_0$ for a Gaussian feature with peak optical depth $\tau_0$ and FWHM $\Delta v_0$. Figure \ref{fig:column_ratios_g} shows the distribution of these column densities. The slope of $N(\cch{})/N(\hcop{})$ becomes shallower for $N(\hcop{})\lesssim10^{12}$ \persc{}---this explains why $N(\cch{})/N(\hcop{})$ decreases at high $N(\hcop{})$. The \hcn{} column density increases faster than linearly with the \hcop{} column density. The \hnc{} and \hcn{} column densities scale linearly with each other. 

We find $\langle N(\cch{})/N(\hcop{}) \rangle = 18.8 \pm 9.5$, $\langle N(\hcn{})/N(\hcop{}) \rangle = 1.7 \pm 0.9$, and $\langle N(\hnc{})/N(\hcn{}) \rangle = 0.22 \pm  0.07$ (or equivalently, $\langle N(\hcn{})/N(\hnc{}) \rangle = 4.9 \pm 1.3$). These ratios are consistent with those measured along the total line of sight (Section \ref{subsec:los_abundances}), including the large dispersion in $N(\cch{})/N(\hcop{})$ and $N(\hcn{})/N(\hcop{})$. Previously, \citet{2010A&A...520A..20G} found $\langle N(\hcn{})/N(\hcop{}) \rangle = 1.9 \pm 0.9$ and $\langle N(\hcn{})/N(\hnc{}) \rangle = 4.8 \pm 1.3$ at low Galactic latitudes ($|b|<1$), and \citet{2001A&A...370..576L} found $\langle N(\hnc{})/N(\hcn{}) \rangle = 4.8 \pm 1.1$ and $\langle N(\hcn{})/N(\hcop{}) \rangle = 1.47 \pm 0.86$ at higher Galactic latitudes ($1.6<|b|<38.2$). Our results, measured in the direction of
sources with latitudes $4.0 < |b| < 38.2$, are consistent with both of these previous works, although two of our sources were also included in \citet{2001A&A...370..576L}\footnote{We find at most marginal differences between the column densities measured in this work with the column densities measured in previous works for any overlapping sources; see Section \ref{sec:temporal_stability}.}.  

The similarity in line ratios across Galactic latitude seen by \citet{2001A&A...370..576L}, \citet{2010A&A...520A..20G}, and this work suggest that gas in the Galactic plane is chemically similar to gas in the solar neighborhood. This trend was noticed by \citet{2010A&A...520A..20G}, but the results here verify that this relationship persists to lower column densities (by a factor of a few) than probed by either of the previous experiments.

\subsubsection{FWHM ratios}
\begin{figure}
    \centering
    \includegraphics[width=\linewidth]{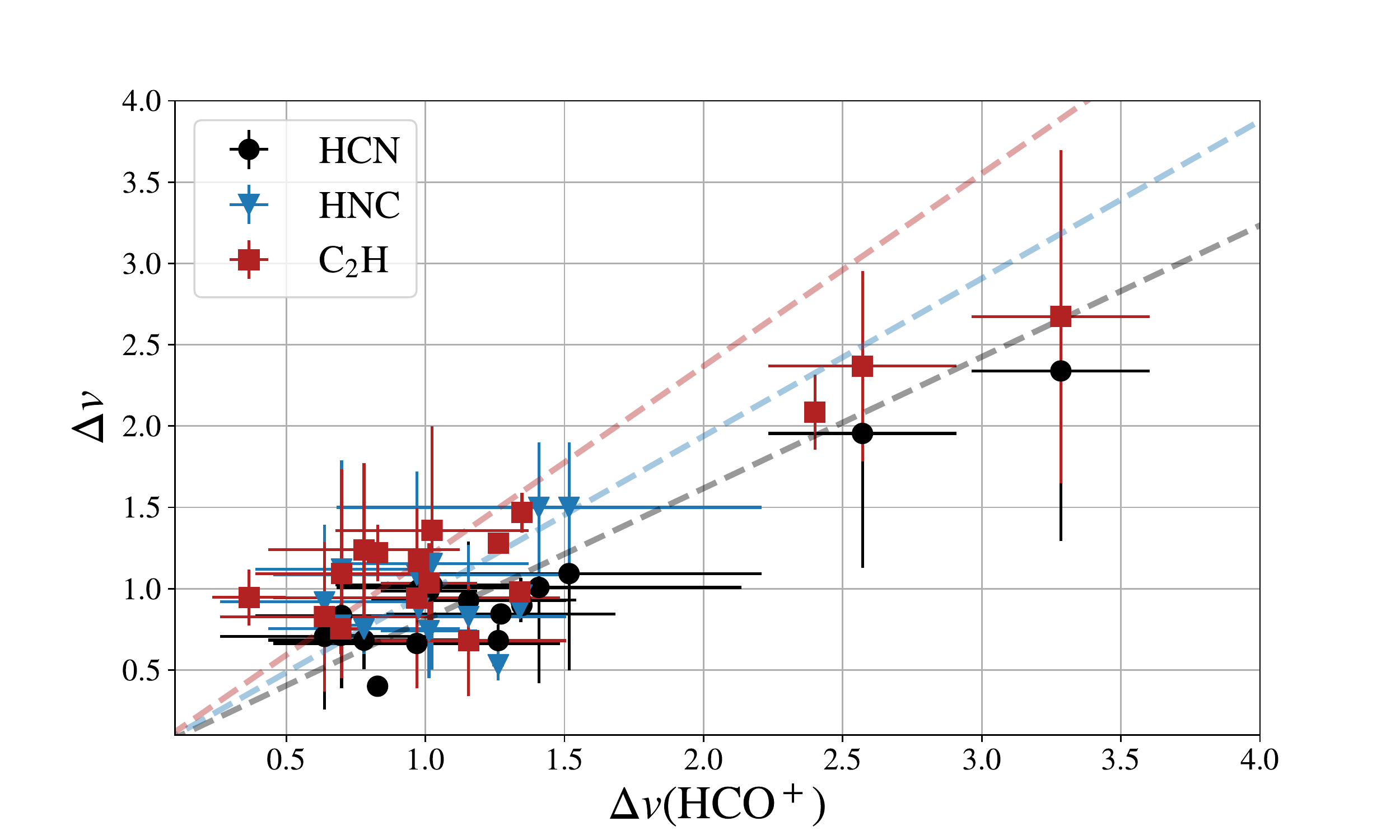}
    \caption{A comparison of the FWHMs of Gaussian fitted functions (Tables \ref{tab:cch_fits}, \ref{tab:hcn_fits}, \ref{tab:hcop_fits}, \ref{tab:hnc_fits}). \hcop{} FWHMs are shown on the $x$-axis. The $y$-axis displays the FWHMs for \hcn{} (black circles), \hnc{} (blue triangles), and \cch{} (red squares).}
    \label{fig:fwhm_comp}
\end{figure}

It has long been known that different molecular species in the diffuse ISM do not share the same kinematics. In particular, differences in the linewidths of absorption/emission features of different species have been suggested as evidence for the formation of these species under different environmental conditions. We compare the FWHMs between different species observed here in Figure \ref{fig:fwhm_comp}. The mean ratio of the \cch{} FWHMs to the \hcop{} FWHMs is roughly $1.2:1$. The mean ratio of the \hcn{} FWHMs to the \hnc{} FWHMs is approximately $1:1$. The mean ratio of the \hcn{} FWHMs to the \hcop{} FWHMs and the mean ratio of the \hnc{} FWHMs to the \hcop{} FWHMs are both roughly $0.9:1$. Our measurement of the relative narrowness of HCN and HNC with respect to \hcop{} is consistent with \citet{2010A&A...520A..20G}, whose data set also included measurements from \citet{1996A&A...307..237L} and \citet{2001A&A...370..576L}. We note, though, that there is considerable scatter in Figure \ref{fig:fwhm_comp} and that these measured ratios have uncertainties of $\sim0.3$, meaning that our estimates for the line ratios are also not statistically significantly different from $1:1$.

\citet{2001A&A...370..576L} and \citet{2010A&A...520A..20G} previously observed that absorption by CN-bearing molecules is narrower than absorption by \hcop{} at mm wavelengths, while \citet{1990ApJ...359L..19L} showed that CN absorption is systematically narrower than CH\textsuperscript{+} absorption at optical wavelengths. \citet{2019A&A...627A..95L} also found that CO absorption is narrower than \hcop{} absorption at mm wavelengths. Molecular absorption lines are turbulently broadened, so it has been suggested that \hcop{} and CH\textsuperscript{+} lines are broader because these species are formed preferentially in turbulent, dynamic environments like TDRs and shocks \citep{2010A&A...520A..20G,2019A&A...627A..95L}. \citet{1995ApJS...99..107C} and \citet{2005ApJ...633..986P} suggested that the differences in linewidths indicated that species like \hcop{} and CH\textsuperscript{+} trace more diffuse regions such as cloud envelopes, while species like CN and HCN trace denser regions or molecular clouds. \citet{2001A&A...370..576L} also note that the molecular absorption linewidths may be produced by the turbulence of many structures along the line of sight rather than a single ``cloud.'' We detect \hcop{} linewidths that are marginally broader than \hcn{} or \hnc{} linewidths, consistent with the idea that \hcop{} forms in more turbulent environments than CN bearing species. Nevertheless, a larger data set is needed to place tighter statistical constraints on the relative widths of different species.

\section{Temporal stability of line profiles}
\label{sec:temporal_stability}
Temporal variations in absorption line profiles against background radio continuum sources have long been used as a probe for small (AU) scale  structure in the ISM \citep[][and references therein]{2018ARA&A..56..489S}.
Such variability has been observed in \hi{} \citep[e.g.,][]{1985A&A...146..223C,1986ApJ...303..702G,1989ApJ...347..302D,2001AJ....121.2706F,2005AJ....130..698B,2012ApJ...749..144R} and several molecular lines \citep{1993ApJ...419L.101M,1995ApJ...452..671M},
although the interpretation of the observed variations has been controversial \citep{1996A&A...312..973T,1997ApJ...481..193H,2000MNRAS.317..199D}. 
The observed AU-scale structure in \hi{} often has densities high enough for the existence of various molecular species. Detecting variability in molecular spectral lines therefore offers an exciting way of probing  internal structure of over-dense \hi{} structures.

In Table \ref{tab:temp_variability}, we compare the integrated optical depths for \hcn{}, \cch{} (considering only the two transitions listed in Table \ref{tab:lines}), \hcop{}, and \hnc{} measured in this work to those measured in previous surveys \citep{1996A&A...307..237L,2000A&A...358.1069L,2001A&A...370..576L}\footnote{The Lucas \& Liszt works do not observe separate components of 3C111, but instead only report one component. We assume that their observations correspond to the A component, as it is the brightest of the 3 components by over an order of magnitude.}.  Observations from these previous works were obtained between 1993 and 1997 with different antenna and receiver setups from those used here and reduced using different software. Our observations were obtained between 2018 and 2020, meaning that we are probing optical depth changes over a span of $\sim25$ years in Table \ref{tab:temp_variability}. 
We find only modest changes in the integrated optical depths, $<3.5\sigma$ differences in all cases. Two cases with over 3$\sigma$ significance are both in the case of 3C111 (\hnc{} and \cch{}) and probe variability on time scales of 25 years. 3C111 is especially interesting as \citet{2020ApJ...893..152R} found significant spatial variations in the \hi{} optical depths between the different components of this source.

\begin{figure}[] 
\gridline{\fig{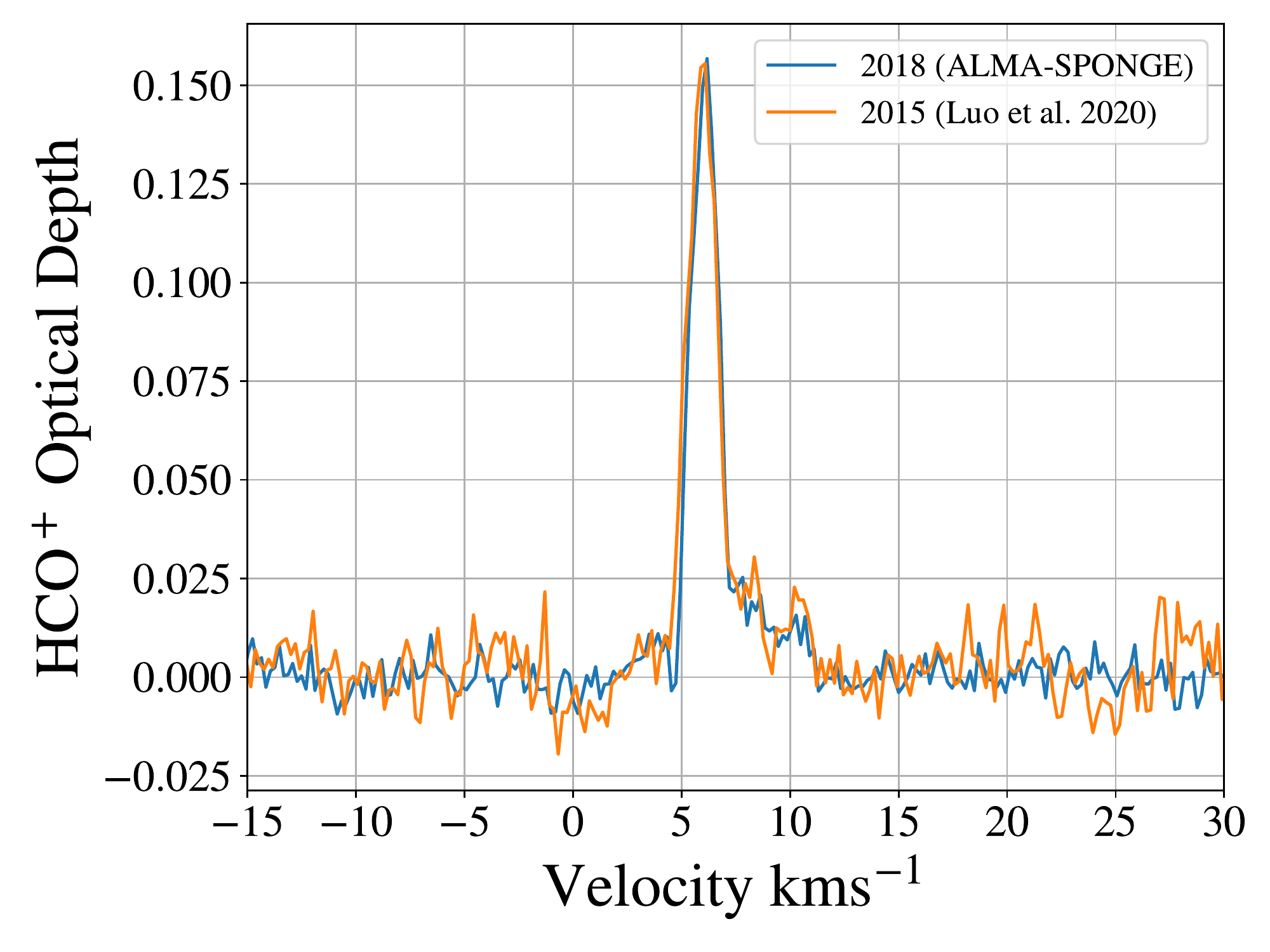}{\columnwidth}{}}
\gridline{\fig{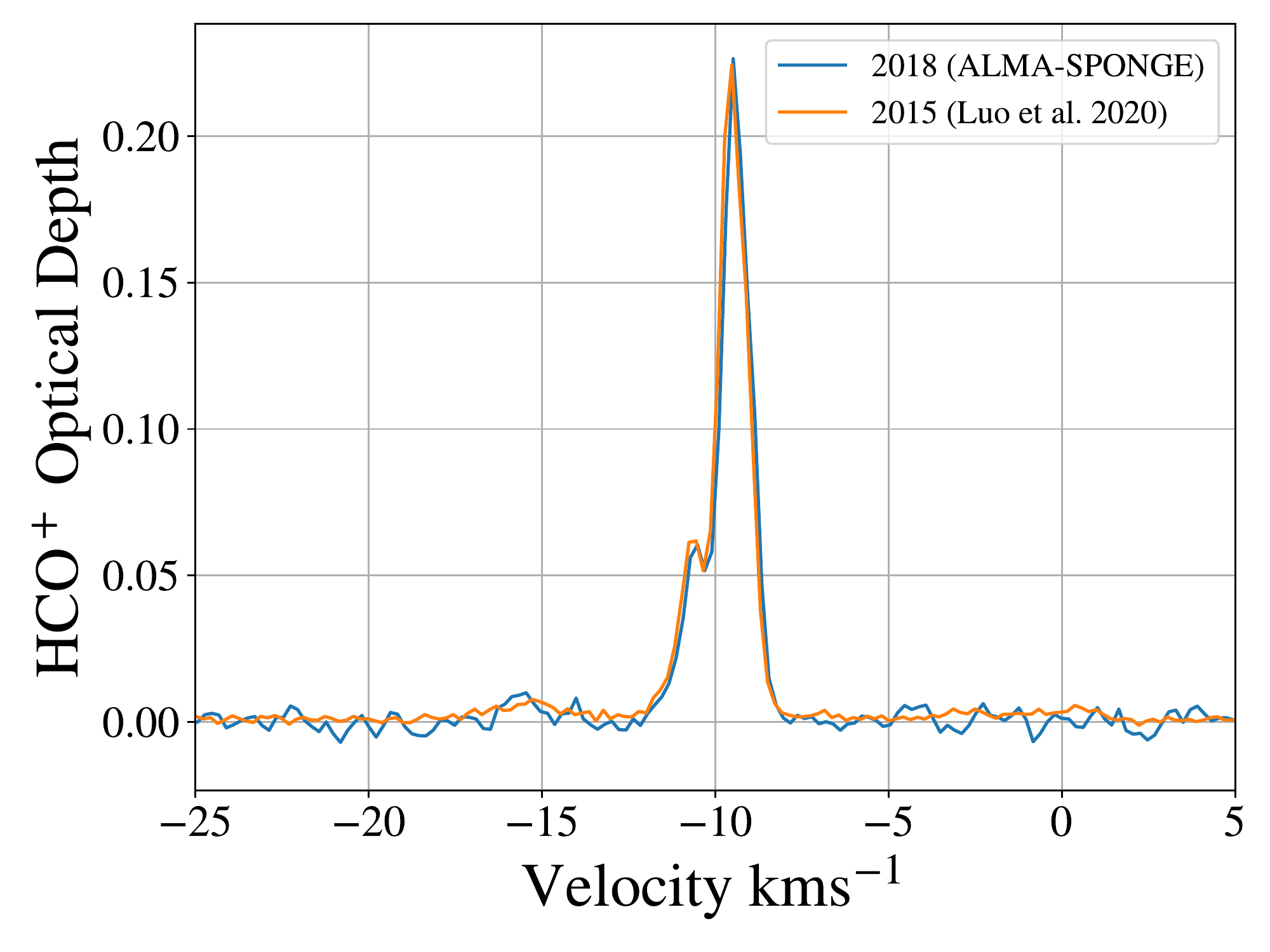}{\columnwidth}{}}
\caption{The \hcop{} optical spectra in the direction of 3C120 (top) and 3C454.3 (bottom). The spectra from \citet{2020ApJ...889L...4L}, obtained by ALMA in 2015, are shown in orange. The spectra from this work, obtained by ALMA in 2018, are shown in blue. \label{fig:luo_comparison}}
\end{figure}

Figure \ref{fig:luo_comparison} further shows the optical depth spectra of \hcop{} in the direction of 3C120 and 3C454.3 measured here and those measured by \citet{2020ApJ...889L...4L}. Both were obtained with ALMA using similar spectral setups; the ALMA-SPONGE spectra were obtained in 2018, while the \citet{2020ApJ...889L...4L} spectra were obtained in 2015, so these sightlines probe optical depth changes of $\sim3$ years.
The \hcop{} optical depth spectra show no significant change with respect to the \citet{2020ApJ...889L...4L} data taken 3 years earlier---the peak optical depths differ by $0.001$--$0.002$, a $<1\%$ difference detected at $<1\sigma$. 

Our sample of sightlines with multi-epoch measurements is limited to five, two of which are non-detections. For these sightlines, the lack of short-term variations in the absorption profiles suggests that the molecular component of the gas is not highly structured on scales $\lesssim100$ AU in these directions\footnote{The motion of the Earth around the Sun and the Sun relative to interstellar clouds can be used to estimate the change in the line of sight over time. See \citet{1993ApJ...419L.101M} for discussion. 100 AU is an order of magnitude estimate. The precise distance depends on the velocity and distance to each cloud, but 100 AU should be appropriate for typical clouds at a distance of order $100~\rm{pc}/\sin |b|$  \citep{1978A&A....70...43C}.}. This is consistent with previous observations that molecular absorption line profiles are generally stable over time. \citet{2000A&A...355..333L} found that \hcop{} and OH absorption line profiles were generally stable over 3--5 year intervals  in the direction of multiple background sources. \citet{1997A&A...324...51W} showed that \hcop{} absorption spectra in the direction of Centaurus A were stable over a roughly 7 year span, and \citet{2017JKAS...50..185H} showed that \hcop{} absorption spectra in the direction of BL Lac and NRAO 150 were stable over $\sim$20 years.
Furthermore, \citet{2000A&A...355..333L} argued that the modest changes detected in molecular absorption line profiles over time were most likely caused by small-scale changes in chemical abundances rather than by the presence of highly overdense, AU-scale structures.

\begin{deluxetable*}{cccccc}
\tablenum{8}
\tablecaption{The integrated optical depths of \hcop{}, \hcn{}, \hnc{}, and \cch{} measured by previous surveys (Column 3) and measured in this work (Column 4). The time elapsed between surveys is listed in Column 5, and the change in integrated optical depth is listed in Column 6. Superscripts in Column 3 indicate references, as follows---1: Lucas \& Liszt 1996; 2: Liszt \& Lucas 2001; 3: Lucas \& Liszt 2000a. \label{tab:temp_variability}}
\tablehead{\colhead{Source} & Species & \colhead{$(\int\tau dv)_1$} &
\colhead{$(\int\tau dv)_{2}$} & \colhead{$\Delta t$} & \colhead{$\Delta(\int\tau dv)$}\\
\colhead{} & \colhead{} & \colhead{$10^{12}$ \persc{}} & \colhead{$10^{12}$ \persc{}} & \colhead{yr} & \colhead{$10^{12}$ \persc{}}}
\startdata
3C111 & \hcop{} & $13.32\pm 0.65^1$ & $>11.054$ & 26 &  \\
 & \hcn{} & $13.572\pm0.276^2$ & $>12.710$ & 25 & \\
 & \hnc{} & $3.149 \pm 0.025^2$ & $3.606 \pm 0.129$ & 25 & $0.457\pm 0.131$ \\
 & \cch{} & $1.913 \pm 0.023^3$ & $2.057 \pm 0.033$ & 25 &  $0.144\pm0.04$\\
3C273 & \hcop{} & nd & nd & 24 & \\
3C345 & \hcop{} & nd & nd & 24 & \\
3C454.3 & \hcop{} & $0.32\pm0.02^1$ & $0.276 \pm 0.004$ & 24 & $0.044\pm0.02$\\
 & \hcn{} & $0.120\pm0.013^2$ & $0.112 \pm  0.004$ & 23 & $0.008 \pm 0.014$ \\
 & \hnc{} & $0.046\pm0.019^2$ & $ 0.024 \pm 0.002$ & 23 & $0.022 \pm 0.019$ \\
 & \cch{} & $0.152\pm0.025^3$ & $0.106 \pm 0.004$ & 23 & $0.046\pm0.025$\\
BL Lac & \hcop{} & $2.30\pm0.07^1$ & $2.260\pm0.004$ & 26 & $0.04 \pm 0.07$ \\
 & \hcn{} & $2.360\pm0.080^2$ & $2.293\pm0.002$ & 25 & $0.067 \pm 0.080$ \\
\enddata
\end{deluxetable*}

\section{Discussion}
\label{sec:discussion}

\subsection{The atomic gas properties necessary for molecule formation}

By comparing the Gaussian components identified in the 21-SPONGE \hi{} absorption spectra \citep{2015ApJ...804...89M,2018ApJS..238...14M} with the Gaussian components identified in the ALMA-SPONGE and NOEMA-SPONGE molecular absorption spectra, we have established \hi{} optical depth, spin temperature, and turbulent Mach number thresholds for the formation of \hcop{}, \hnc{}, \hcn{}, and \cch{} (Section \ref{sec:hi_molecules}). We find that molecular gas forms only if the \hi{} optical depth of a gas structure exceeds $0.1$, the spin temperature---approximately equal to the kinetic temperature---is less than 80 K, and the turbulent Mach number is greater than $\sim2$. In our sample, these conditions appear necessary but not sufficient for the formation of \hcop{}, \cch{}, \hcn{}, and \hnc{}, as several \hi{} absorption features satisfy these criteria but do not have a molecular component.
We do not find significantly different thresholds for the different species, although our sample size is modest. 

The optical depth threshold of 0.1 is similar to that established by \citet{1985A&A...148...83D} for CO formation at $|b|<2^\circ{}$. However, our $T_s$ threshold ($T_s\lesssim 80$ K) is significantly lower than theirs ($T_s\sim100\text{--}350$ K). We note, though, that \citet{1985A&A...148...83D} derived $T_s$ using an isothermal approximation of channels with CO emission 
which is known to result in higher spin temperatures relative to 
Gaussian decomposition \citep{2015ApJ...804...89M}. They also used CO emission observations at $\sim4\arcmin{}.3$ beamwidth, which probe more gas than the \hi{} pencil beam and are less sensitive to low and even moderate column density gas than CO absorption observations.
Observations of 21 cm \hi{} and mm wavelength molecular absorption in the same direction have been otherwise limited to a very small number of sightlines \citep[$N=1$--3;][]{1995ApJ...452..671M,2000A&A...355..333L,2004A&A...428..445L,2018A&A...610A..49L}. For example \citet{2000A&A...355..333L} established that strong \hcop{} absorption was associated with lower $\langle T_s \rangle$, $46$--$103$ K, in three directions (see Section \ref{sec:broad_hcop}). 

The CNM associated with molecule formation in the diffuse ISM is thus systematically colder and more optically thick than the mean of the total cold \hi{} cloud population. CNM structures with a molecular component also have higher turbulent Mach numbers than the general \hi{} cloud population, which is a reflection of the colder temperature in most cases, as the 1D turbulent velocities are not systematically higher than the general cloud population.
It is also interesting to note that the \hi{} temperature threshold we find for molecule formation, 80 K, is near the mean kinetic temperature estimated from FUV observations of \htwo{} \citep[$55-80$ K;][]{1977ApJ...216..291S,Rachford2002,Rachford2009}.
This temperature agreement is expected if \hi{} and H$_2$ are co-spatial, as is the case for H$_2$ forming out of the CNM on the surface of dust grains and being injected back to the CNM. 
The high turbulent Mach numbers seen here in the direction of diffuse molecular gas is consistent with previous observations of cold \hi{} in the direction of Perseus \citep[e.g.,][]{Burkhart2015}.

A relationship between the \hi{} column density and H$_2$ formation has long been known.
The \hi{}-to-\htwo{} transition has been extensively studied in theory and using observations of both entire galaxies and individual molecular clouds. The total gas column density threshold for \htwo{} formation has been well established \citep[e.g.,][and references therein]{1977ApJ...216..291S,2009ApJ...693..216K,2020arXiv200905466B}, and many studies have also noticed a threshold for CO formation \citep[e.g.,][]{Lombardi2006,Wong2009,Leroy2009,Lee2014,ImaraBurkhart2016} which is higher \citep[$A_V \sim1-2$, e.g.,]{Lee2014} than that of \htwo{} ($A_V \sim 0.25$). Observationally established thresholds for the formation of other molecular species have been rare. Our results are consistent with the previous work of \citet{1996A&A...307..237L} and \citet{2000A&A...358.1069L} for \hcop{} and \cch{}, although our observations are more sensitive than theirs by a factor of $2$--$3$. Both simulations from \citet{Gong2020} (we discuss this more in \papertwo{}) and our ALMA/NOEMA observations show a transition from atomic to molecular gas at $N(\hi{})\sim 5 \times 10^{20}$ cm$^{-2}$. This threshold appears consistent with that of \htwo{} and somewhat lower than that of CO. This clearly demonstrates the importance of shielding in building molecular abundances.

\subsection{The significance of the UNM in molecule formation}

\citet{2018ApJS..238...14M} quantified the fraction of thermally unstable \hi{} along each line of sight. Although we find that the UNM column density is not directly correlated with the total quantity of molecular gas, it is nevertheless clearly important in setting the \hcop{} column densities. As we have shown in Figure \ref{fig:NX_v_var}, positions with the highest \cch{}, \hcop{}, \hcn{}, and \hnc{} column densities also tend to have a significant quantity of UNM gas, $10^{20}$--$10^{21}$ \persc{}, comprising $\sim30$--$70\%$ of all atomic gas along those lines of sight. The UNM plays a role in shielding but also may reveal mixing of the CNM and WNM, which is suggested to enhance the abundances of certain molecular species, including \hcop{} \citep{Valdivia2016H2,Lesaffre2020}.

Previous observational and theoretical work has highlighted the role of the multiphase atomic ISM in molecule formation.
For example, \citet{2015Bialy} investigated the \hi{}-to-\htwo{} transition in several regions in the Perseus molecular cloud, where \citet{Lee2012} have found a well-defined threshold for \htwo{} formation, $(8\text{--}14)\times10^{20}$ \persc{}. By fitting the \citet{Sternberg2014} model for \htwo{} formation in equilibrium and assuming that \hi{} gas is multiphase and in pressure equilibrium, they constrained the volume density of \hi{} in the atomic shielding envelope of Perseus. As the  estimated densities were in the range $\sim 2-10$ cm$^{-3}$, they concluded that much of this \hi{} is thermally unstable, possibly in a cooling transition from the WNM to CNM phases.
The behavior in Perseus suggests that in addition to CNM, less dense UNM and perhaps some diffuse WNM, are important in controlling the \hi{}-to-\htwo{} transition. Our observations indicate that the UNM is important for enhancing the abundances of species other than \htwo{} and CO, as well.

\subsection{Interpretation of broad, faint \hcop{} absorption}
In a majority of sightlines, we find a broad, faint component to the \hcop{} absorption that spans a large portion of the velocity space where \hi{} absorption is observed (Section \ref{sec:broad_hcop}).
Our study suggests that this broad HCO$^+$ component is likely to be frequently seen in high-sensitivity observations. 
The \hi{} coincident with this broad \hcop{} absorption component has higher $\langle T_S \rangle$ than the \hi{} at velocities where strong, narrow \hcop{} absorption is observed, indicating that the fraction of \hi{} in the CNM is lower at these velocities \citep[e.g.,][]{Kim2014}. If we assume that $N(\hcop{})/N(\htwo{})$ is constant \citep{2000A&A...355..333L,2018A&A...617A..54L}, then we also find that the fraction of hydrogen in its molecular form is lower for velocities with broad absorption ($f_\mathrm{mol}\lesssim0.1$, versus $f_\mathrm{mol}\gtrsim0.2$ at velocities with a strong, narrow absorption component).

The fact that the broad component of \hcop{} absorption is associated with gas that has a lower CNM fraction and a lower molecular fraction may indicate that it is probing the outer layers of molecular structures, where the turbulent mixing between the CNM and WNM allows for the formation of CH\textsuperscript{+} and \hcop{}. This would explain why the broad \hcop{} absorption spans roughly the same velocity range as the cold \hi{} observed by 21-SPONGE in nearly all sightlines (Figure \ref{fig:broadHCOp}). The relatively broad linewidths of the faint component of \hcop{} absorption may also suggest formation in warmer, more diffuse interstellar environments \citep{2006A&A...452..511F}. For example, \citet{Valdivia2017} and \citet{Lesaffre2020} showed that CH\textsuperscript{+}---which contributes to the formation of \hcop{}---is formed efficiently in turbulent dissipation regions, at the edge of molecular clumps, and in slow shocks, where the \hi{} is thermally unstable and the \htwo{} formation is out of equilibrium. In the \citet{Valdivia2017} simulations, most of the CH$^+$ is produced in regions with \htwo{} fractions $\sim0.003$--$0.3$ and with gas temperatures as high as $\sim10^{2-3}$ K. Given the role CH\textsuperscript{+} plays in \hcop{} formation, the broad \hcop{} may trace the active regions of CH\textsuperscript{+} production.

We see no CO emission at velocities where the broad, faint \hcop{} absorption is detected, although the resolution and sensitivity of CO emission observations from \citet{2001ApJ...547..792D} do not match those of our ALMA/NOEMA absorption spectra. Moreover, due to low excitation temperature, CO may not be detectable in 
emission  \citep[e.g.,][]{1996A&A...307..237L,2000A&A...355..333L}. The possibility that the broad \hcop{} absorption component is CO dark warrants further investigation, as it may contribute to the explanation of why \hcop{} is a better tracer of the total gas content than CO in the diffuse ISM \citep{2018A&A...617A..54L}. \citet{2019A&A...627A..95L} have already shown that \hcop{} linewidths are broader than CO linewidths, which is consistent with the idea that \hcop{} is enhanced in diffuse, turbulent regions of the ISM relative to CO. Moreover, Figure 1 of \citet{2020ApJ...889L...4L} suggests that there is indeed broad, faint \hcop{} absorption in the direction of 3C454.3---only marginally detected in our less sensitive spectra---that is predominately CO dark. However, a quantitative comparison of the amount of \hcop{} occupied by this broad component with the total amount of CO dark molecular gas has not yet been made.

\subsection{The similarity of gas at different latitudes}

We show that the relative abundances of different species are similar between this study and several previous surveys, even though these measurements collectively span Galactic latitudes between $<1^\circ$ and $\sim40^\circ$ \citep[Table \ref{tab:line_ratios};][]{2001A&A...370..576L,2010A&A...520A..20G}. Similarly, we do not find significant differences between the FWHM ratios of different species with those measured in previous experiments over the same range of latitudes \citep{2010A&A...520A..20G}, although these ratios are less tightly constrained.

The similarity in the relative abundances and linewidths of different species suggests that similar chemical pathways operate across the Galaxy, both in the plane and in the solar neighborhood.  This is consistent with the interpretation of \citet{2010A&A...520A..20G}, but we have added significantly to the total sample of observations of diffuse molecular gas, including gas at column densities lower than those probed by \citet{2001A&A...370..576L} and \citet{2010A&A...520A..20G} by a factor of a few. \citet{2010A&A...520A..20G} suggested that the lack of dependence on Galactic latitude indicated that the formation mechanism of different species was weakly dependent on the ambient UV field, perhaps pointing to a ubiquitous mechanism like turbulent dissipation. As discussed further in \papertwo{}, it will be important to observe tracers of non-equilibrium processes like CH\textsuperscript{+} or SiO and test PDR predictions to verify this hypothesis.

\subsection{Is there a molecular component to tiny scale atomic structure?} \label{subsec:tsas_discussion}

In Section \ref{sec:temporal_stability}, we show that over periods of 3--25 years, \hcop{}, \cch{}, \hcn{}, and \hnc{} absorption spectra are  generally stable for our small sample of sightlines with multi-epoch absorption observations.

\citet{1993ApJ...419L.101M} and \citet{1995ApJ...452..671M} detected temporal variations in the H\textsubscript{2}CO optical depth in the direction of 3C111,  arguing in favor of dense clouds ($n\sim10^6$ \percc{}) in this direction at a scale of $\sim10$--$100$ AU. However, \citet{1996A&A...312..973T} showed that the H\textsubscript{2}CO optical depth variations were most probably a result of changes in the relative abundance or excitation of H\textsubscript{2}CO rather than tiny dense structures.
We find marginal variations ($3.5\sigma$) in the integrated optical depths of \hnc{}, and \cch{} in the direction of 3C111 (Section \ref{sec:temporal_stability}), which suggests that there are not extremely dense structures on scales of $\sim10$ AU in this direction as suggested by \citet{1993ApJ...419L.101M} and \citet{1995ApJ...452..671M}, consistent with the interpretation of \citet{1996A&A...312..973T}.

No other sources show temporal variations in the \hcop{}, \hcn{}, \hnc{}, or \cch{} optical depths at a level of $>2.2\sigma$. 
BL Lac---in the direction of the Lacerta molecular cloud---shows remarkably stable absorption spectra, $<1\sigma$ changes over $\sim25$ years \citep{1996A&A...307..237L,2001A&A...370..576L}.  \citet{2017JKAS...50..185H} previously found no difference in \hcop{} absorption spectra obtained with the Korean VLBI Network to the \hcop{} absorption spectra from \citet{1996A&A...307..237L} 20 years earlier in the direction of BL Lac. Similarly, the integrated and peak optical depths measured here in the direction of 3C120 and 3C454.3 show $<1\sigma$ changes compared to those obtained by \citet{2020ApJ...889L...4L} 3 years earlier.

Spatial variations in the \hi{} optical depths toward the multiple-component sources 3C111 and 3C123 have revealed TSAS at a scale of $\sim10^4$.
3C111 is a particularly intriguing case, as \citet{1978A&A....70..415D}, \citet{1986ApJ...303..702G}, \citet{2008MNRAS.388..165G}, and \citet{2020ApJ...893..152R} have all measured optical depth variations of $\sim0.3$--0.4 between its different components, corresponding to column density changes of $\gtrsim10^{20}$ \persc{} at scales of $\sim10^4$ AU.
\citet{2008MNRAS.388..165G} and \citet{2020ApJ...893..152R} also previously found \hi{} optical depth variations at a level of $\sim0.1$ toward the different components of 3C123. Unfortunately, due to a combination of poor sensitivity toward the fainter components and line saturation, we can only put very weak upper limits on the spatial molecular optical depth variations in these directions. Future sensitive observations of unsaturated molecular absorption lines are required to test if the these lines vary with the \hi{} on scales of $10^4$ AU.

We note that the \hi{} optical depth in the direction of 3C138 has been observed to vary dramatically, both spatially ($\sim$few$\times10$ AU) and temporally ($\lesssim7$ yr), making it a regular subject of TSAS investigations. \citet{1989ApJ...347..302D} and \citet{1996MNRAS.283.1105D} both found spatial variations in the \hi{} optical depth measured across resolved images of 3C138 obtained using very long baseline interferometry (VLBI) techniques. Later, \citet{2001AJ....121.2706F}, \citet{2005AJ....130..698B}, and \citet{2012ApJ...749..144R} constructed \hi{} optical depth maps against 3C138 using the Very Long Baseline Array (VLBA) across three different  epochs. They detected milliarcsecond \hi{} optical depth variations (corresponding to spatial scales of $\sim25$ AU) in each epoch, as well as significant variations across epochs. These measurements have been attributed to \hi{} structures with densities $\sim10^4$ \percc{}. Moreover, \citet{2000ApJ...543..227D}, who were critical of the interpretation of optical depth variations as evidence of dense structures, acknowledged that the optical depth variations in the direction of 3C138 cannot be explained by geometrical effects and are likely due to true tiny, dense structures. 
We see no \hcop{}, \cch{}, \hcn{}, or \hnc{} absorption in the direction of 3C138. The full sky maps in \citet{Pelgrims2020}, made from the \citet{Lallement2019} 3D dust maps, suggest that this line of sight intersects the Local Bubble wall at $\lesssim150$ pc, while the \hi{} associated with Taurus lies at a distance of $\lesssim140$ pc \citep{2014ApJ...786...29S}. Considering that the absorbing gas is located inside the Local Bubble wall \citep[e.g.,][]{2014A&A...561A..91L}, the lack of molecules associated with TSAS in this direction could be caused by the physical conditions associated with Local Bubble \citep[e.g.,][]{SS2010}.

\section{Conclusion} \label{sec:conclusions}
We have complemented  \hi{} observations \citep[VLA;][]{2018ApJS..238...14M} with new \hcop{}, \hcn{}, \hnc{}, and \cch{} (ALMA, NOEMA) observations in the direction of 20 background radio continuum sources with $4^\circ\leq|b|\leq81^\circ$ to constrain the atomic gas conditions that are suitable for the formation of diffuse molecular gas. In our sample, we find that these molecular species form along sightlines where $A_V\gtrsim0.25$, corresponding to $N_\mathrm{H}\gtrsim5\times10^{20}$ \persc{} \citep[e.g.,][]{2017MNRAS.471.3494Z}, consistent with a threshold for the \hi{}-to-\htwo{} transition at solar metallicity \citep{1977ApJ...216..291S}. This confirms the trend noticed by \citet{1996A&A...307..237L} and \citet{2000A&A...358.1069L} for \hcop{} and \cch{}, with absorption spectra $\sim2$--$3$ times more sensitive than theirs. Moreover, we find that molecular gas is associated only with structures that have an \hi{} optical depth $>0.1$, a spin temperature $<80$ K, and a turbulent Mach number $\gtrsim 2$.

We identify a broad, faint component to the \hcop{} absorption in a majority of sightlines. In several cases this component spans nearly the entire range of velocities where \hi{} is seen in absorption. The \hi{} at velocities where only the faint, broad \hcop{} absorption is observed has systematically higher $\langle T_s \rangle$ than the \hi{} at velocities where strong, narrow \hcop{} absorption is observed, indicating that the \hi{} at these velocities has a lower CNM fraction \citep{Kim2014}. We also do not detect CO emission at these velocities, whereas CO emission is detected at nearly all velocities where strong \hcop{} absorption is observed. 
The broad \hcop{} may therefore be a good tracer of the CO dark molecular gas and deserves further investigation.

The relative column densities and linewidths of the different molecular species observed here are similar to those observed in previous experiments over a range of Galactic latitudes, suggesting that gas in the solar neighborhood and gas in the Galactic plane are chemically similar.
For a select sample of previously-observed sightlines, we show that the absorption line profiles of \hcop{}, \hcn{}, \hnc{}, and \cch{} are stable over periods of $\sim3$ years and $\sim25$ years, likely indicating that molecular gas structures in these directions are at least $\gtrsim100$ AU in size. 

In \papertwo{}, we compare the observational results presented here to predictions from the PDR chemical model in \citet{Gong2017} and the ISM simulations in \citet{Gong2020} to investigate in detail the local physical conditions (density, radiation field, cosmic ray ionization rate) needed to explain the observed column densities.

\acknowledgements{
We would like to thank Ningyu Tang for providing useful supplementary data. We are grateful to Antoine Gusdorf, Benjamin Godard, and Carl Heiles for useful discussions about this work.

Support for this work was provided by the NSF through award SOSPA6-023 from the NRAO. S.S. acknowledges the support by the Vilas funding provided by  the  University  of  Wisconsin  and  the  John  Simon Guggenheim fellowship.
This work is based on observations carried out under project number W19AQ and S20AB with the IRAM NOEMA Interferometer.  IRAM is supported by INSU/CNRS (France), MPG (Germany) and IGN (Spain).
This paper makes use of the following ALMA data: ADS/JAO.ALMA\#2018.1.00585.S and ADS/JAO.ALMA\#2019.1.01809.S.. ALMA is a partnership of ESO (representing its member states), NSF (USA) and NINS (Japan), together with NRC (Canada), MOST and ASIAA (Taiwan), and KASI (Republic of Korea), in cooperation with the Republic of Chile. The Joint ALMA Observatory is operated by ESO, AUI/NRAO and NAOJ. The National Radio Astronomy Observatory is a facility of the National Science Foundation operated under cooperative agreement by Associated Universities, Inc. }

\software{Astropy \citep{astropy:2013, astropy:2018}, CASA \citep{2007ASPC..376..127M}, GILDAS \citep{2005sf2a.conf..721P,2013ascl.soft05010G}, dustmaps \citep{2018JOSS....3..695M}, Scipy \citep{JonesSciPy,2020NatMe..17..261V}.}

\bibliography{refs}{}
\bibliographystyle{aasjournal}

\begin{deluxetable*}{ccccc}
\tablenum{3}
\tablecaption{Gaussian fits to the \cch{} absorption spectra. Column 1: Source name; Column 2: peak optical depth, $\tau_0$; Column 3: FWHM, $\Delta v_0$; Column 4: central velocity, $v_0$; Column 5: integrated optical depth. . \label{tab:cch_fits}}
\tablehead{\colhead{Source} & \colhead{$\tau_0$} & \colhead{$\Delta v_0$} & \colhead{$v_0$} & \colhead{$\int\tau_0 dv$}\\
\colhead{} &  & \colhead{\kms{}} & \colhead{\kms{}} & \colhead{\kms{}}}
\startdata
3C111A   & $0.752 \pm 0.036$  & $ 1.0 \pm  0.0$ & $-0.9 \pm  0.0$ & $ 0.78 \pm  0.05$ \\
3C111A   & $0.332 \pm 0.021$  & $ 1.5 \pm  0.1$ & $-2.5 \pm  0.0$ & $ 0.52 \pm  0.05$ \\
3C120    & $0.069 \pm 0.001$  & $ 1.3 \pm  0.0$ & $ 6.6 \pm  0.0$ & $ 0.09 \pm 0.01$ \\
3C120    & $0.016 \pm 0.002$  & $ 0.9 \pm  0.2$ & $ 5.1 \pm  0.1$ & $ 0.02 \pm 0.01$ \\
3C120    & $0.014 \pm 0.001$  & $ 2.1 \pm  0.2$ & $ 9.3 \pm  0.1$ & $ 0.03 \pm 0.01$ \\
3C120    & $0.007 \pm 0.002$  & $ 0.6 \pm  0.2$ & $ 3.7 \pm  0.2$ & $ 0.00 \pm 0.01$ \\
3C123A   & $0.372 \pm 0.095$  & $ 1.2 \pm  0.5$ & $ 4.3 \pm  0.3$ & $ 0.49 \pm  0.25$ \\
3C123A   & $0.313 \pm 0.104$  & $ 1.4 \pm  0.6$ & $ 5.3 \pm  0.3$ & $ 0.45 \pm  0.26$ \\
3C123A   & $0.276 \pm 0.070$  & $ 0.7 \pm  0.3$ & $ 3.1 \pm  0.2$ & $ 0.20 \pm  0.11$ \\
3C123B   & $0.584 \pm 0.140$  & $ 0.9 \pm  0.6$ & $ 4.2 \pm  0.6$ & $ 0.59 \pm  0.37$ \\
3C123B   & $0.437 \pm 0.081$  & $ 1.1 \pm  0.6$ & $ 5.4 \pm  0.8$ & $ 0.51 \pm  0.31$ \\
3C123B   & $0.351 \pm 0.072$  & $ 0.8 \pm  0.5$ & $ 3.0 \pm  0.7$ & $ 0.31 \pm  0.18$ \\
3C154    & $0.449 \pm 0.050$  & $ 0.8 \pm  0.2$ & $-2.0 \pm  0.1$ & $ 0.38 \pm  0.07$ \\
3C154    & $0.221 \pm 0.042$  & $ 1.0 \pm  0.3$ & $-0.9 \pm  0.1$ & $ 0.23 \pm  0.12$ \\
3C154    & $0.087 \pm 0.021$  & $ 2.3 \pm  0.6$ & $-4.0 \pm  0.3$ & $ 0.21 \pm  0.10$ \\
3C154    & $0.050 \pm 0.015$  & $ 2.5 \pm  1.0$ & $ 2.7 \pm  0.8$ & $ 0.14 \pm  0.06$ \\
3C454.3  & $0.046 \pm 0.002$  & $ 1.2 \pm  0.0$ & $-9.2 \pm  0.0$ & $ 0.06 \pm 0.01$ \\
3C454.3  & $0.021 \pm 0.001$  & $ 1.2 \pm  0.2$ & $-10.5 \pm  0.1$ & $ 0.03 \pm 0.01$ \\
J2136    & $0.008 \pm 0.001$  & $ 2.4 \pm  0.0$ & $ 5.3 \pm  0.1$ & $ 0.02 \pm 0.01$ \\
\enddata
\end{deluxetable*}

\begin{deluxetable*}{ccccc}
\tablenum{4}
\tablecaption{Gaussian fits to the \hcn{} absorption spectra. Column 1: Source name; Column 2: peak optical depth, $\tau_0$; Column 3: FWHM, $\Delta v_0$; Column 4: central velocity, $v_0$; Column 5: integrated optical depth. Components for sightlines with saturated absorption are superscripted by a dagger ($\dagger$). \label{tab:hcn_fits}}
\tablehead{\colhead{Source} & \colhead{$\tau_0$} & \colhead{$\Delta v_0$} & \colhead{$v_0$} & \colhead{$\int\tau_0 dv$}\\
\colhead{} &  & \colhead{\kms{}} & \colhead{\kms{}} & \colhead{\kms{}}}
\startdata
3C111A   & $3.639^\dagger \pm 0.039$  & $ 1.0 \pm  0.0$ & $-0.9 \pm  0.0$ & $ 3.78^\dagger \pm  0.05$ \\
3C111A   & $1.959^\dagger \pm 0.009$  & $ 0.9 \pm  0.0$ & $-2.5 \pm  0.0$ & $ 1.87^\dagger \pm  0.01$ \\
3C111B   & $2.884^\dagger \pm 1.381$  & $ 1.0 \pm  0.6$ & $-1.0 \pm  0.1$ & $ 3.20^\dagger \pm  2.49$ \\
3C111B   & $1.678^\dagger \pm 0.787$  & $ 1.1 \pm  0.6$ & $-2.5 \pm  0.1$ & $ 1.89^\dagger \pm  1.42$ \\
3C120    & $0.025 \pm 0.003$  & $ 0.7 \pm  0.1$ & $ 6.4 \pm  0.0$ & $ 0.02 \pm  0.00$ \\
3C120    & $0.008 \pm 0.002$  & $ 1.2 \pm  0.1$ & $ 5.4 \pm  0.2$ & $ 0.01 \pm  0.00$ \\
3C123A   & $2.594 \pm 0.826$  & $ 0.7 \pm  0.2$ & $ 4.4 \pm  0.1$ & $ 1.89 \pm  0.78$ \\
3C123A   & $0.926 \pm 0.201$  & $ 1.0 \pm  0.3$ & $ 5.2 \pm  0.2$ & $ 1.01 \pm  0.39$ \\
3C123A   & $0.587 \pm 0.125$  & $ 0.9 \pm  0.4$ & $ 3.4 \pm  0.2$ & $ 0.58 \pm  0.26$ \\
3C123B   & $2.753^\dagger \pm 0.948$  & $ 0.7 \pm  0.2$ & $ 4.2 \pm  0.5$ & $ 1.94^\dagger \pm  0.82$ \\
3C123B   & $1.265^\dagger \pm 0.383$  & $ 0.8 \pm  0.4$ & $ 5.2 \pm  0.7$ & $ 1.12^\dagger \pm  0.69$ \\
3C123B   & $0.622^\dagger \pm 0.195$  & $ 0.7 \pm  0.4$ & $ 3.1 \pm  0.7$ & $ 0.47^\dagger \pm  0.33$ \\
3C154    & $1.734 \pm 0.182$  & $ 0.7 \pm  0.1$ & $-2.0 \pm  0.0$ & $ 1.31 \pm  0.17$ \\
3C154    & $0.394 \pm 0.046$  & $ 1.0 \pm  0.2$ & $-1.0 \pm  0.1$ & $ 0.41 \pm  0.09$ \\
3C154    & $0.056 \pm 0.016$  & $ 2.0 \pm  0.8$ & $-4.5 \pm  0.7$ & $ 0.12 \pm  0.06$ \\
3C154    & $0.063 \pm 0.017$  & $ 2.3 \pm  1.0$ & $ 2.9 \pm  0.7$ & $ 0.16 \pm  0.08$ \\
3C454.3  & $0.068 \pm 0.000$  & $ 1.0 \pm  0.0$ & $-9.3 \pm  0.0$ & $ 0.07 \pm  0.00$ \\
3C454.3  & $0.010 \pm 0.000$  & $ 0.4 \pm  0.0$ & $-10.5 \pm  0.0$ & $ 0.00 \pm  0.00$ \\
BLLac    & $0.923 \pm 0.053$  & $ 0.8 \pm  0.1$ & $-1.4 \pm  0.0$ & $ 0.83 \pm  0.07$ \\
BLLac    & $0.410 \pm 0.043$  & $ 0.9 \pm  0.1$ & $-0.6 \pm  0.0$ & $ 0.41 \pm  0.07$ \\
\enddata
\end{deluxetable*}

\begin{deluxetable*}{ccccc}
\tablenum{5}
\tablecaption{Gaussian fits to the \hcop{} absorption spectra. Column 1: Source name; Column 2: peak optical depth, $\tau_0$; Column 3: FWHM, $\Delta v_0$; Column 4: central velocity, $v_0$; Column 5: integrated optical depth. Components for sightlines with saturated absorption are superscripted by a dagger ($\dagger$). \label{tab:hcop_fits}}
\tablehead{\colhead{Source} & \colhead{$\tau_0$} & \colhead{$\Delta v_0$} & \colhead{$v_0$} & \colhead{$\int\tau_0 dv$}\\
\colhead{} &  & \colhead{\kms{}} & \colhead{\kms{}} & \colhead{\kms{}}}
\startdata
3C111A   & $4.172^\dagger \pm 0.074$  & $ 1.3 \pm  0.0$ & $-0.9 \pm  0.0$ & $ 5.94^\dagger \pm  0.14$ \\
3C111A   & $3.296^\dagger \pm 0.043$  & $ 1.3 \pm  0.0$ & $-2.5 \pm  0.0$ & $ 4.72^\dagger \pm  0.08$ \\
3C111B   & $2.393^\dagger \pm 0.918$  & $ 1.4 \pm  0.6$ & $-1.0 \pm  0.1$ & $ 3.54^\dagger \pm  2.12$ \\
3C111B   & $1.749^\dagger \pm 0.775$  & $ 1.6 \pm  0.8$ & $-2.5 \pm  0.1$ & $ 2.94^\dagger \pm  1.95$ \\
3C120    & $0.154 \pm 0.002$  & $ 1.3 \pm  0.0$ & $ 6.2 \pm  0.0$ & $ 0.21 \pm  0.00$ \\
3C120    & $0.082 \pm 0.013$  & $ 0.4 \pm  0.1$ & $ 5.3 \pm  0.0$ & $ 0.03 \pm  0.01$ \\
3C120    & $0.020 \pm 0.001$  & $ 2.4 \pm  0.0$ & $ 8.6 \pm  0.1$ & $ 0.05 \pm  0.00$ \\
3C120    & $0.011 \pm 0.001$  & $ 1.2 \pm  0.2$ & $ 3.7 \pm  0.1$ & $ 0.01 \pm  0.00$ \\
3C123A   & $3.214 \pm 1.100$  & $ 0.8 \pm  0.3$ & $ 4.3 \pm  0.2$ & $ 2.66 \pm  1.49$ \\
3C123A   & $2.123 \pm 0.469$  & $ 1.0 \pm  0.3$ & $ 5.3 \pm  0.2$ & $ 2.31 \pm  0.93$ \\
3C123A   & $1.138 \pm 0.261$  & $ 1.2 \pm  0.3$ & $ 3.4 \pm  0.2$ & $ 1.40 \pm  0.53$ \\
3C123B   & $3.737^\dagger \pm 1.137$  & $ 1.0 \pm  0.5$ & $ 4.4 \pm  0.7$ & $ 3.85^\dagger \pm  2.36$ \\
3C123B   & $2.110^\dagger \pm 0.689$  & $ 0.7 \pm  0.3$ & $ 5.2 \pm  0.6$ & $ 1.57^\dagger \pm  0.86$ \\
3C123B   & $1.203^\dagger \pm 0.331$  & $ 0.6 \pm  0.4$ & $ 3.0 \pm  0.5$ & $ 0.82^\dagger \pm  0.53$ \\
3C154    & $1.709 \pm 0.195$  & $ 0.7 \pm  0.1$ & $-2.2 \pm  0.1$ & $ 1.29 \pm  0.16$ \\
3C154    & $0.847 \pm 0.078$  & $ 1.0 \pm  0.2$ & $-1.1 \pm  0.1$ & $ 0.92 \pm  0.11$ \\
3C154    & $0.139 \pm 0.016$  & $ 2.5 \pm  0.3$ & $-4.1 \pm  0.3$ & $ 0.37 \pm  0.07$ \\
3C154    & $0.147 \pm 0.008$  & $ 3.1 \pm  0.3$ & $ 2.2 \pm  0.2$ & $ 0.49 \pm  0.04$ \\
3C454.3  & $0.212 \pm 0.000$  & $ 1.0 \pm  0.0$ & $-9.4 \pm  0.0$ & $ 0.22 \pm  0.00$ \\
3C454.3  & $0.055 \pm 0.000$  & $ 0.8 \pm  0.0$ & $-10.7 \pm  0.0$ & $ 0.05 \pm  0.00$ \\
3C78     & $0.039 \pm 0.012$  & $ 1.3 \pm  0.7$ & $-7.9 \pm  0.5$ & $ 0.05 \pm  0.03$ \\
BLLac    & $0.898 \pm 0.277$  & $ 1.3 \pm  0.4$ & $-1.4 \pm  0.0$ & $ 1.22 \pm  0.54$ \\
BLLac    & $0.747 \pm 0.018$  & $ 1.3 \pm  0.2$ & $-0.5 \pm  0.1$ & $ 1.07 \pm  0.16$ \\
\enddata
\end{deluxetable*}

\begin{deluxetable*}{ccccc}
\tablenum{6}
\tablecaption{Gaussian fits to the \hnc{} absorption spectra. Column 1: Source name; Column 2: peak optical depth, $\tau_0$; Column 3: FWHM, $\Delta v_0$; Column 4: central velocity, $v_0$; Column 5: integrated optical depth. Components for sightlines with saturated absorption are superscripted by a dagger ($\dagger$).  \label{tab:hnc_fits}}
\tablehead{\colhead{Source} & \colhead{$\tau_0$} & \colhead{$\Delta v_0$} & \colhead{$v_0$} & \colhead{$\int\tau_0 dv$}\\
\colhead{} &  & \colhead{\kms{}} & \colhead{\kms{}} & \colhead{\kms{}}}
\startdata
3C111A   & $3.039 \pm 0.303$  & $ 0.9 \pm  0.1$ & $-0.9 \pm  0.0$ & $ 2.84 \pm  0.33$ \\
3C111A   & $0.675 \pm 0.031$  & $ 1.0 \pm  0.0$ & $-2.5 \pm  0.0$ & $ 0.69 \pm  0.04$ \\
3C111B & $1.825 \pm 1.734^\dagger$ & $0.81\pm 0.55$ & $-1.15\pm 0.38$ & $1.57 \pm 1.83^\dagger$  \\
3C111B & $0.493 \pm 0.560^\dagger$ & $0.82\pm 0.99$ & $-2.71\pm 0.50$ & $0.43 \pm 0.49^\dagger$  \\
3C120    & $0.018 \pm 0.003$  & $ 0.5 \pm  0.1$ & $ 6.7 \pm  0.0$ & $ 0.01 \pm  0.00$ \\
3C123A   & $0.775 \pm 0.159$  & $ 0.8 \pm  0.2$ & $ 4.4 \pm  0.1$ & $ 0.62 \pm  0.18$ \\
3C123A   & $0.311 \pm 0.089$  & $ 1.2 \pm  0.7$ & $ 5.1 \pm  0.3$ & $ 0.38 \pm  0.24$ \\
3C123A   & $0.183 \pm 0.051$  & $ 0.8 \pm  0.4$ & $ 3.2 \pm  0.4$ & $ 0.16 \pm  0.10$ \\
3C123B   & $0.989 \pm 0.235$  & $ 1.1 \pm  0.6$ & $ 4.2 \pm  0.6$ & $ 1.14 \pm  0.72$ \\
3C123B   & $0.396 \pm 0.094$  & $ 1.1 \pm  0.7$ & $ 5.2 \pm  0.7$ & $ 0.47 \pm  0.30$ \\
3C123B   & $0.214 \pm 0.038$  & $ 0.9 \pm  0.5$ & $ 3.1 \pm  0.7$ & $ 0.21 \pm  0.11$ \\
3C154    & $0.825 \pm 0.064$  & $ 0.8 \pm  0.1$ & $-2.2 \pm  0.0$ & $ 0.68 \pm  0.09$ \\
3C154    & $0.194 \pm 0.039$  & $ 0.7 \pm  0.3$ & $-1.2 \pm  0.2$ & $ 0.15 \pm  0.07$ \\
3C454.3  & $0.018 \pm 0.001$  & $ 0.9 \pm  0.0$ & $-9.4 \pm  0.0$ & $ 0.02 \pm  0.00$ \\
3C454.3  & $0.006 \pm 0.001$  & $ 1.4 \pm  0.1$ & $-10.0 \pm  0.1$ & $ 0.01 \pm  0.00$ \\
\enddata
\end{deluxetable*}

\end{document}